# Toward Reliable $0\nu\beta\beta$-decay Nuclear Matrix Elements: Exploring the potential of measuring $\gamma\gamma$-transitions

Beatriz Romeo Zaragozano

*Supervisors:*
Dr. Javier Menéndez Sánchez
Dr. Carlos Peña Garay

*Tutor:*
Dr. Pilar Hernández Gamazo

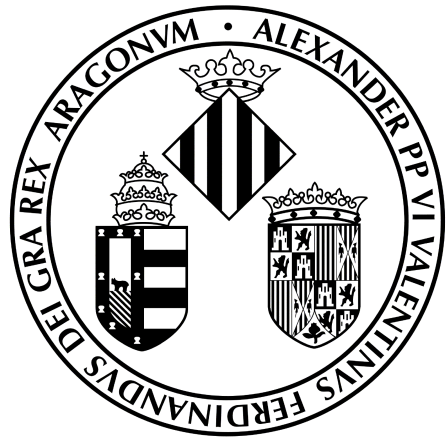

UNIVERSITAT DE VALÈNCIA

# Vniver§itat de València · Facultat de Física

## INFORME DEL DIRECTOR

**Javier Menéndez Sánchez**, investigador Ramón y Cajal en el Departmento de Física Cuántica de la University of Barcelona,

**Carlos Peña Garay**, Director del Laboratorio Subterráneo de Canfranc,

CERTIFICAN:

Que la memoria **Toward Reliable $0\nu\beta\beta$-decay Nuclear Matrix Elements: Exploring the potential of measuring $\gamma\gamma$-transitions** aquí detallada ha sido realizada bajo su dirección, por la estudiante **Beatriz Romeo Zaragozano** y constituye el trabajo de tesis doctoral por la que la estudiante opta al grado de Doctor en Física por la Universidad de Valencia.

Valencia, a 31 de octubre del 2023.

Javier Menéndez Sánchez                Carlos Peña Garay

The following individuals certify that they have read, and recommend to the Doctoral School of the Universitat de València for acceptance, the thesis entitled: **Toward Reliable 0νββ-decay Nuclear Matrix Elements: Exploring the potential of measuring γγ-transitions** submitted by Beatriz Romeo Zaragozano in partial fulfillment of the requirements for the degree of Doctor of Philosophy in Physics.

**Examining Committee:**

Dr. Juan José Gómez-Cadenas, Ikerbasque Research Professor, Donostia International Physics Center.

Dr. Jacobo López Pavón, Research Scientist, Department of Theoretical Physics, Universitat de València.

Dr. Jenni Kotila, Academy Researcher, Finnish Institute for Educational Research, University of Jyväskylä.

**Additional Supervisory Committee:**

Dr. Arnau Rios Huguet, Research Fellow, Department of Quantum Physics and Astrophysics, Universitat de Barcelona.

Dr. José Javier Valiente-Dobón, Research Scientist, Laboratori Nazionali di Legnaro, Istituto Nazionale di Fisica Nucleare.

Dr. Ana Isabel Morales López, Research Scientist, Instituto de Física Corpuscular, CSIC.

# Abstract

In this work, a physics process known since quite long ago, double-gamma decay ($\gamma\gamma$), has been revisited from a new perspective: providing valuable insights into neutrinoless double-beta decay ($0\nu\beta\beta$) nuclear matrix elements (NMEs). The yet undiscovered $0\nu\beta\beta$ decay has a unique physics potential since it is sensitive to the way of describing neutrinos and more fundamentally to the symmetries of nature.

At the same time that an eager experimental search of $0\nu\beta\beta$ decay is underway, the nuclear and particle physics communities have made huge progress during the last years. Such improvements have been centered developing the effective field theory (EFT) framework that describes this decay from the high-energy scale at which the new physics emerges to the low-energy scale where the nuclear $0\nu\beta\beta$ decay takes place. In parallel, the first calculations of more reliable $0\nu\beta\beta$-decay NMEs from *ab initio* many-body methods have recently become feasible. The EFT framework also provides a way to systematically improve interaction Hamiltonians rooted in the fundamental theory of quantum chromodynamics and when combined with an *ab initio* many-body method they allow one to quantify the uncertainty of $0\nu\beta\beta$-decay NMEs, a key theoretical aspect.

Currently, the uncertainty quantification of $0\nu\beta\beta$-decay NMEs is a thriving field, where several approaches are under active research. The main goal of this thesis has been to investigate one of those approaches which is the computation of nuclear observables related to $0\nu\beta\beta$ decay as a way to help in determining and reducing the theoretical uncertainties in $0\nu\beta\beta$-decay NMEs. This way, we have proposed that the measurement of the double magnetic dipole $\gamma\gamma$ decay from the double isobaric analog state in the $\beta\beta$-decay final nucleus could establish the value and reduce the uncertainty in $0\nu\beta\beta$-decay NMEs, because the NMEs of the two processes are very well correlated. We have explored the validity of this approach to predict and quantify NMEs uncertainties in the case where data is available for the nuclear observable related to $0\nu\beta\beta$ decay, that is for the Standard Model allowed $2\nu\beta\beta$ decay.

A further objective has been to start with the theoretical characterization of the first steps toward the measurement of the proposed $\gamma\gamma$ double magnetic dipole decay from the double isobaric analogue state. In particular, we have studied the main decaying modes that can compete with this process: single gamma decay and proton emission, and we have calculated the corresponding branching ratios.

Most of the computations presented in this thesis have been done in the nuclear shell model framework. In addition, we have also given the first steps in the study of the relation between $\gamma\gamma$ and $0\nu\beta\beta$-decay NMEs in the *ab initio* valence space in-medium similarity renormalization group. Therefore, apart of providing a new perspective of our study by adding new elements to investigate, such as isospin breaking effects, the valence space in-medium similarity renormalization group represents a new method to systematically improve the approximated solution of the many-body problem.

# Preface

The work presented in this thesis has been carried out fundamentally at the Laboratorio Subterráneo de Canfranc within the NEXT collaboration, and also during research visits to the Barcelona's Hadron, Nuclear and Atomic Physics group at the Departament de Física Quàntica i Astrofísica of the Universitat de Barcelona.

First, I would like to express my gratitude to my supervisors, Prof. Javier Menéndez and Dir. Carlos Peña, for their guidance throughout this work. To Carlos, for welcoming to the laboratory and for awaking in me the curiosity in the nuclear many-body problem and neutrinoless double beta decay. To Javier, for everything I have learned about nuclear physics and double beta decay, and also to propose interesting topics to explore and collaborations that have enriched this work. To both, for encouraging me during this journey and make me grow as a scientist, not to mention their infinite patience. I would like to thank to the members of the supervisory committee for they comments and suggestions to improve this work, and their encouragement at the different points of my thesis.

I am deeply grateful to Juanjo and Francesc, and to all the NEXT collaboration for hosting me, and to make possible I could return to theoretical physics at the same time I learned the hard work behind experimental physics. I owe big thanks to the ZULO's engineers team: Sara, Alberto, Vicente, Javi Rodriguez and Marc, and also to Raúl, José María Benlloch, Vicente Carrion and Ander, since everything I know of NEXT-White detector hardware, DAQ and proper running is a small part of their kindly shared knowledge. I would like to specially thanks Marc for the trust he put in me from the beginning, for having taught me how to operate the NEW gas system and Slow Controls, and for his support. With all of them I never feel alone while underground.

I would like to express my gratitude to the members of the Nuclear & Atomic Physics group at the Departament de Física Quàntica i Astrofísica for all their support throughout my research, and also to TRIUMF's theory group for wellcoming me during my visits and for involving me in their group seminars and meetings.

I want to thank Carmen Romo, for everything I have learned about sensor calibrations, but mainly for her valuable help and encouragement during the last period of my PhD.

To Lotta Jokiniemi and Pablo Soriano, for the meaningful discussions during our collaboration. To Takayuki Miyagi for helping me with VS-IMSRG code and speed up my learning during my visit to TRIUMF.

I am very grateful to all the LSC staff members for their kindness and help whenever I asked for it. Specially to Iulian and Javi with whom I passed many hours underground and introduced me into the radiopurity detection techniques and assessment. Always willing to help and advise me. To Javi, with whom I lived many NEXT adventures and who sometimes played a joke on me. To Mari, for her smile and *goizean* greetings. To Quim and Jorge, for that memorable trip to Gran Sasso. Quim, I will never forget that storm at Campo Imperatore and the lightning striking quite close to us! To both of you and also to Daria, Ana and Vicente, for the advise and to pushing me to write the thesis and not delaying it more.

Huge thanks to all my friends. I enjoyed very much all the moments I shared during the last years with all friends I made here in Canfranc, Bea, Toni, Pedro, Mabel, Marta, Nestor, Carlota and Asier. I am very lucky to have you in my life.

I would like to thank my family, my mother and father for their unconditional support, and my brother and sisters for being the motivation to keep me on this path.

Thank you Alex, one way or another I got here also because of you. There is no light without the dark, nor dark without the light, the force is in the balance.

Canfranc-Estación, October 2023.

# Contents

# List of Figures

# List of Tables

# Chapter 1

# Introduction

The description of the ordinary matter content of the universe is based on the Standard Model of particle physics [1], a high precision theory of the fundamental laws that govern particle interactions. Although all its particle content has been discovered and it has demonstrated to be an extremely successful theory pushing its capabilities up to distances $l \sim 10^{-17}$cm, some experimental data and observations can not be explained within this theory. For instance the discovery of neutrino oscillations [2–5] proved that neutrinos are massive particles in contrast to the predictions of the Standard Model. This has sparked our interest into the long standing problem of the nature of neutrinos, namely whether they are Dirac or Majorana particles, a puzzle that involves questioning fundamental symmetries.

Among the most promising processes to answer this question is neutrinoless double-beta decay ($0\nu\beta\beta$), a second order weak transition where two neutrons inside an atomic nucleus are transformed into two protons with the emission of two electrons and no antineutrinos. Therefore, an observation of this decay would show that lepton number is not conserved, violating a Standard Model's symmetry. Theoretical models that predict this hypothetical decay require that the neutrino mass has a Majorana component, and therefore that the neutrino is its own antiparticle. Thus, the discovery of $0\nu\beta\beta$ decay could shed light into the matter-antimatter asymmetry in the universe through leptogenesis [6, 7]. At the same time it could give valuable information on the neutrino mass.

These implications make clear the physics potential of $0\nu\beta\beta$ decay as a groundbreaking discovery, which has motivated an eager experimental search [8]. Although $0\nu\beta\beta$ has not been observed yet, with current experimental half-life lower limit of about $10^{26}$yr, there is an ambitious program that aims to push the limit up to $10^{28}$yr thanks to the development of detector technology, radiopurity of the materials and by increasing the mass of the $\beta\beta$ decaying isotope to the ton scale and beyond.

Meanwhile more data of the two-neutrino double beta decay ($2\nu\beta\beta$), the Standard Model allowed double beta decay, is starting to be accumulated mainly because $2\nu\beta\beta$ decay is an inevitable background for $0\nu\beta\beta$ searches. With this high statistics, measurements of the individual and summed electron spectra are making feasible tests of new physics through $2\nu\beta\beta$ decay.

Apart from the experimental effort, there is a pressing need for improving the theoretical predictions of the $0\nu\beta\beta$ rate. The half-life of this decay depends quadratically on the nuclear matrix element associated to the $0\nu\beta\beta$ transition operator, an input that must be calculated within a nuclear structure model. Although remarkable progress has been made in the computation of the nuclear matrix elements, still there is a considerable discrepancy on the predicted values with no clear guess of which are the true values and their uncertainty. All these aspects motivate our introductory review of the field in Chapter 2.

The discrepancy on the calculated $0\nu\beta\beta$-decay nuclear matrix elements reflects the difficulty of solving the nuclear many-body problem and the quantification of the uncertainties introduced by its approximate solution. As a physics process that occurs inside atomic nuclei, it is fundamental having a good description of such quantum system or at least having control over our ignorance about it.

Atomic nuclei are composed of protons and neutrons which are the effective degrees of freedom used in nuclear physics. However, they are not elementary particles, but as other hadrons they are complex bound systems of quarks and gluons which interact by the $SU(3)_c$ gauge theory of quantum chromodynamics (QCD). Although QCD is the fundamental theory of strong interactions, the strong coupling constant is of order 1 at the low energies where nuclear physics operates, making the perturbative treatment inapplicable.

A fundamental change in nuclear physics came when Steven Weinberg proposed to apply the concept of an effective field theory to QCD [9]. Taking nucleons and pions as degrees of freedom, he constructed the most general Lagrangian compatible with the symmetries of QCD, known as chiral effective field theory. This framework allows the construction of high-precision Hamiltonians rooted in QCD. On the other hand, the problem of solving the Schrödinger equation becomes increasingly difficult as the number of nucleons grow. We address these points in Chapter 3, where we discuss the nuclear many-body problem and the different nuclear Hamiltonians and many-body methods employed in this thesis.

The main result of this work is to propose a new way to improve the theoretical predictions of $0\nu\beta\beta$-decay nuclear matrix elements, namely through the measurement of double magnetic dipole transitions, $\gamma\gamma(M1M1)$, from the $0^+$ double isobaric analog state. In Chapter 4 we derive in detail the transition amplitude and the differential decay width for a general two photon decay as previously done in literature [10], and apply this formalism to $\gamma\gamma(M1M1)$ transitions from the $0^+$ double isobaric analog state. Furthermore, we discuss there the $\gamma\gamma$ decay experimental measurements and review the main experimental features of such second order electromagnetic decay.

The search of nuclear observables related to the $0\nu\beta\beta$ operator with the aim to reduce the uncertainty in $0\nu\beta\beta$-decay nuclear matrix elements has been previously studied theoretically for example with double Gamow-Teller transitions [11]. In Chapter 5 we discuss in detail the work published in Ref. [12], where we found a good linear correlation between $\gamma\gamma(M1M1)$ and $0\nu\beta\beta$-decay nuclear matrix elements calculated within the nuclear shell model for tens of nuclei with different many body interactions. Additionally, the first steps on the study of $\gamma\gamma(M1M1)$ using the *ab initio* valence space in-medium similiarity renormalization group method are presented in this chapter.

Moreover, Chapter 5 also presents the possible relation between $2\nu\beta\beta$- and $0\nu\beta\beta$-decay nuclear matrix elements, published in Ref. [13]. The focus is placed on the good linear correlation that could be used in combination with experimental data of $2\nu\beta\beta$ decay half-lives to give a prediction of $0\nu\beta\beta$-decay nuclear matrix elements. Further, one can associate a theoretical uncertainty that includes systematic calculations in tens of nuclei using different interactions. Further, Chapter 5 discusses briefly our predictions for the half-life of $2\nu\beta\beta$ decay of $^{136}$Xe to the first excited $0_2^+$ state of $^{136}$Ba published in Ref. [14].

Finally, Chapter 6 concludes with a summary of the main results presented in this thesis and an outlook over the possible future projects that could add further insights into the observed connection between $\gamma\gamma(M1M1)$-decay and $0\nu\beta\beta$-decay nuclear matrix elements. This chapter also discusses the way to complete the theoretical predictions valuable for the experimental realization of a $\gamma\gamma(M1M1)$ measurement.

# Chapter 2

# Double-Beta Decay

With the experimental observation of neutrino oscillations [2–5, 15–21] a solid path for new physics beyond the Standard Model of particle physics (SM) unfold. Neutrino masses and mixing play a key role in our understanding of flavour dynamics with relevant consequences in the content and evolution of the Universe. The neutrino mass origin and absolute scale are intrinsically related with the neutrino nature- whether they are Dirac or Majorana fermions-and violation of SM exact global symmetries. They represent fundamental open questions that have awaked an unprecedented interest in neutrinoless double-beta decay ($0\nu\beta\beta$), the lepton number violating (LNV) two nucleon decay $nn \rightarrow ppe^-e^-$, as the most feasible process to unravel this puzzle. Its observation would represent a matter-creation process, probing models that explain the matter-antimatter asymmetry in the Universe.

Since its theoretical proposal in 1939 [22] a challenging experimental search for this rare decay begun and currently is one of the most lively research programs in low energy physics. Experiments with a variety of double-beta decay isotopes and technologies have been developed and tested with half-live sensitivities exceeding $10^{26}$ yr. In the most simple mechanism the $0\nu\beta\beta$ decay is caused by the exchange of light Majorana neutrinos. In this case the half-life depends quadratically on what is know as the *effective neutrino mass* $m_{\beta\beta}$, a parameter which combines the neutrino masses $m_i$, the elements of the Pontecorvo-Maki-Nakawa-Sato (PMNS) mixing matrix ($U_{ei}$) and two Majorana phases. Although the parameters that describe massive neutrinos within the three-flavor mixing, such as square mass differences and mixing angles, had been determined by oscillation experiments at the percent level, it is still not clear what the ordering of neutrino masses is. There are two possibilities, either the masses follow a normal ordering (NO), that is $m_1 \ll m_2 < m_3$, or an inverted mass ordering (IO) with $m_3 \ll m_1 < m_2$. There are distinct allowed regions for $m_{\beta\beta}$ depending on the mass ordering being NO or IO, that experiment can probe if $0\nu\beta\beta$ decay is driven by light-neutrino exchange.

Future detectors at the ton scale aim to cover the entire IO mass region, $m_{\beta\beta} > (18.4 \pm 1.3)$meV. Therefore, a discovery will be possible if the spectrum is inverted or if $m_{\text{lightest}} >$ 50meV, independently of the ordering. In the case of a positive signal at the ton-scale, techniques developed in these experiments or at their parallel research and development (R&D) programs would be of paramount importance for a precision measurements in order to establish the decaying mechanism [23].

Along with this experimental plan there is an ambitious theoretical program with two main goals, one is to compute $0\nu\beta\beta$ rates with minimal model dependence and quantifiable theoretical uncertainties. This demands further progress in particle and nuclear effective field theories (EFTs), lattice quantum chromodynamics (LQCD) and nuclear few- and many-body *ab initio* methods. In parallel, the development of simplified models of LNV is encouraged, exploring the possibility of going beyond the Majorana neutrino-mass paradigm, and test them against the results of current and future $0\nu\beta\beta$ experiments, the Large Hadron Collider (LHC), and astrophysics and cosmology observations [24].

This chapter is organized as follows: first in Sec. 2.1 an overview of the neutrino problem is presented, focused mainly on the proposals of a neutrino mass term and their physics consequences as LNV that is allowed in models where neutrinos are Majorana particles. Then in Sec. 2.2 we introduce the $2\nu\beta\beta$ decay, and describe how current experiments could provide test of new physics beyond the SM. Also this overview is motivated by its connection with $0\nu\beta\beta$. In Sec. 2.3 we will discuss $0\nu\beta\beta$ theoretically for a better understanding of the experimental and theoretical challenges that physics community face today as described in Sec. 2.4. Finally, in Sec. 2.5 we explain the main difficulties and possible paths that are open for making more reliable predictions of $0\nu\beta\beta$-decay NMEs.

## 2.1   Neutrino nature: an elusive and ambitious puzzle

With the discovery of neutrino oscillations, the dynamical origin of neutrino masses became a pressing problem for the neutrino physics community. Today, fundamental questions such as neutrino mass ordering, determination of $CP$ violation phase $\delta$, how many neutrinos with definite masses ($\nu_i$) exist, and what is the nature of neutrino (Dirac or Majorana), remain to be answered. The present status of different neutrino parameters can be found in the latest global fit analysis in [25]. In our way of describing the fundamental laws of nature we have learned that it chooses the simplest possibility. This suggests the preferred answer to the last of these questions is that neutrinos are Majorana particles. However, there is not compelling evidence in favour of this hypothesis yet from current experimental measurements [25].

The Standard Model of particle physics is a quantum field theory based on the gauge symmetry group $SU(3)_C \times SU(2)_L \times U(1)_Y$ [1], where $C$ stand for color, $L$ for left-handedness, and $Y$ for hypercharge. This gauge group includes the symmetry group of strong interactions $SU(3)_C$, and the symmetry group of electroweak interactions $SU(2)_L \times U(1)_Y$ with the electromagnetic $U_{EM}$ as a subgroup. After the spontaneuos electroweak symmetry breaking the SM gauge group is $SU(3)_C \times U(1)_{EM}$.

In the SM, matter is composed by fermions which in relativistic quantum field theory can be described by two component Weyl spinors of opposite chirality $\psi_L$ and $\psi_R$ ($\gamma_5 \psi_{L,R} = \pm \psi_{L,R}$). Like other elementary particles, they transform under an irreducible representation of the Poincaré group, which in this case are the left and right chiral representation respectively.

Neutrinos, as neutral leptons are singlets under the subgroup $SU(3)_C \times U(1)_{EM}$. Usually, a distinction between *active* and *sterile* neutrinos is made. Active neutrinos, defined as those that take part in weak charged and neutral current interactions, are a component of the lepton doublet $\psi_{lL}^T = (\nu_l, l)_L$ ($l = e, \mu, \tau$). Sterile neutrinos do not have SM gauge interactions and therefore are not included in the SM. They would be singlets under the SM gauge group.

### 2.1.1 Extending Standard Model: neutrino masses

Fermion masses in the SM are generated via the Higgs mechanism, which is based on the assumption of Yukawa $SU(2)_L \times U(1)_Y$ invariant interactions. For leptons, such a term is [1]

$$\mathcal{L}_Y = -\sum_{l_i l_j} \overline{\psi}_{l_i L}(x) Y_{l_i l_j} l'_{l_j R}(x) \phi(x) + \text{h.c},  \tag{2.1}$$

where $\phi$ is the scalar Higgs doublet, $\phi(x)^T = (\phi_+(x), \phi_0(x))$, $\psi_{lL}^T(x) = (\nu'(x)_{lL}, l'(x)_L)$ and $l'(x)_R$ are the left and right-handed charged lepton fields. After spontaneous symmetry breaking this term leads to

$$\mathcal{L}_Y = -\sum_{l_i l_j} \overline{l}'_{l_i L}(x) Y_{l_i l_j} l'_{jR}(x)(v + h(x)) + \text{h.c},  \tag{2.2}$$

with the Higgs vacuum expectation value $v = (\sqrt{2} G_F)^{-1/2} \simeq 246 \text{GeV}$ and $h$ the Higgs field in unitary gauge. The $3 \times 3$ complex Yukawa matrix can be diagonalized by the unitary transformations $V_{L,R}$

$$Y = V_L y V_R^\dagger,  \tag{2.3}$$

and $y_{ll'} = y_l \delta_{ll'}$ ($l, l' = e, \mu, \tau$). Therefore the mass term is

$$\mathcal{L}_{m_l} = -\sum_{l=e,\nu,\tau} m_l \overline{l}(x) l(x), \qquad m_l = y \frac{v}{\sqrt{2}},  \tag{2.4}$$

with $l(x) = l_L(x) + l_R(x) = \sum_{l_i} (V_L)_{ll_i} l'_{iL}(x) + \sum_{l_i} (V_R)_{ll_i} l'_{iR}(x)$. The Yukawa constants are determined from the Higgs boson decays into fermion-antifermion pairs (as given by the second term in Eq. (2.2)) which is in good agreement with the SM predicted values from Eq. (2.4). The same mechanism works for quarks masses.

In principle, neutrino masses can also be generated by the standard Higgs mechanism

$$\mathcal{L}_Y^\nu = -\sum_{l_i l_j} \bar{\phi}_{l_i L}(x) Y_{l_i l_j}^\nu \nu_{l_j R}(x) \widetilde{\phi} + \text{h.c.},  \tag{2.5}$$

where $\widetilde{\phi}(x) = i\tau_2 \phi^*(x)$ is the conjugate Higgs doublet, $\nu_{lR}(x)$ are right-handed singlets (sterile-neutrinos), and $Y^\nu$ the $3 \times 3$ matrix of Yukawa couplings. After electroweak symmetry breaking this term can be written as

$$\mathcal{L}_Y^\nu = -\sum_{l_i l_j} \bar{\nu}_{l_i L}(x) Y_{l_i l_j}^\nu \nu_{l_j R}(x)(v + h(x)) + \text{h.c.}.  \tag{2.6}$$

Again, the term proportional to $v$ is the neutrino mass term, which after diagonalization of $Y^\nu$ leads to

$$\mathcal{L}_Y^\nu = -\sum_{i=1}^{3} m_i \bar{\nu}_{iL}(x) \nu_{iR}(x) + \text{h.c.} = -\sum_{i=1}^{3} m_i \bar{\nu}_i(x) \nu_i(x),  \tag{2.7}$$

with $\nu_i(x) = \nu_{iL}(x) + \nu_{iR}(x)$ and $m_i = y_i^\nu v / \sqrt{2}$ the neutrino mass. $\nu_i(x)$ is a Dirac field and the mass term in Eq. (2.7) is known as Dirac mass term. The Lagrangian $\mathcal{L}_{SM} + \mathcal{L}_Y^\nu$ is invariant under the global transformation of lepton fields $\nu_i(x) \rightarrow e^{i\alpha}\nu_i(x)$ and $l_i(x) \rightarrow e^{i\alpha}l_i(x)$. Hence, the model preserves total lepton number $L$ with $L(\nu_i) = 1$ for neutrinos and $L(\bar{\nu}_i) = -1$ for antineutrinos.

Although absolute neutrino masses are currently unknown, from existing neutrino oscillation experiments and cosmological observations a conservative bound can be set

$$5 \cdot 10^{-2} \text{eV} \simeq \sqrt{|\Delta m_{3l}^2|} \leq m_3 \leq \sum_i m_i \simeq 0.1 \text{eV},  \tag{2.8}$$

where $\Delta m_{3l}^2 \equiv m_3^2 - m_l^2 \simeq 2.510(-2.490) \cdot 10^{-3} \text{eV}^2$ is the neutrino mass squared difference for the NO ($l = 1$) and IO ($l = 2$) as reported in Ref. [25]. The sum $\sum m_i < 0.12(0.15)\text{eV}$ for NO(IO) constitute the stronger but model dependent[1] upper bound obtained from cosmological observations such as cosmic microwave (CMB), baryon acoustic oscillation (BAO) and supernovae Ia luminosity data [26, 27]. Then we have for the Yukawa coupling $y_3 \sim 3 \cdot 10^{-13}$, which is about 10 orders of magnitude smaller than the Yukawas of the third generation of SM charged fermions: $y_t \simeq 7 \cdot 10^{-1}$, $y_b \simeq 2 \cdot 10^{-2}$, and $y_\tau \simeq 7 \cdot 10^{-3}$. This could be a possibility technically natural in the 't Hooft [28] sense, since making $y_3 \rightarrow 0$ increases the symmetry of the theory, but it is not completely satisfactory because explaining so tiny neutrino masses suggests that their generation mechanism should be different that the rest of fermions. Despite this, there are theoretical models that aim to understand the dynamical origin of a very light Dirac neutrino mass, among others Dirac seesaw mechanism [29], radiative models [30, 31] or extra-dimension theories[32].

---

[1] All the bounds to $\sum m_i$ that come from cosmological observations are based on the cosmological $\Lambda$CDM model.

### 2.1.2 On the smallness of neutrino mass

A general mechanism to explain the smallness of neutrino mass beyond the SM is the introduction of the Weinberg effective operator [33] or higher dimension operators. A different mass term for neutrinos can be constructed without the introduction of right-handed neutrinos and it is based on the observation that the conjugate field $\nu_{lL}^c = C\bar{\nu}_{lL}^T$, transforms as a right-handed field [34] with $C$ the charge conjugation matrix. This fact allows one to write a different mass term built only from left-handed neutrino fields

$$\mathcal{L}^M = -\sum_{l',l} \bar{\nu}_{l'L} M_{l'l}^M \nu_{lL}^c + \text{h.c.,} \qquad (2.9)$$

where $M^M$ is a complex and non-diagonal $3 \times 3$ matrix. $\mathcal{L}^M$ is known as the *Majorana mass term* and violates total lepton number by two units. The matrix $M^M$ can be diagonalized by a unitary transformation $M^M = U m U^T$ with $m_{ij} = m_i \delta_{ij}$, so

$$\mathcal{L}^M = -\sum_{i=1}^{3} m_i \bar{\nu}_i \nu_i, \qquad (2.10)$$

with the neutrino field

$$\nu_i = \sum_l U_{il}^\dagger \nu_{lL} + \sum_l (U_{il}^\dagger \nu_{lL})^c, \qquad (2.11)$$

satisfying the Majorana condition $\nu_i = \nu_i^c$. The flavor neutrino fields are the mixed fields

$$\nu_{lL} = \sum_{i=1}^{3} U_{li} \nu_{iL}, \qquad (2.12)$$

and $U$ is the Pontecorvo-Maki-Nakagawa-Sakata (PMNS) mixing matrix [1]. At this point neutrino masses $m_i$ are parameters and there is no explanation for their small values. Here is where the Weinberg operator could provide a mechanism to explain the smallness of $\nu$-masses within the framework of effective field theory (EFT). In the effective Lagrangian [33]

$$\mathcal{L}_{\text{eff}} = -\frac{1}{\Lambda} \sum_{l',l} \bar{\psi}_{l'L} \widetilde{\phi} O'_{l'l} \widetilde{\phi}^T \psi_{lL}^c + \text{h.c.,} \qquad (2.13)$$

the operator within the sum has dimension 5, and therefore $\Lambda$ has dimensions of energy. The parameter $\Lambda$ represents the scale of new physics, and after the spontaneous symmetry breaking the neutrino mass term is

$$\mathcal{L}_{\text{eff}}^M = -\frac{v^2}{2\Lambda} \sum_{l',l} \bar{\nu}_{l'L}' N_{l'l}' \nu_{lL}'^c + \text{h.c.,} \qquad (2.14)$$

or in terms of flavour fields $\nu_{lL} = \sum_{l'}(V_L^+)_{ll'}\nu_l'$ and after diagonalization of the mass matrix $N = V_L^+ N'(V_L^+)^T$ through the unitary PMNS mixing matrix U, $N = UnU^T$, one has

$$\mathcal{L}_{\text{eff}}^M = -\sum_{i=1}^3 m_i \bar{\nu}_i \nu_i,\tag{2.15}$$

with $m_i = v^2/(2\Lambda)n_i$ and $\nu_{lL} = \sum_{i=1}^3 U_{li}\nu_{iL}$. Then, neutrino masses generated by the effective Lagrangian of Eq. (2.13) are naturally suppressed with respect to SM masses by a factor $v/\Lambda$. As before, an order of magnitude estimation for the new physics parameter gives $\Lambda \sim 10^{15}$GeV which makes lepton number violation physics far beyond experimental reach.

There are many models which can explain the origin of the Weinberg operator. For example through the introduction of heavy Majorana leptons with invariant $SU(2)_L \times U(1)_Y$ interaction. This mechanism is known as the type-I seesaw mechanism [35–39]. Another possible interaction that can generate this dimension-5 operator is the interaction of a heavy triplet scalar boson field with a pair of lepton double and pair of Higgs doublet fields (type-II seesaw [40–43]). Finally, the Weinberg effective Lagrangian can be explained through the exchange of heavy virtual Majorana triplet leptons between lepton-Higgs pairs (type-III seesaw [44]).

In all these models $\Lambda$ is equal to the masses of the corresponding seesaw counterpart. However, the fact that the new physics scale $\Lambda \sim 10^{15}$GeV needs to be very large motivates non trivial radiative neutrino mass models where the new physics can be detected at much lower energy scales [45]. This can be bypassed if instead of tree-level seesaw realizations one introduces new symmetries which are softly broken in such a way that the Weinberg operator is forbidden at tree level but it is allowed at certain loop level $n_l$ with a suppression factor $\sim 1/(16\pi^2)^{n_l}$. Therefore, with a suitable $n_l$ one can shift the new phenomenology scale down to TeV scale. Another possibility is to maintain the symmetries that prevent the appearance of dim-5 operator but allow higher dimension operators $(5 + n)$ $(n = 2, 4, ...)$ which give a suppression factor $(v/\Lambda)^n$.

A completely different approach to explain neutrino mass comes from the spontaneous breaking of chiral symmetry via quark-antiquark condensate [46, 47]. These models assume the existence of new scalar interactions between Majorana neutrinos and quarks in vacuum as well as in nuclear medium, and the formation of the quark condensate transform the interaction term into a mass term for Majorana neutrinos.

## 2.2  Two-neutrino Double-Beta decay

Double beta-decay processes are sensitive probes to new physics beyond the SM. The non observation of $0\nu\beta\beta$ decay put a lower bound to the half-life $T_{1/2}^{0\nu}$ which can be translated to an upper limit range on $m_{\beta\beta}$ if we assume that the dominant mechanism is the light neutrino exchange and the extreme predictions of $0\nu\beta\beta$ NMEs. At the same time, the eager search for the hypothetical $0\nu\beta\beta$ decay is making possible to explore high precision properties of its allowed SM counterpart, the two-neutrino double beta decay ($2\nu\beta\beta$). Furthermore, our recent work in [48] shows for the first time that $2\nu\beta\beta$ and $0\nu\beta\beta$ nuclear matrix elements (NMEs) follow a good linear correlation. When this correlation is combined with the measurement of the $2\nu\beta\beta$ decay, it is possible to make a prediction of $0\nu\beta\beta$ decay with theoretical uncertainties based on the systematics of many NMEs calculations.

Two-neutrino double beta decay is one of the rarest processes observed with half-lives $T_{1/2} \sim 10^{21}$ yr [49]. It is a second-order weak process where a nucleus with mass and atomic number $(A, Z)$ changes into its isobar with $(A, Z + 2)$ by emitting two electrons and two electron-type antineutrinos, conserving the lepton number:

$$(2\beta^-)_{2\nu}: \quad (A, Z) \rightarrow (A, Z + 2) + 2e^- + 2\bar{\nu}_e \,. \tag{2.16}$$

It was proposed by Maria Goeppert-Mayer in 1935 [50] who also estimated its half-life $T_{1/2}^{2\nu} > 10^{17}$ yr. Soon after the Fermi weak interaction theory appeared, Eugene Wigner suggested to consider this process as Maria Goeppert-Mayer pointed out in [50]. For a detailed historic discussion of double-beta decay see for example Ref. [51]. Due to nuclear pairing interaction, this decay can be measured in even-even nuclei in which other decay modes such as single beta decay or electron capture, are energetically forbidden or strongly suppressed by a big change in angular momentum. Among the 35 $2\nu\beta\beta$ unstable nuclei [52], only a few are interesting for experimental searches (see Sec. 2.4 for details).

The first detection of $2\nu\beta\beta$ decay was reported for $^{130}$Te in 1950 by a geochemical experiment which determined an isotopic excess of $^{130}$Xe in an old tellurium-ore [53]. It was not until 1987 when a direct counting experiment reach enough sensitivity to measure this decay in $^{82}$Se [54] using a Time Projection Chamber (TPC). Currently, the half-life of this decay has been measured in 11 nuclei [49]: $^{48}$Ca, $^{76}$Ge, $^{82}$Se, $^{96}$Zr, $^{100}$Mo, $^{116}$Cd, $^{130}$Te, $^{136}$Xe, $^{150}$Nd, $^{238}$U by direct counting experiments, and in $^{128}$Te, and $^{238}$U by geochemical and radiochemical indirect searches. It also has been measured in the $0_1^+$ excited states of $^{100}$Ru, $^{150}$Sm and $^{136}$Ba in the double beta decay of $^{100}$Mo, $^{150}$Nd and $^{136}$Xe respectively [49, 55].

Despite being one of the rarest processes ever observed, precision measurements allow one to test nuclear structure predictions and also to search for physics beyond the SM, such as violation of Lorentz and CPT symmetries [56, 57] or the existence of right-handed neutrinos since these extensions could affect the energy distribution and angular correlation of the outgoing electrons [58, 59].

### 2.2.1   The $2\nu\beta\beta$-decay rate and nuclear matrix elements

Within the SM, the $2\nu\beta\beta$ is mediated by the weak V-A current-current effective interaction Hamiltonian

$$H_W^{SM} = \frac{G_F \cos\theta_C}{\sqrt{2}} j_{L\mu}(x) J_L^{\mu\dagger}(x) + \text{h.c.} \qquad (2.17)$$

where $G_F$ is the Fermi coupling constant ($G_F = 1.16637 \times 10^{-5} \text{GeV}^{-2}$) and $\theta_C$ is the mixing angle of Cabibbo-Kobayashi-Maskawa mechanism for mixing quark flavours ($\cos\theta_C = 0.97373 \pm 0.00031$) [1]. Implicit in Eq. (2.17) is the assumption that the momentum transfer between hadrons and leptons is small compared with the W-boson mass ($m_W$), which is a good approximation for nuclear $\beta$ decay.

In the case of $(2\beta^-)_{2\nu}$ the left-handed leptonic current $j_{L\mu}$ is given by

$$j_{L\mu}(x) = \bar{e}(x)\gamma_\mu(1 - \gamma_5)\nu_e(x), \qquad (2.18)$$

where at this point the light electron neutrino can be thought of as a 4-component spinor field. Although the treatment of light neutrino as a Dirac (a spinor constructed from a combination of the SM $\nu_L$ and a new SM-sterile right-handed neutrino $\nu_R$) or Majorana fermion (a spinor constructed from the SM active left-handed neutrino $\nu_L$ and its charge-conjugate $\bar{\nu}_L^c$) has important consequences for the underlying theory, it does not play any role in the dominant contribution to $2\nu\beta\beta$ decay.

One is used to see the description of the weak interaction involving hadrons at the level of quarks and gluons, however the genuine description of weak interaction of hadrons would need the complete knowledge of the wave function of quarks inside them, a complex problem which has not been solved yet. Currently, state of the art LQCD calculations have demonstrated a per-cent level determination of the nucleon axial coupling $g_A$ in neutron decay with controlled uncertainties [60]. Moreover, as we will see in Sec.3.3, the exact solution of the A-body nuclear bound system for $A \gtrsim 12$ depending on the computed observable, is even a Dantesque task. For this reason in the usual treatment of nuclear beta decay one assumes that at the moment of the decay, the decaying nucleon feels only the weak interaction and does not interact strongly with the rest of the nucleons inside the nucleus [61]. Therefore, at the moment of the interaction this *'active'* nucleon is free and only in the initial and final nuclear states interacts strongly. This approximation is known as the *impulse approximation*.

A usual description of the hadronic current comes from a phenomenological approach guided by symmetry principles. The most general current operator written in terms of Dirac fermion bilinears, $\gamma^\mu$-matrices and the transferred 4-momenta $p$, that transform as

a Lorentz vector is

$$J_L^{\mu\dagger}(x) = \Psi(x)\tau^- \left[ g_V(p^2)\gamma^\mu + ig_M(p^2)\frac{\sigma^{\mu\nu}}{2m_N}p_\nu - g_A(p^2)\gamma^\mu\gamma^5 - g_P(p^2)p^\mu\gamma^5 \right] \Psi(x),$$
(2.19)

where $p^\mu = (p_n - p_p)^\mu$ ($p_n$ and $p_p$ are the four momenta of neutron and proton, respectively) is the momentum transferred from hadrons to leptons and $\Psi$ is a nucleon isospin-doublet and $\tau^-$ is the isospin lowering operator. This approximation can be improved by adding two-body or meson-exchange currents as it is done systematically by chiral EFT [62]. In Eq.(2.19) invariance under parity and time-reversal has been also imposed.

The functions $g_V$, $g_M$, $g_A$ and $g_P$ could be any real functions of a Lorentz scalar constructed from the momenta $p_n$ and $p_p$, but the usual choice is in terms of $p^2$. They are known as the vector, magnetic, axial and pseudoscalar form factors, and their values at zero momentum transfer limit, $p^2 \to 0$, are the vector, magnetic, axial and pseudoscalar couplings. The magnetic and pseudoscalar couplings are related to the vector and axial ones by means of the conserved vector current (CVC) and the partially conserved axial currrent (PCAC) hypothesis [63, 64]. Further, the CVC hypothesis yields $g(0) = 1$. Both in nuclear $\beta$ or $2\nu\beta\beta$ decay, only the vector and axial form factors are considered due to the small momentum transfer at the weak vertex of a few MeV, which is the typical Q-value for these nuclear transitions.

The momentum dependence of the form factors assumed in neutrino involved processes takes the dipolar [65] form

$$g_V(p^2) = \frac{g_V(0)}{\left(1 + \frac{p^2}{\Lambda_V^2}\right)^2}, \quad g_A(p^2) = \frac{g_V(0)}{\left(1 + \frac{p^2}{\Lambda_A^2}\right)^2},$$
(2.20)

which is a reasonable good approximation below $p^2 < 1\,\mathrm{GeV}^2$ [66]. The dipole form factors result from the identification in the non-relativistic limit of the form factors with the Fourier transform of an exponential charge and magnetic moment distribution, $\rho(r) = \rho_0 e^{-\Lambda_X r}$, with free adjustable parameters $\Lambda_{V,A}$ called vector and axial masses (or dipole masses). From nuclear $\beta$ decay $g_A(0) = 1.2756$ [67] and from $eN$ and $\nu N$ scattering data dipole masses lie in the range $1.1m_N \lesssim \Lambda_A \lesssim 1.7m_N$ and $\Lambda_V = 0.84m_N$ [65, 68, 69], where $m_N$ is the nucleon mass.

Finally, as we have said earlier, in $2\nu\beta\beta$ the momenta of both electrons and neutrinos are restricted to the $Q$-value (of a few MeV) and it is a good approximation to take the leading terms in the non-relativistic approximation of the weak hadronic current

$$J_L^{\mu\dagger}(x) = \sum_{n=1}^A \tau_n^- \delta(\mathbf{x} - \mathbf{r}_n) \left( g_V(p^2)g^{\mu 0} + g_A(p^2)\sigma_n^k g^{\mu k} \right).$$
(2.21)

As a second order electroweak process the matrix element for the transition amplitude can be calculated using perturbation theory as

$$\mathcal{M}^{2\nu} \equiv \langle e_1 e_2 \bar{v}_1 \bar{v}_2 N_f | \mathcal{S}^2 | i \rangle = \frac{(-i)^2}{2} \int d^4 x \, d^4 y \, \langle e_1 e_2 \bar{v}_1 \bar{v}_2 N_f | \mathcal{T}[H_I(x) H_I(y)] | N_i \rangle, \quad (2.22)$$

with $\mathcal{T}[H_I(x) H_I(y)] = \theta(x_0 - y_0) H_I(x) H_I(y) + \theta(y_0 - x_0) H_I(y) H_I(x)$ the time-ordered product, and $H_I$ is the effectve interaction defined in Eq. (**??**). Further, $|N_{i,f}\rangle$ are the initial and final nuclear states and $e_{1,2}$, $\bar{v}_{1,2}$ the emitted electrons and electron-antineutrinos.

Taking into account the temporal evolution of the wave functions and currents, one finds, after time integration of Eq.(2.22), that

$$\mathcal{M}^{2\nu} = 2\pi\delta(E_{e_1} + E_{e_2} + E_{\bar{v}_1} + E_{\bar{v}_2} + E_f - E_i) \left( \frac{G_F \cos\theta_C}{\sqrt{2}} \right)^2 \left( \frac{1}{\sqrt{2}} \right)^2 [1 - P_{e_1, e_2}] [1 - P_{\bar{v}_1, \bar{v}_2}]$$
$$\times \int d\mathbf{x} \, d\mathbf{y} \, [\bar{\psi}(p_{e_1}, \mathbf{x}) \gamma_\mu (1 - \gamma_5) \psi^c(p_{\bar{v}_1}, \mathbf{x})] \, [\bar{\psi}(p_{e_2}, \mathbf{y}) \gamma_\nu (1 - \gamma_5) \psi^c(p_{\bar{v}_2}, \mathbf{y})]$$
$$\times \left[ \sum_n \frac{\langle N_f | J^\mu(0, \mathbf{x}) | N_n \rangle \langle N_n | J^\nu(0, \mathbf{y}) | N_i \rangle}{E_n - E_i + E_{e_2} + E_{\bar{v}_2}} + \frac{\langle N_f | J^\nu(0, \mathbf{y}) | N_n \rangle \langle N_n | J^\mu(0, \mathbf{x}) | N_i \rangle}{E_n - E_i + E_{e_2} + E_{\bar{v}_2}} \right].$$
$$(2.23)$$

Here, $E_i$, $E_f$ and $E_n$ are the energies of the initial, final and intermediate nuclear states, $E_{e_i} = \sqrt{\mathbf{p}_{e_i}^2 + m_e^2}$ and $E_{\bar{v}_i} = \sqrt{\mathbf{p}_{\bar{v}_i}^2 + m_v^2}$ correspond to the energy of the emitted electrons and antineutrinos, respectively. The operator $P_{ab}$ permutes the particles $a$ and $b$ and is introduced to fulfill the antisymmetry requirements for identical fermions, i.e. for electrons and antineutrinos in the final state. Moreover, $\psi(p, \mathbf{x})$ refers to the spinor wave function with four momentum $p = (E, \mathbf{p})$ at the spacial coordinate $\mathbf{x}$.

Defining $\mathcal{M}^{2\nu} = 2\pi\delta(E_{e_1} + E_{e_2} + E_{\bar{v}_1} + E_{\bar{v}_2} + E_f - E_i)\mathcal{R}^{2\nu}$ and focusing on $0^+ \to 0^+$ transitions one has [70]

$$\mathcal{R}^{2\nu} = i\frac{1}{2} \left( \frac{G_F \cos\theta_C}{\sqrt{2}} \right)^2 [1 - P_{e_1, e_2}] [1 - P_{\bar{v}_1, \bar{v}_2}]$$
$$\times [\bar{\psi}(p_{e_1}) \gamma^\mu (1 - \gamma_5) \psi^c(p_{\bar{v}_1}) \bar{\psi}(p_{e_2}) \gamma^\nu (1 - \gamma_5) \psi^c(p_{\bar{v}_2})]$$
$$\times \left[ g_V^2 g_{\mu 0} g_{\nu 0} \sum_n \frac{M_F(n)}{E_n - E_i + E_{e_2} + E_{\bar{v}_2}} + \frac{1}{3} g_A^2 \sum_n \frac{M_{GT}(n)}{E_n - E_i + E_{e_2} + E_{\bar{v}_2}} \right] \quad (2.24)$$

where the $M_F(n)$ and $M_{GT}(n)$ are the nuclear matrix elements for a given intermediate state

$$M_F(n) = \langle 0_f^+ || O_F || 0_n^+ \rangle \langle 0_n^+ || O_F || 0_f^+ \rangle, \quad (2.25)$$

$$M_{GT}(n) = \langle 0_f^+ || O_{GT} || 1_n^+ \rangle \cdot \langle 1_n^+ || O_{GT} || 0_f^+ \rangle. \quad (2.26)$$

Here $O_F$ and $O_{GT}$ are the Fermi and Gamow-Teller operators defined as

$$O_F = \sum_{i=1}^{A} t_i^-, \quad O_{GT} = \sum_{i=1}^{A} t_i^- \boldsymbol{\sigma}_i, \tag{2.27}$$

which are written in terms of the isospin-lowering $t_i^-$ and spin $\boldsymbol{\sigma}_i$ operators for the $i$-nucleon.

Finally, the differential decay rate for this decay can be calculated using the second order Fermi's Golden Rule

$$d\Gamma^{2\nu} = 2\pi \sum_{spin} |\mathcal{R}^{2\nu}|^2 \delta(E_{e_1} + E_{e_2} + E_{\bar{\nu}_1} + E_{\bar{\nu}_2} + E_f - E_i) d\Omega_{e_1} d\Omega_{e_2} d\Omega_{\nu_1} d\Omega_{\nu_2}, \tag{2.28}$$

with $d\Omega_i = d^3\boldsymbol{p}_i / (2\pi)^3$. After integrating over the phase-space of the outgoing neutrinos, the differential decay rate as a function of the electrons energies ($0 \leq E_{e_i} \leq Q + m_e$) and the angle $\theta$ between the electrons momenta [70]

$$\frac{d\Gamma^{2\nu}}{dE_{e_1} dE_{e_2} d\cos\theta} = c_{2\nu} \left( A^{2\nu}(E_{e_1}, E_{e_2}) + B^{2\nu}(E_{e_1}, E_{e_2}) \cos\theta \right), \tag{2.29}$$

with

$$c_{2\nu} = \frac{G_F^4 \cos^2\theta_C m_e^9}{8\pi^7}. \tag{2.30}$$

The functions $A^{2\nu}(E_{e_1}, E_{e_2})$ and $B^{2\nu}(E_{e_1}, E_{e_2})$ are phase-space integrals depending on electron energies obtained by integrating over the neutrino allowed energies, $E_{\nu_i} \in (m_{\nu_i}, E_i - E_f - E_{e_1} - E_{e_2})$ (see App. A.1 for details). These phase space factors depend on products of rather involved combinations of fermion radial wave functions and the nuclear matrix elements $M_\alpha^{K,L}$, $\alpha = $ F, GT (see App. A.2), which are defined as

$$M_\alpha^{K,L} = m_e \sum_{n,\eta} M_\alpha(n) \frac{\epsilon_{\alpha,n}}{\epsilon_{\alpha,n}^2 + (-1)^\eta \epsilon_{K,L}^2}, \tag{2.31}$$

where $\eta = 1, 2$ and $J_n^\pi = 0_n^+ (1_n^+)$ for Fermi and Gamow-Teller operators respectively, and the energies $\epsilon_{\alpha,n}$ and $\epsilon_{K(L)}$ are defined by

$$\epsilon_{\alpha,n} = E_n(J_n^\pi) - \frac{1}{2}(E_i + E_f), \tag{2.32}$$

$$\epsilon_{K(L)} = \frac{1}{2}(E_{e_2(e_1)} + E_{\nu 2} - E_{e_1(e_2)} - E_{\nu_1}), \tag{2.33}$$

with $\epsilon_{K(L)} \in (-Q/2, Q/2)$.

The Fermi and Gamow-Teller operators are generators of isospin SU(2) and spin-isospin

SU(4) symmetries, respectively. If both symmetries were exact in nuclei, then $2\nu\beta\beta$ decay would be forbidden as ground states of initial and final nuclei would belong to different multiplets. However, the nuclear interaction has an approximately good isospin symmetry but the SU(4) spin-isospin is not. This is the reason why the Fermi matrix element is usually neglected [71] in comparison with the Gamow-Teller matrix element. The total isospin-lowering operator $T^- = \sum_{i=1}^{A} t_i^-$ connects states in the same isospin multiplet, since the initial and final $\beta\beta$ decay nuclei have different isospin then $M_F \simeq 0$ to the amount of the mixing caused by the Coulomb isotensor term in the interaction. For a recent quantification of this effect see [72].

### 2.2.2   A simplified $2\nu\beta\beta$ decay half-life

In order to simplify the calculation of the differential $2\nu\beta\beta$ decay angular distributions and total rate, the energies $\epsilon_{K,L}$ are usually neglected. This allow one to completely decouple the nuclear matrix elements (NMEs) from the leptonic phase-space integral. Due to the isospin symmetry, the Fermi transition is highly suppressed and only the Gamow-Teller NME is considered

$$M^{2\nu} = M_{GT} = m_e \sum_n \frac{\langle 0_f^+ || \sum_i t_i^- \boldsymbol{\sigma}_i || 1_n^+ \rangle \cdot \langle 1_n^+ || \sum_j t_j^- \boldsymbol{\sigma}_j || 0_f^+ \rangle}{E_n(1_n^+) - \frac{1}{2}(E_i + E_f)}. \tag{2.34}$$

In this case $M^{2\nu}$ does not depend on the lepton energies and the phase-space integral, $G_{2\nu}$ [73, 74], can be factorized out, and the decay half-life is

$$\left[ T_{1/2}^{2\nu} \right]^{-1} = G_{2\nu} g_A^4 |M^{2\nu}|^2. \tag{2.35}$$

Note that this is a convenient way to write the half-life since the units of the phase-space factor $G^{2\nu}$ are $y^{-1}$, the NME $|M^{2\nu}|^2$ as deffined in Eq. (2.34) is dimensionless and the the axial vector coupling has been factorized out making more explicit the distinction between the hadron coupling and the many-body physics present in the NME $M^{2\nu}$. Then, from the experimental measurement of $2\nu\beta\beta$ decay one can extract the so-called effective matrix element [49]

$$M_{\text{eff}}^{2\nu} = g_A^2 M^{2\nu}, \tag{2.36}$$

which incorporates all the nuclear information. In $M^{2\nu}$ there is a sum over all $1^+$ virtual intermediate states of the intermediate nucleus. A fundamental question here is how much is the contribution of the higher-lying intermediate states to the $M^{2\nu}$, which are not so well known. In the 1980s, Abad and collaborators suggested in Ref. [75] that for those $2\nu\beta\beta$ transitions where the ground state of the intermediate nucleus is a $J^\pi = 1^+$ state, the transition could be dominated by virtual transitions through this state, therefore driven by Gamow-Teller transitions. First studies of this hypothesis in [76] showed that single-state dominance (SSD) hypothesis is realized in two different ways, either through a true $1_1^+$ dominance or by a cancellation among contributions of higher lying $1^+$ states of the intermediate nucleus. This hypothesis can be extended to include lower-lying states (LLD) [77].

Experimental studies of the SSD hypothesis were done either in $2\nu\beta\beta$ decay and in GT strenght measurements coming from charge-exchange reactions in $^{100}$Mo and $^{116}$Cd [78, 79]. Furthermore, in Ref. [80] it was shown that one could confirm or rule out experimentally the SSD hypothesis by measuring the single electron energy distribution and the angular correlation of the two outgoing electrons in $2\nu\beta\beta$ decay. Motivated by this study, experimental evidence of SSD for $^{82}$Se and $^{100}$Mo has been found in the CUPID-0 [81], the CUPID-Mo [82] and the NEMO-3 [83] experiments.

The presence of the energy denominator in Eq. (2.34) reduces the dependence on the not so well known higher energy intermediate states. This effect is clearly observed when one calculates the cumulative contribution to $M^{2\nu}$ as a function of the intermediate state excitation energy in the intermediate nucleus and compare it with its matrix element without the energy denominator both in the quasiparticle random phase approximation(QRPA) [84] and in the nuclear shell model [85]. From this running sum, one also sees that there are states that have a positive contribution at first, and above some excitation energy between $5 - 10$ MeV negative contributions decrease markedly the NMEs. Such higher-lying $1^+$ states which contribute with opposite sign could lead to cancellations and therefore could validate the SSD or LLD hypothesis. This behaviour of the cumulative sums seems to be universal for different nuclei and also when computed with different many-body methods [84]. The origin of such negative contributions is studied in Ref. [86] within QRPA and is associated with the enhancement of the ground-state correlations that produce a change in the sign in GT transitions of the higher-lying states.

### 2.2.3 Theoretical over-predictions of $2\nu\beta\beta$ matrix elements

It has long been known that predictions of the $\beta$-decay half-lives from calculated GT NMEs sistematically underestimate experimental observations, making evident the need of a correction [87]. A way to bypass this problem was to introduce a factor $q$, called "quenching factor", in the GT operator ($\sigma\tau \rightarrow q\sigma\tau$) to fix the mismatching between the overestimation of the NME and the observation. An alternative approach was to attribute the quenching to the axial coupling constant $g_A \rightarrow g_A^{\text{eff}} = qg_A$. However, it is not clear that this over-estimations of the NMEs can be encoded in a single parameter and further investigation is needed to identify possible systematic effects of the approximated calculations.

The origin of the quenching factor has been a subject of long and lively debate. The physics behind this factor could be the missing nuclear correlations due to the truncated configuration space used to calculate nuclear wave-functions and the neglected contributions to the transition operator from meson-exchange currents (two-body currents (2BC)) [62]. In Ref. [88] GT transitions in heavy nuclei $\beta$ decay were addressed with *ab initio* methods for the first time, showing that when one includes nuclear correlations and 2BC, experimental and theoretical GT NMEs are in good agreement. The authors also found the relative importance of both effects.

Since the GT operator in $\beta$ decay requires a quenching factor for phenomenological models, it is reasonable to assume that $M^{2\nu}$ is also quenched. The naive approximation for this quenching in $2\nu\beta\beta$ decay is $q^2$. This approach was used to correct shell model calculations and the results are in good agreement with the experimental measurements [89–94].

### 2.2.4 Extensions to the standard $2\nu\beta\beta$ decay NMEs

Today, we live in a stimulating time for double beta decay physics since the improvements in the precision of current and future experiments allow us both to test well established theories and models at the same time that demand more accurate theoretical predictions not only within the SM but also beyond.

First, inside the SM, an improvement in the accuracy in the calculation of NMEs can be obtained making a Taylor expansion of the energy denominator of $M_{GT}^{K,L}$ in Eq.(A.11) in terms of the small ratio $\delta_{GT,n}^{K,L} = \epsilon^{K,L}/\epsilon_{GT,n}$ [95]. Truncating the series expansion of the NMEs in Eq. (A.11) at fourth order in $\delta_{GT,n}^{K,L}$ one gets

$$M_{GT}^{K,L} = m_e \sum_n \frac{M_{GT}(n)}{\epsilon_{GT,n}} \left\{ 1 + (\delta_{GT,n}^{K,L})^2 + (\delta_{GT,n}^{K,L})^4 + \mathcal{O}(\delta_{GT,n}^{K,L})^6) \right\}. \tag{2.37}$$

In Ref. [95], contributions to the total $2\nu\beta\beta$-decay rate were calculated, and the results show that corrections to the $2\nu\beta\beta$ decay rate, although dependent on the isotope can be up to 25%. The authors also found that matrix elements with increasing power of the energy denominators decrease in magnitude: $M_{GT,-1} > M_{GT,-3} > M_{GT,-5}$, with $M_{GT,-q} = \sum_n M_{GT}(n)2^{q-1}m_e^q \epsilon_{GT,n}^{-q}$. Further, when analysing the running sum of $M_{GT,-3}$ the largest contributions come from the low-energy states of the intermediate nucleus contrary to $M_{GT,-1}$ which also has important contributions of high-energy states, but for the latter there are cancellations among them. These corrections also induce modifications in the shape of the single and summed electron energy distributions making possible the determination of the ratio $M_{GT,-3}/M_{GT,-1}$ if the $2\nu\beta\beta$-decay experiment reach enough statistics.

On the other hand, with the increasing precision in $2\nu\beta\beta$-decay experiments, it starts to be feasible making tests of its decay rate due to physics beyond the SM. One of such examples is the test of the violation of the fundamental symmetry of Lorentz invariance (LIV). LIV is a characteristic feature that can be accommodated by many candidate theories of quantum gravity [96–98]. The minimal (or conventional) realization of such spacetime symmetry violation is by means of the spontaneous symmetry breaking mechanism. The general EFT that incorporates operators that break Lorentz invariance to the SM is the Standard Model Extension (SME) [99].

The high sensitivity of neutrino oscillation experiments have drawn the attention to study neutrino sector of SME [56]. However, there are some LIV effects known in the literature as *countershaded* effects, that can not be observed using neutrino mixing observables and in that cases, weak decays offer a valuable tool [100]. In particular, these effects are governed by the LIV coefficient $a_{\text{of}}^{(3)}$. This coefficient enters in a modified neutrino 4-momentum and its corresponding effect is a distortion of the standard electron summed spectrum. Since in $2\nu\beta\beta$-decay experiments the two neutrinos are not detected, only the isotropic coefficient $\mathring{a}_{\text{of}}^{(3)}$ survives in the decay rate.

## 2.3 Neutrinoless Double-Beta decay

Neutrinoless double beta decay ($0\nu\beta\beta$) is a weak second-order decay analogous to $2\nu\beta\beta$ but without the neutrino emission:

$$(A, Z) \rightarrow (A, Z + 2) + 2e^-, \tag{2.38}$$

which change lepton number (L) in two units while keeping baryon number (B) unchanged. Since B-L is conserved in the SM, an observation of $0\nu\beta\beta$ decay would necessarily imply new physics beyond the SM.

Although the detection of $0\nu\beta\beta$ decay, as showed by Schechter and Valle [101], would imply that neutrinos are massive Majorana particles, several lepton-number violating mechanisms could be the dominant cause of this decay. Therefore, the connection between the $0\nu\beta\beta$ decay half-live or its limit and neutrino mass can only be made under a particular assumption for the source of LNV. The general way to write all the *i*-mechanism contributions to the decay half-life is [62, 87]

$$\left[T_{1/2}^{0\nu}\right]^{-1} = \sum_i G_i g_i^4 |\mathcal{M}_i|^2 f_i(\Lambda) + \text{interference terms}, \tag{2.39}$$

factorized as the direct product of four factors: **i)** the phase-space factor $G_i$ which takes into account the kinematics of the decay and depends mainly on $Q_{\beta\beta}$ and the atomic number of the initial nucleus; **ii)** $g_i$ is the hadronic matrix element which describes the coupling to the nucleon; **iii)** the nuclear matrix element $\mathcal{M}$ which encodes the nuclear structure of the initial and final nuclear states as well as the transition operator written in terms of nucleons degrees of freedom and finally, **iv)** $f_i$ is the dimensionless new-physics parameter distinctive of each LNV mechanism.

The standard assumption is that the dominant source of LNV at low energies is the Majorana neutrino mass $(-\overline{\Psi} M_\nu \Psi^C / 2)$. In the minimal version of this model we only have the SM neutrinos $(\nu_{L,i})$ and for the light neutrino exchange one usually defines [62]

$$m_{\beta\beta} = \left| \sum_{i=1}^{3} |U_{ei}|^2 e^{i\varphi_i} m_i \right|, \tag{2.40}$$

as the *effective Majorana mass* of $\nu_e$, which is the *ee*-element of the mass matrix $M_\nu = U \text{diag}(m_1, m_2, m_3) U^T$. Here $U$ is the Pontecorvo-Maki-Nakagawa-Sakata (PMNS) neutrino mixing matrix and $\varphi = \{\varphi_1, \varphi_2, 1\}$ are known as Majorana phases.

The $0\nu\beta\beta$ decay half-life in this case is [62]

$$\left[ T_{1/2}^{0\nu} \right]^{-1} = G_{01} g_A^2 |\mathcal{M}_{\text{ligth}}^{0\nu}|^2 \frac{m_{\beta\beta}^2}{m_e^2}, \tag{2.41}$$

where $\mathcal{M}_{\text{ligth}}^{0\nu}$ is the light neutrino exchange NME and $G_{01}$ is the phase space factor. It is clear from Eq. (2.41) that if the experimental physics community can measure neutrinoless double beta decay together with the improvements in the measurement of PMNS mass matrix, and if the nuclear theory community can determine both the phase space factor and the NME, then one can determine the absolute neutrino mass. However, both the experimental and theory sides face great challenges as we will in Sec. 2.4. In fact, one of the most pressing problems that the nuclear theory community is dealing with in the last decade is the discrepancy of a factor of about 3-5 in the NMEs predictions from different many-body methods and by different groups.

### 2.3.1   $0\nu\beta\beta$ NMEs

Assuming that the light Majorana-neutrino exchange is the dominant $0\nu\beta\beta$ decay mechanism, the NME has two components [102]

$$\mathcal{M}_{\text{ligth}}^{0\nu} = M_L^{0\nu} + M_S^{0\nu},$$

where for any mechanism responsible for the decay $M_L^{0\nu}$ consists of three parts: Fermi (F), Gamow-Teller (GT) and tensor (T) (see for example [87] and references therein)

$$M_L^{0\nu} = M_{\text{GT}}^{0\nu} - \frac{g_V^2}{g_A^2} M_{\text{F}}^{0\nu} + M_{\text{T}}^{0\nu}, \tag{2.42}$$

each one associated with the spin-space operators $\mathcal{O}_{ij}^{\text{F}} = \mathbb{I}$, $\mathcal{O}_{ij}^{\text{GT}} = \boldsymbol{\sigma}_i \cdot \boldsymbol{\sigma}_j$, $\mathcal{O}_{ij}^{\text{T}} = 3(\boldsymbol{\sigma}_i \cdot \hat{\mathbf{r}}_{ij})(\boldsymbol{\sigma}_j \cdot \hat{\mathbf{r}}_{ij}) - \boldsymbol{\sigma}_i \cdot \boldsymbol{\sigma}_j$ acting on the $(i,j)$<sup>th</sup>-nucleon space and spin variables which come in

the definition

$$
\begin{aligned}
M_\alpha^{0\nu} &= \sum_{ij} \langle 0_f^+ || \mathcal{O}_{ij}^\alpha \, t_i^- t_j^- \, H_\alpha(r_{ij}) \, f_{\text{SRC}}^2(r_{ij}) || 0_i^+ \rangle, \\
&= \sum_{abcd} \sum_{JJ'} \langle ab : J || \mathcal{O}_{ij}^\alpha \, t_i^- t_j^- \, H_\alpha(r_{ij}) \, f_{\text{SRC}}^2(r_{ij}) || cd : J' \rangle \\
&\quad \times \langle 0_f^+ || \left[ [c_a^\dagger \otimes c_b^\dagger]_J \otimes [\widetilde{c}_c \otimes \widetilde{c}_d]_{J'} \right]_0 || 0_i^+ \rangle,
\end{aligned}
\tag{2.43}
$$

being $r_{ij} = |\boldsymbol{r_i} - \boldsymbol{r_j}|$ the distance between two nucleons and $\hat{\boldsymbol{r}}_{ij} = \boldsymbol{r}_{ij}/r_{ij}$. The neutrino potentials, $H_\alpha(r_{ij})$, are integrals over the neutrino exchange momenta $q$, and $f_{\text{SRC}}$ represent the effects of short range correlations [103], which are only needed if the initial and final wave-functions are obtained with a many-body method which neglects such corrections. The second line of Eq. (2.43) is the NME in the occupation number representation, with the first term the two-body NME (TBNME) while the second term is the two-body transition density (TBTD).

Implicit in Eq. (2.43) is the *closure approximation*. The general expression for $M_\alpha^{0\nu}$ involves a sum over intermediate states $|k\rangle = |J_k^\pi\rangle$ with a given angular momentum $J_k$, parity $\pi$ and energy $E_k$. Then instead of Eq. (2.43) one has

$$
\begin{aligned}
M_\alpha^{0\nu} &= \sum_{k,abcd} \langle ab : J || \mathcal{O}_{ij}^\alpha \, t_i^- t_j^- \, H_\alpha(r_{ij}, E_k) \, f_{\text{SRC}}^2(r_{ij}) || cd : J \rangle \langle 0_f^+ || [c_a^\dagger \otimes c_b^\dagger]_J || J_k^\pi \rangle \\
&\quad \times \langle J_k^\pi || [\widetilde{c}_c \otimes \widetilde{c}_d]_J || 0_i^+ \rangle.
\end{aligned}
\tag{2.44}
$$

However, corrections beyond the closure approximation appear at higher order according to chiral EFT [102] in agreement with actual calculations [104].

The neutrino potentials, $H_\alpha(r_{ab}, E_k)$ are defined as

$$
H_\alpha(r_{ij}, E_k) = \frac{2R}{\pi} \int_0^\infty dq \, \frac{q^2 j_\lambda(qr_{ij}) h_\alpha(q^2)}{q(q + E_k + (E_i + E_f)/2)},
\tag{2.45}
$$

where $R = 1.2 A^{1/3}$fm and the functions $j_\lambda(qr_{ij})$ are the spherical Bessel's functions with $\lambda = 0$ for Fermi and Gamow-Teller components, while $\lambda = 2$ for the tensor component. The functions $h_\alpha(q^2)$ include the form factors that incorporate the finite-size effects and higher-order currents of nucleons [105],

$$
h_{\text{F}}(q^2) = \frac{g_V^2(q^2)}{g_V^2},
\tag{2.46}
$$

$$
h_{\text{GT}}(q^2) = \frac{g_A^2(q^2)}{g_A^2} \left[ 1 - \frac{2}{3}\frac{q^2}{q^2 + m_\pi^2} + \frac{1}{3}\left( \frac{q^2}{q^2 + m_\pi^2} \right)^2 \right] + \frac{1}{6}\frac{g_M^2(q^2)}{g_A^2}\frac{q^2}{m_p^2},
\tag{2.47}
$$

$$
h_{\text{T}}(q^2) = \frac{g_A^2(q^2)}{g_A^2} \left[ \frac{2}{3}\frac{q^2}{q^2 + m_\pi^2} - \frac{1}{3}\left( \frac{q^2}{q^2 + m_\pi^2} \right)^2 \right] + \frac{1}{12}\frac{g_M^2(q^2)}{g_A^2}\frac{q^2}{m_p^2}.
\tag{2.48}
$$

The $g_{V,A,M}$ form factors are usually parameterized using the dipole approximation

$$g_V(q^2) = \frac{g_V}{\left(1 + \frac{q^2}{\Lambda_V^2}\right)^2}, \quad g_A(q^2) = \frac{g_A}{\left(1 + \frac{q^2}{\Lambda_A^2}\right)^2}, \quad g_M(q^2) = 1 + \frac{\mu_p - \mu_n}{\mu_N} g_V(q^2), \quad (2.49)$$

and their values in the zero momentum limit are known as the vector, axial-vector and weak-magnetism coupling constants, respectively. In the way they are written in Eq. (2.47)-(2.48), terms proportional to $q^2/(q^2 + m_\pi^2)$ come from the pseudoscalar coupling $g_P(q^2)$ which is given by the partially conserved axial-vector current hypothesis (PCAC) [106]

$$g_P(q^2) = 2m_N g_A \frac{q^2}{q^2 + m_\pi^2} \quad (2.50)$$

In Eqs. (2.46)-(2.48), $m_{p,\pi}$ are the proton and pion masses, while the values for the other constants are $g_A = 1.27$, $g_V = 1$, $\Lambda_V = 850$ MeV, $\Lambda_A = 1086$ MeV and $\mu_p - \mu_n = 3.7\mu_N$.

A way of simplifying the computation of the TBNMEs is to make the transition operators independent of the intermediate nuclear state, a dependence that enters through their energy denominator. Therefore making the replacement in Eq. (2.45) $E_k + (E_i + E_f)/2 \rightarrow E_{cl}$ the operators in Eq.(2.44) become energy independent as in Eq. (2.43). This approximation is well justified since the momentum transfer to the neutrino $q$ is of the order of $100 - 200$ MeV, while the relevant transition energies are of the order of 10 MeV [107]. In [107], they calculated $0\nu\beta\beta$ NMEs in $^{48}$Ca both with and without the closure approximation and found a $\sim 10\%$ smaller NMEs within the second approximation. The authors also point out that variations in $E_{cl}$ slightly yield a small change in the NME with an estimated uncertainty around 10% [108]. In the results presented in this thesis the values chosen for the closure energy are either 0 or $E_{cl} = q(q + 1.12A^{1/2}\text{MeV})$ [71]. This approximation also avoids the computation of a large set of intermediate states in the intermediate nucleus which could be particularly challenging in heavy nuclei.

The additional function $f_{SRC}$ in Eq. (2.43) is introduced in the radial wave function to take into account the short-range correlations that are missing in a truncated Hilbert space within the many-body calculation. Short-range correlation can be calculated using realistic NN interactions, usually the charge-dependent Bonn potential (CD-Bonn) [109] and the Argonne V18 potential (Argonne) [110]. In [103] they use the coupled-cluster method (CCM) to derive the set of parameters that enter in the analytic form of a Jastrow-like function

$$f_{SRC}(r_{ij}) = 1 - ce^{-ar^2}(1 - br^2), \quad (2.51)$$

with parameters $a = 1.52(1.59)$ fm$^{-2}$, $b = 1.88(1.45)$ fm$^{-2}$ and $c = 0.46(0.92)$ for CD-Bonn and Argonne NN interactions respectively. However, for the case of light neutrino mixing mechanism the effect of the short-range correlations in $0\nu\beta\beta$-NMEs deviates within a few percent with respect to the bare wave functions [103, 111]. Nonetheless, a more recent work suggests a higher effect from short-range correlations [72]. This issue will be further investigated in the near future.

Finally, a short-range operator induced by light Majorana neutrinos was introduced quite recently in Ref. [102]. Within the framework of chiral EFT, the authors of Ref. [102] demonstrate using two different regularization schemes the need to promote this contact operator to leading order only for spin-single S-wave transitions [112], because otherwise the observables depend on the regulator. The new two-body operator NME is given by

$$M_S^{0\nu} = \frac{2R}{\pi g_A^2} \langle 0_f^+ || \sum_{ij} \tau_i^- \tau_j^- \int j_0(qr_{ij}) h_S(q^2) q^2 dq || 0_i^+ \rangle \tag{2.52}$$

with neutrino potential

$$h_S(q^2) = 2g_\nu^{NN} e^{-q^2/(2\Lambda_S)}. \tag{2.53}$$

The coupling $g_\nu^{NN}$ depends on the Gaussian regulator $\Lambda_S$ and can only be calculated from first principles in lattice QCD (see [113] and references therein) because of the unavailable lepton-number violating data. There are other approaches to estimate its value as in [112] using symmetry arguments which relate LNV interactions and electromagnetic charge-independence breaking (CIB). A similar approach was followed in [114] to compute $M_S^{0\nu}$ with $g_\nu^{NN}$ taken from the CIB term of different effective Hamiltonians [115, 116]. First values for $g_\nu^{NN}$ from approximate QCD calculations have been obtained recently [113].

## 2.4 Experimental and theoretical challenges

The long-term $0\nu\beta\beta$ experimental program that has been planned and evaluated over the years [8, 23, 62] highlights the potential of a new discovery that would push forward our understanding of the fundamental principles of nature.

The major challenge that the next generation of $0\nu\beta\beta$ experiments is committed to face is to fully cover the IO region of the neutrino mass spectrum, that is, going below $(m_{\beta\beta}^{min})_{IO} = 18.4 \pm 1.3$ meV. For this purpose, several detection techniques applied to different isotopes are being developed and tested in order to identify the best approaches to pass from a few hundred of kilograms to ton-scale detectors over longer running times. Furthermore, the use of different isotopes, apart of enriching the experimental approaches, could help in establishing the underlying $0\nu\beta\beta$ decay mechanism and also avoiding accidental false signals due to unknown background.

A simple order of magnitude estimation of the signal rate to explore the inverted mass ordering region gives approximately $\sim 10^{-3}$counts yr$^{-1}$kg$^{-1}$. Therefore, the main requirements that guide experimentalists in selecting an isotope and a particular technology in order to have enough sensitivity to reach the above-mentioned goal are: i) very low background index, below $10^{-4}$counts keV$^{-1}$yr$^{-1}$kg$^{-1}$, ii) an excellent energy resolution in order to discriminate between signal and background[2] at the region of interest[3] (ROI), typically of a few keV and iii) a large isotopic mass and scalability [8]. Some of these requirements are interconnected and also related with three properties of a double-beta decay isotope: the $Q$-value, the isotopic abundance and the ease of enrichment [117]. The above considerations make that among the isotopes that can undergo $0\nu\beta\beta$ decay only a few have been selected for experimental searches. The current experimental limits to the lower bounds for $T^{0\nu}_{1/2}$ at 90% confidence level (CL) are in the range $10^{21} - 10^{26}$ y [62]. Among the main experimental techniques we find high-purity Ge (HPGe) detectors [118–120], loaded organic liquid scintillators [121–123], croyenic bolometers [124–126], time projection chambers (TPCs) [127, 128] and tracking chambers [129, 130]. A detailed overview of recent and future experiments can be found in the reviews [8, 62].

The physics reach of an experiment is usually specified in terms of the sensitivity to discover a signal or *discovery sensitivity*. In [131] this parameter is defined as the value of $T_{1/2}$ or $m_{\beta\beta}$ for which the experiment has a 50% chance to measure a signal above the background with a statistical significance of $3\sigma$. In turn, the *exclusion sensitivity* is defined as the expected number of signal events that an experiment has 50% probability of excluding at 90% CL assuming a null observation. In fact, first one computes the sensitivity for $T^{0\nu}_{1/2}$ given an expectation for the background counts in the ROI, and then the result is converted to a range of $m_{\beta\beta}$ using Eq. (2.40) for the different ranges of NMEs.

A simplified statistical analysis is based on two parameters, which are the *sensitive exposure* ($\mathcal{E}$) and the *sensitive background* ($\mathcal{B}$). $\mathcal{E}$ is defined as the product of the source mass and the exposure time, but it is corrected by the signal detection efficiency. $\mathcal{B}$ is determined from the expected number of background events in the ROI divided by $\mathcal{E}$. Therefore, the sensitivity to $T^{0\nu}_{1/2}$, $\mathcal{S}(T^{0\nu}_{1/2})$, is given by [131, 132]

$$\mathcal{S}(T^{0\nu}_{1/2}) = \ln 2 \frac{N_A \mathcal{E}}{W S_{3\sigma}(b)},\qquad(2.54)$$

where $N_A$ is the Avogadro constant, $W$ the atomic weight and $S_{3\sigma}(b)$ is the Poisson signal expectation at which 50% of the measurements would report a $3\sigma$ positive fluctuation above $b = \mathcal{E}\mathcal{B}$.

---

[2]The most relevant contribution to the background is the intrinsic $2\nu\beta\beta$ decay whose discrimination demands an energy resolution below few percent. Natural radiation from $^{238}$U and $^{232}$Th chains can also produce events with similar energy deposition, particularly important are the $\beta$ decays from $^{214}$Bi with a Q-value of $Q_\beta = 2.25$ MeV and the maximum energy of the $\gamma$ radioactivity at 2.615 MeV. Other important irreducible background are neutrinos coming from the sun.

[3]The ROI is usually the 1FWHM around $Q_{\beta\beta}$, because the finite energy resolution of a detector is modeled so that single events are fitted to gaussians.

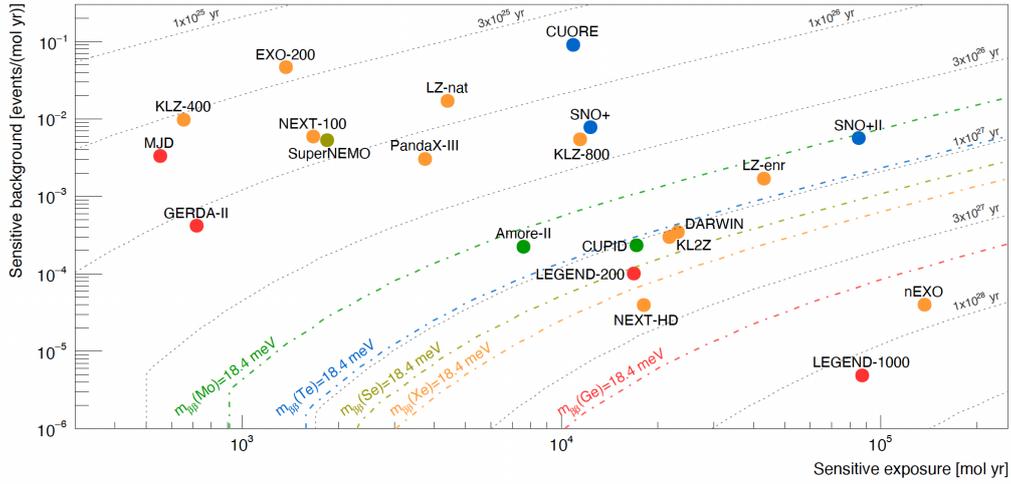

FIGURE 2.1: Sensitive background and exposure for recent and future experiments. The grey dashed lines indicate some discovery sensitivity values on the $0\nu\beta\beta$ decay half-life. The colored dashed line indicate the half-life sensitivities required to test the bottom of the inverted ordering scenario for $^{76}$Ge (red), $^{136}$Xe (orange), $^{130}$Te (blue) $^{100}$Mo (green), and $^{82}$Se, assuming for each isotope the largest NME value among the QRPA calculations listed in Tab. I in [62]. A lifetime of 10 yr is assumed except for completed experiments, for which the final reported exposure is used. Figure taken from [62].

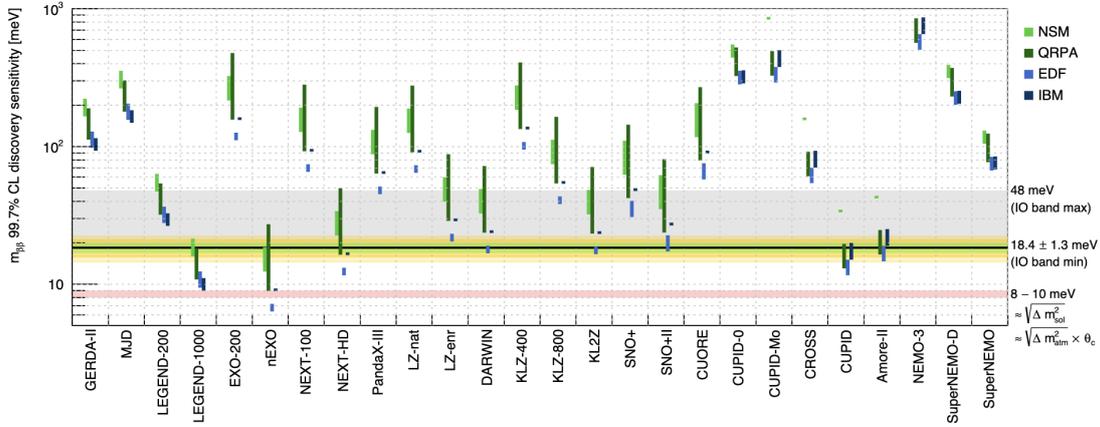

FIGURE 2.2: $m_{\beta\beta}$ discovery sensitivity of current and next-generation $0\nu\beta\beta$ decay experiments assuming dominance of the light-neutrino exchange mechanism. The minimum value of $m_{\beta\beta}$ for the IO and its $1\sigma$, $2\sigma$, and $3\sigma$ uncertainty bands are indicated by the black, green, orange, and yellow bands, respectively. Figure taken from [62].

Figure 2.1 taken from [62] shows an enlightening way to compare recent, ongoing and future experimental expectations. This figure clearly displays the evolution of each technological approach as well as the huge challenge that represents going beyond the inverted mass ordering (IO) region. Improvements in the $\mathcal{E}/\mathcal{B}$ ratio depend on the isotope and technology. For example, increasing the exposure is easier for $^{130}$Te with a high natural abundance, while decreasing the background rate is easier for $^{76}$Ge detectors due to the higher energy resolution. Figure 2.2 from [62] also shows the updated $m_{\beta\beta}$ sensitivity of current and next-generation $0\nu\beta\beta$ decay experiments assuming the dominance of the light-neutrino exchange mechanism. It is clear from this figure that a positive signal will be observed by future experiments if the inverted mass ordering is the scenario realized in nature. Then the next step would be to identify the $0\nu\beta\beta$ decay mechanism. For this, precision measurements of energy and angular distribution of the emitted electrons in different isotopes would be needed.

On the theoretical side, both the particle physics and nuclear theory communities also face a challenging program since a proper interpretation of experimental results demands reliable theoretical predictions. The particle physics community is focused on reducing the lepton number violating processes responsible for the decay. Searches of other probes such as low-energy neutrino experiments, high energy colliders, astrophysics and cosmology observations are proposed to complement the results of $0\nu\beta\beta$ experiments. Meanwhile the nuclear theory community focuses on the hadronic and many-body part which enters in the nuclear matrix element. These two pieces are linked but their main improvements can be worked separately.

Although both theoretical ingredients are fundamental, with no $0\nu\beta\beta$ observation and as long as theoretical predictions of the new physics parameter are unknown, NMEs are an input that must be calculated within a nuclear structure model. Hence, either a $0\nu\beta\beta$ observation or a limit will demand an accurate calculation of the NME with a minimal model dependence and quantified uncertainty to extract a quantitative information of neutrino masses or the LNV parameters of other BSM models. To accomplish this multi-scale goal, efforts have been directed in the advance of lattice QCD, effective field theories (EFTs), and *ab-initio* many-body methods.

In Ref. [62] the authors show an updated comparison of the long-range contribution for the light-neutrino exchange NMEs within several many-body methods and variants of them as shown in Figure 2.3. The methods range from the long established nuclear shell model (NSM), the quasiparticle random-phase approximation (QRPA) method, the interacting boson model (IBM) or the energy-density functional theory (EDF). New ab-initio results in $^{48}$Ca with in-medium generator coordinate method (IM-GCM), multi-reference similarity renormalization group (IMSRG), and coupled cluster (CC) theory are also shown. Finally, Fig. 2.3 presents the results from valence-space IMSRG (VS-IMSRG) method for $^{48}$Ca, $^{76}$Ge and $^{82}$Se.

The discrepancy in the calculated NMEs of a factor around three translates to a factor nine (nearly an order of magnitude) in the mass of the isotope needed to overcome the inverse mass ordering limit on the effective neutrino mass parameter. Moreover, there is no clear guess of which calculation is near the true value. This uncertainty also makes difficult the selection of the $\beta\beta$-isotope since a higher matrix element corresponds to a lower half-life and there is no clear difference due to the spread in the NMEs. This stresses the uncertainties introduced by the approximate solution to the nuclear many-body problem as well as the need for developing an error quantification program. However, there is a general pattern in Fig. 2.3: shell model NMEs tend to be the smallest, while EDF theory results are the largest, being IBM and QRPA generally in between. On the other hand, different ab initio predictions are compatible in $^{48}$Ca, while in $^{76}$Ge and $^{82}$Se they are smaller than the rest of the calculations. Thus, one can conclude that phenomenological NMEs could be overestimated. This deficiency can be traced back to missing correlations not captured by phenomenological approaches with a truncated model space or without the inclusion of non-nucleonic degrees of freedom such as many-nucleon currents [87]. Moreover, in $^{48}$Ca and $^{76}$Ge the NSM and some QRPA NMEs are not so far from ab initio results. Since ab initio methods when applied to $^{48}$Ca predict quite well $\beta$-decay matrix elements without any adjustments, this suggest that the overestimation of the NMEs in these phenomenological models attributed to a quenching in axial-vector coupling is less important.

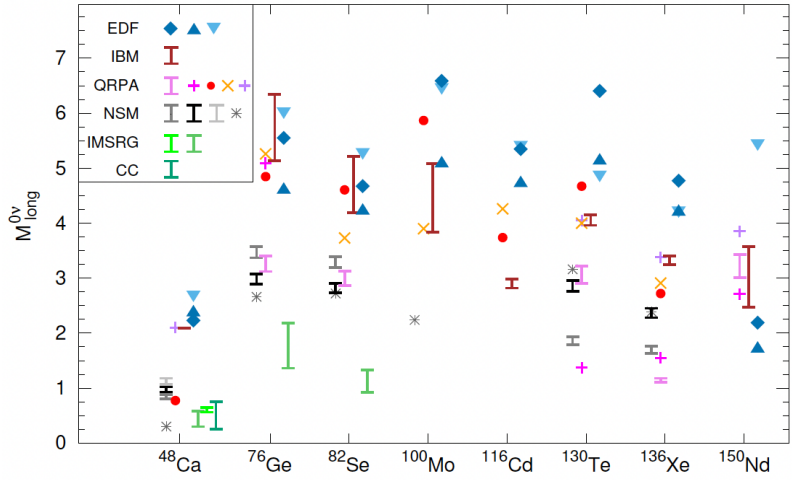

FIGURE 2.3: Nuclear matrix elements of $M^{0\nu}$ for the standard component of the light neutrino exchange from different many-body methods. Figure taken from [62].

This problem shows us that uncertainty quantification (UQ) is fundamental for many reasons mentioned in this section, and efforts towards developing such field within $0\nu\beta\beta$ decay started almost a decade ago [133, 134]. As stated in Ref. [135] the goal of the UQ is not a precise evaluation of the physics that is missing but estimating the range of probable values for the matrix element and doing it in a statistically meaningfull way.

The first method one can think of UQ is just taking the parameters of the nuclear model, estimate their errors, and then consistently propagate them through the model to asses the uncertainty in its predictions [135]. For example, in phenomenological models such as NSM or EDF theory, uncertainties can be partially estimated [62] by usually varying the nuclear Hamiltonian. Other, though smaller, uncertainties, come from the inclusion of SRCs. A first statistical model within the NSM has been recently proposed in [136] for analysing the distribution $0\nu\beta\beta$-NMEs and has been applied to $^{48}$Ca, proposing a mean value and a range at 90% CL for its NME. Ab initio approaches are able to give a UQ in a systematic way thanks to their connection with EFT methods (see the promising results in Ref. [137]).

## 2.5  Paths towards reliable NMEs

A complete understanding of $0\nu\beta\beta$ experiments demands accurate predictions of the decay rate which in turn depends quadratically on the nuclear matrix elements independently of the new physics mechanism that could originate such decay. The problem of computing the nuclear matrix elements entails different aspects of our physics description of the whole process that by their own are not well determined, like the nuclear interaction or the physics mechanism that causes $0\nu\beta\beta$ decay. The work of calculating nuclear matrix elements demands having control of the physics at a wide energy range, from high energies where the LNV originates to the low energies where the bound nucleons undergo $0\nu\beta\beta$ decay. Although further development of particle-physics models of LNV is needed, an accurate prediction of the nuclear matrix element with controlled uncertainties is imperative to test all the models against current and future data. This pressing problem has originated an intense and enlightening study of ways in which experiments could help and provide guidance in predicting more reliable $0\nu\beta\beta$ NMEs through the measurement of other nuclear observables.

A proper description of the nuclear states involved in the transition is fundamental to be confident that the structure of the wave functions can account for different spectroscopic properties [138–141]. For instance, several experiments have studied the occupancies of valence orbitals in the initial and final ground states, analysed through single-nucleon adding or removing reactions. In fact, these experimental results have been used to improve the description of the initial and final $\beta\beta$ decay states, and the effect is to bring a bit closer QRPA and NSM NMEs [62]. There are many other spectroscopic measurements that are helpful not only to benchmark the low-lying spectrum against predictions of many-body methods, but also to provide new angles and hints about nucleon pairing correlations that are relevant for $0\nu\beta\beta$ NMEs [142].

Going further, processes such as ordinary muon capture (OMC) to the intermediate nuclei of $\beta\beta$ decay can access the region of neutrino momentum transfer similar to $0\nu\beta\beta$ decay ($p \sim 100\text{MeV}$) [143–145], making OMC a useful tool to study $0\nu\beta\beta$ NMEs. Both transition operators contain the spin-isospin operator associated with axial-vector $g_A$ coupling, and for quite long ago a systematic overestimation of $\beta$ decay was amended by a quenching to the corresponding matrix elements. A solution to this problem came by ab initio studies of $\beta$ decay including many-body correlations and two-body currents [88]. However, a possible quench in the matrix elements at high momentum transfer has not been addressed yet. This makes OMC an interesting process since comparison between OMC prediction rates with measurements could help to answer this question. Additionally, neutrino-nucleus scattering and charge exchange reactions can be promising tools to test $0\nu\beta\beta$ NMEs [146], because they also occur at similar momentum transfer $p \sim 100\,\text{MeV}$.

Despite being of great help to test nuclear many-body calculations, the above observables do not show apparent correlation with $0\nu\beta\beta$ decay. In this respect, two observables not measured yet show good correlation with $0\nu\beta\beta$ NMEs: double Gamow-Teller (DGT) [147–150] and double magnetic dipole transitions, $\gamma\gamma(M1M1)$ [12, 150]. Moreover, quite recently the unnoticed correlation to date between $2\nu\beta\beta$- and $0\nu\beta\beta$-decays NMEs was revealed by the study in Ref. [151].

Cross sections measurements of double charge exchange reactions (DCEs) called the attention of experimentalists as a path to access $0\nu\beta\beta$ NMEs, in the same way as single-charge exchange reactions do for $\beta$ decay (for a comprehensive summary see [62]). The NUMEN project [152] started with the goal of accessing the NMEs $0\nu\beta\beta$ decay by measuring the cross sections of DCE reactions. Different many-body methods found a good linear correlation between DGT transitions to the ground state and $0\nu\beta\beta$-decay NMEs, although not all of them are fully consistent [150].

In the same way, second order electromagnetic transitions could also open the door to test $0\nu\beta\beta$-decay NMEs. As we will see in Chp. 5 were the results in Ref. [12] are explained, we found a good linear correlation in NSM between $\gamma\gamma(M1M1)$ decays from $0^+$ double isobaric analog state of $\beta\beta$ $0^+_{\text{gs}}$ of the initial nuclei to the $\beta\beta$ $0^+_{\text{gs}}$ of the final nuclei and $0\nu\beta\beta$ NME. In Ref. [150], they also found a good linear correlation within QRPA, but differ from the one found in [12]. The authors of [150] suggest that these discrepancies could be related with the different contributions of intermediate states to $0\nu\beta\beta$-decay NMEs in QRPA. Currently, we are working in the study of this correlation in the VS-IMSRG method, but the role of isospin-symmetry breaking must be understood before, as suggested in Sec. 5.2. On the other hand, although in its beginning phase, there is a promising experimental program to measure $\gamma\gamma(M1M1)$ from DIAS in $^{48}$Ti [153].

Finally, the relation found between $2\nu\beta\beta$- and $0\nu\beta\beta$-decays NMEs both in NSM and in QRPA [151] allowed the estimation of $0\nu\beta\beta$-decay NMEs from the measured values of $2\nu\beta\beta$ half-life. Within this approach, $0\nu\beta\beta$ NME can be predicted using both data and systematic calculations across dozens of nuclei using different nuclear interactions. This way, it also allows one provide their theoretical uncertainties based on the systematic errors that are captured by all the NMEs that follow the same correlation.

# Chapter 3

# The Nuclear Many-Body problem

The description of nuclear phenomena in terms of the elementary particles that constitute nuclei, quarks and gluons, is a long-standing goal of nuclear physics. Throughout this chapter, a brief introduction to the main ingredients of this problem will be given. First, in Sec. 3.1 we introduce the problem we face when describing the low energy domain of nuclear physics from the fundamental theory of strong interactions. Then, in Sec. 3.2 we introduce the difficulty one has to construct the nuclear interaction and give the relevant approximations to construct interaction Hamiltonians. In Sec. 3.3 and Sec. 3.4 we will review the main aspects of the nuclear shell model (NSM), and valence-space in-medium similarity renormalization group (VS-IMSRG) as the nuclear many-body methods we have used in our work.

## 3.1 From QCD to Low-Energy Nuclear Physics

The description of the nucleus from the fundamental theory of strong interactions between quarks and gluons, QCD, is a daunting task for nuclear physics. On one side QCD does not admit a simple solution at low energy where typical nucleon momentum scales $Q$ are such that $Q \ll \Lambda_{QCD} \sim 1\,\mathrm{GeV}$, the chiral-symmetry breaking scale. Indeed, even the interaction in the two-nucleon system is not yet under control within Lattice QCD [154], a framework in which the strong interactions can be studied from first principles, from high to low energy scales. This is a consequence of the non-abelian character of the gauge group of the strong interactions, which at low energies are highly non-perturbative.

On the other hand, over the years we have learned that there are complex phenomena which can be described properly without using the fundamental degrees of freedom of the underlying theory, like in nuclear physics where we can use nucleons as degrees of freedom instead of quarks and gluons. Additionally to the proper description of the interaction between nucleons, the technical problem of solving exactly the Schrödinger equation, the quantum many-body calculation, becomes unfeasible as $A$ increases. Nevertheless, the wealth of physics processes occurring within the nucleus and that can be used as tests of fundamental symmetries like $0\nu\beta\beta$ decay [155], demands a precise and accurate description of nuclei pushing forward the development of nuclear theory.

Over the last decades a remarkable progress has been made in techniques that allow one to face the above issues from the *ab initio* or first principles perspective. During this time a better understanding of how to construct consistent nucleon-nucleon (NN), three-nucleon (3N) and higher order interactions grounded in QCD using Effective Field Theory (EFT), Chiral Perturbation Theory and Lattice QCD has been developed thanks to Renormalization Group (RG) ideas. At the same time many-body frameworks such as the Green functions Monte-Carlo, no-core shell model, Coupled Cluster, In-Medium similarity renormalization group, etc. have improved their capabilities. In parallel, the computing demands of these numerical calculations have pushed forward a tremendous progress both in algorithm development and hardware resources of high performance computing systems.

## 3.2 Nuclear Hamiltonians

Despite its fundamental nature, determining the strong interaction between the relevant degrees of freedom at the low energy scales of nuclear structure phenomena still represents a challenge. Although the fundamental theory of strong interactions is QCD, the strong coupling constant $\alpha_S$ is of order 1 at low energy, which makes the perturbative treatment in this "strong QCD" regime inapplicable. The nuclear force that binds protons and neutrons within the nucleus is a residual manifestation of the non-abelian color $SU(3)_c$ interaction of quarks, in a similar way as the van der Waals forces between neutral molecules. Nowadays the link between QCD and its symmetries (particularly the chiral symmetry) with the nuclear forces has strengthened thanks to the development of theoretical tools such as LQCD, and effective field theory.

However, before this new period of high-precision Hamiltonians rooted in QCD more *phenomenological interactions* were used. These Hamiltoninas fitted to rather customized parametrizations were built to reproduce few-body observables such as two-nucleon *scattering phase shifts*, or fitted to reproduce low energy spectra and transition probabilities in heavy nuclei. This kind of interactions were the first to be used in nuclear shell model calculations and today a wide variety of high-quality phenomenological interactions exists with $\chi^2/\mathrm{datum} \sim 1$. There are other types of interactions used in nuclear structure calculations but here are only reviewed the ones used in the present work.

### 3.2.1 Microscopic phenomenological interactions

The first model of nuclear forces was proposed by Hideki Yukawa in 1935 [156]. Yukawa, guided by the success of the Coulomb interaction and the experimental observation of the finite range of the nuclear interaction, suggested that the nuclear force was mediated by a boson, called at that time the *mesotron*. Afterwards, with its discovery in 1947, this boson would be identified as the pion.

Yukawa's idea proved to be very successful in describing the *long-range* part of the nuclear interaction, but fails to explain its *short-range* part. In the 1960s heavier mesons such as $\eta$, $\rho$, $\omega$ and $\sigma$ were systematically added to the nuclear interaction, leading to the one-boson-exchange NN potentials. However, multi-meson exchange potentials were disfavoured due to the energy dependence of the NN interaction.

In the 1990s several meson-exchange potentials appeared such as the CD-Bonn [157, 158] or Argonne $v_{18}$ [110] charge-dependent NN interactions. They were fitted to proton-proton and proton-neutron Nijmegen partial wave scattering database [159] below 350 MeV with $\chi^2/\text{datum} \simeq 1$. In order to make such interactions applicable in the nuclear medium some kind of regularization must be applied (for example Brueckner's G matrix formalism by Kuo, Brown and collaborators [160]). This type of interactions are categorized as *realistic interactions* from meson-exchange potentials. However, all of these NN interactions when used alone showed some deficiencies in the description of nuclei with $A \leq 12$, particularly a systematic underbinding of energies which increase with $A$ [161]. Therefore, several phenomenological $3N$ potentials were developed such as the Urbana interaction [162, 163] and more recently the Illinois interaction [164, 165], showing a good agreement between predicted and experimental ground and low-energy states for light nuclei $A \leq 12$ [161].

An alternative approach commonly used in shell-model calculations is to begin from some renormalized realistic interaction, and then to adjust the two-body matrix elements (TBMEs) of the effective interaction within a particular model space in order to reproduce experimental low energy spectra and transition probabilities of nuclei that can be modeled by such configuration space. Usually an effective interaction is obtained for some mass value and then one can use this interaction in a broader mass region. The use of a mild scaling is needed in the TBMEs as $TBME(A) = TBME(A_c)(A/A_c)^{-0.3}$ to take into account the radial dependence of the wave functions with the mass number $A$.

This family of interactions also falls into the the category of *realistic interactions* but as *fitted effective interactions*. They can also be called *empirical* or *phenomenological* interactions. Some examples are the interactions we have used in the study of double gamma decay in this thesis, such as the USD family interactions [166–168] valid in the *sd* configuration space, the KB3G [169] and GXPF1B [170] interactions used for nuclei well described by the *pf*-shell, the GCN2850 [171], JUN45 [172] and JJ4BB [173] interactions for nuclei in the *pfg* valence space, and the GCN5082 [171] and QX [174] interactions for nuclei in the *sdgh* configuration space.

For bigger model spaces the number of TBMEs is quite large to be determined by existing data. Particularly, data is more sensitive to certain linear combinations of single-particle energies (SPEs) and TBMEs which can be determined by a single-value decomposition method, while other combinations are left as they are in the realistic interaction. This is the case of GXPF1, JUN45, JJ44B. Other approach for larger model spaces is again starting from a realistic effective Hamiltonian to decompose it as a sum of a monopole and a multipole term. The monopole combination of TBMEs determines the bulk properties of the effective interaction and the evolution of the SPEs as a function of the valence proton and neutron numbers. Shell-model calculations suggest that realistic interactions can give a good description of the multipole part but not of the monopole term, and this seems to be caused by the lack of three-body forces [175]. We find effective interactions where this monopole terms are fitted to spectroscopic data, such as KB3G and QX interactions. These kinds of fitting methods provide global interactions in such a way that a particular set of SPEs and TBMEs can be applied to all nuclei that fit within a given model space.

A major limitation of these Hamiltonians is the lack of a systematic way to improve them and to quantify theoretical uncertainties related to each particular model. Instead, what is usually done is to give a compound uncertainty of the many-body method with several phenomenological interactions. Conversely, the interactions derived within the framework of chiral EFT ($\chi$EFT) can be systematically improved at the same time that this framework provides a method to estimate uncertainties related with the truncation of the chiral expansion in a reliable way.

### 3.2.2 Chiral interactions

The non-abelian nature of the gauge theory of strong interactions has important consequences. One of those is that these interactions are strong at low momentum-transfer (or long distances $\gtrsim 1$fm) which leads to quark confinement into colorless systems, hadrons. Nuclear structure and phenomena live in such regime where the non-perturbative analysis is needed to describe the dynamics properly.

Deriving the nuclear force in terms of quarks and gluons is a complicated problem. Although it is extremely challenging it could be addressed with enough computing power by lattice QCD. A huge effort is in progress to study light nuclei within this framework, whose current limits are set to $A \leq 4$ with non-physical pion mass [176]. Therefore, even for nuclear structure of light nuclei a more technically feasible approach must be applied. This more suited approach is based on effective field theory. The idea of EFT was proposed by Weinberg [9, 177, 178], and is based on constructing a low-energy theory including all the interactions consistent with the symmetries of the theory in terms of the relevant degrees of freedom at that low-energy scale. Such an effective field theory of QCD valid at nuclear structure energy scales is known as chiral EFT. First applications of $\chi$EFT to nuclear systems were done by Ordóñez, Ray and van Kolck, who constructed a NN potential based on $\chi$EFT at third order in the chiral expansion or next-to-next-to-leading order ($N^2$LO) [179–182].

Starting from the $\chi$EFT Lagrangian, one can construct a systematic low-momentum expansion in terms of $Q/\Lambda$, where $Q$ (soft scale) is a typical momentum of the interacting system (typical momenta of nucleons or the pion mass) and $\Lambda$ (hard scale) is the breakdown scale for which the physics that is not explicitly considered in the EFT starts to be important. A typical breakdown scale for $\chi$EFT with explicit pions and nucleons is $\Lambda \simeq m_\rho (776)$, which corresponds to the chiral symmetry breaking scale. Also, depending on the system under study and the separation of scales that best represent it, one can construct other kinds of EFTs in nuclear physics, such as pion-less EFT or halo EFT.

In $\chi$EFT the nuclear interaction is analyzed as an order by order expansion in terms of $(Q/\Lambda)^\nu$ and organized according to a power-counting scheme sketched in Fig. 3.1. For example, the leading order (LO) $\nu = 0$ corresponds to one-pion exchange plus a nucleon-nucleon contact interaction. In general, all chiral interactions between nucleons consist of one or multi pion exchanges between nucleons as well as nucleon contact interactions. The contact interaction vertices are proportional to the low-energy constants (LECs), which are not constrained by the symmetries of the theory and contain the unresolved short-range physics (integrated out degrees of freedom). These EFT parameters in principle can be calculated by the underlying theory of QCD, which is equivalent to solving it at low energies. However, direct calculation from lattice QCD is limited to LECs of single- and two- nucleon diagrams [183], and LECs are typically determined by fits to experimental data [184–186] for each resolution scale $\Lambda$.

Another advantage of $\chi$EFT is that the power counting scheme naturally yields to NN, 3N and higher order many-body interactions with decreasing relevance, justifying the phenomenological hierarchy of many-body forces. Likewise, it provides a consistent treatment of other operators such as electroweak currents which contain the same LECs and therefore represent additional observables to fit them [183, 187, 188]. Furthermore, the power counting scheme also provides a quantitative way to systematically asses the theoretical uncertainties that come from working at a given expansion order.

In general, the EFT expansion of an observable can be written as [189]

$$\mathcal{O} = \mathcal{O}_0 \sum_{\nu=0}^{\infty} c_\nu \left( \frac{Q}{\Lambda} \right)^\nu, \qquad (3.1)$$

where $c_\nu$ are dimensionless expansion coefficients, and the leading order result $\mathcal{O}_0$ comes from the first term in the expansion. It is factorized out to set the overall scale, but other choice could have been used. However, if the expansion coefficients are of $\sim \mathcal{O}(1)$ one should expect that each term in this expansion to be $Q/\Lambda$ smaller than the previous one. Therefore, if the expansion is truncated at some order $k$, a truncation error can be determined from

$$\delta_k = \mathcal{O}_0 \sum_{\nu=k+1}^{\infty} c_\nu \left( \frac{Q}{\Lambda} \right)^\nu. \tag{3.2}$$

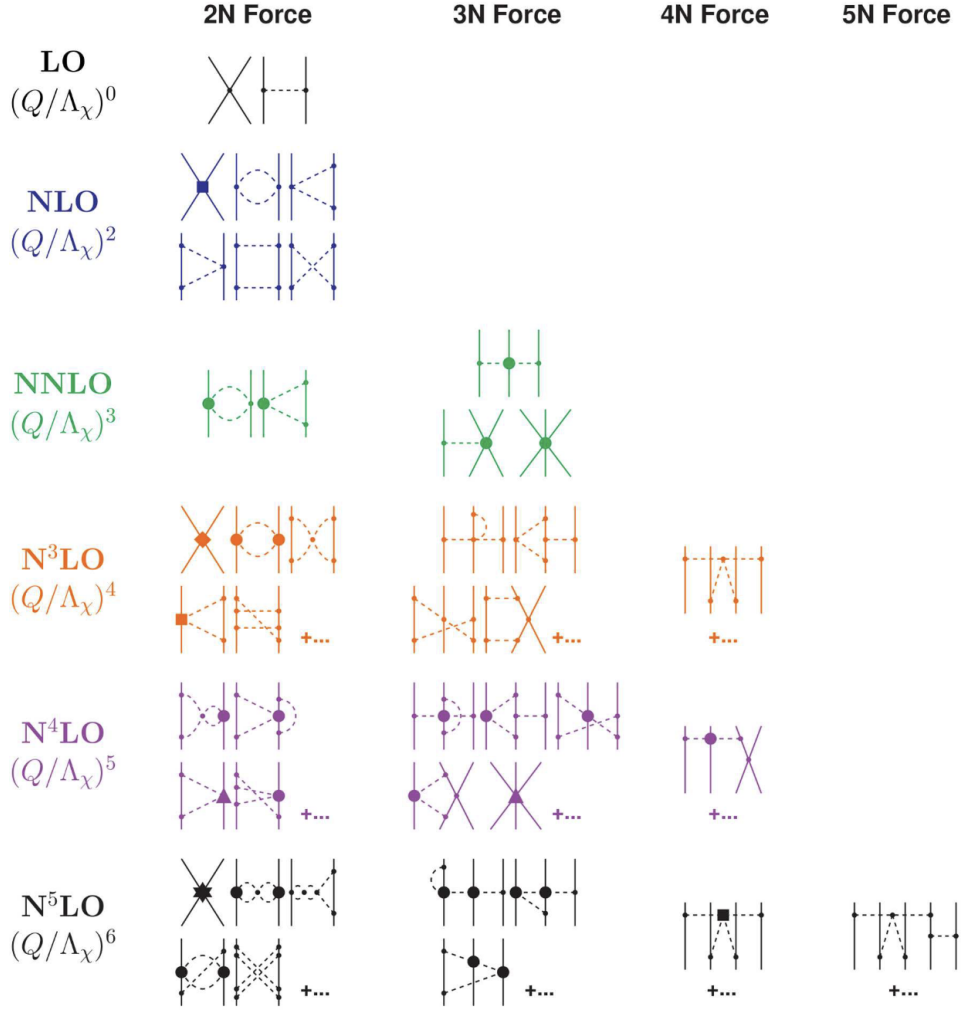

FIGURE 3.1: Hierarchy of nuclear forces up to N$^5$LO or sixth order of the chiral expansion. Dashed lines represent pion exchanges. Large solid squares, circles and a star, represent vertices which are proportional to LECs. Figure taken from [186].

There is still work to do to quantify the confidence of such uncertainty estimates of observables computed using these chiral interactions within *ab initio* many-body methods, and how they are connected to missing physics at the level of the effective Lagrangian [189].

### 3.2.3    Softened interactions

A decade ago, another class of interactions emerged to deal with the difficulties associated with the strong repulsive short-range ("hard core") and tensor components of some nuclear potentials. A consequence of the short-range character is that when one sees the interaction in momentum space its matrix elements present a strong coupling between low and high momenta (large off diagonal matrix elements) [190]. This translates into a highly suppressed probability density at short distances (known as "short range correlations" in this context). From the numerical calculation perspective the convergence is slowed down by the off diagonal matrix elements, limiting the convergence of the many-body calculation.

The solution to this problem is to decouple in some way the high-momentum components in the nuclear potential from the low-momentum components, or equivalently to soften the repulsive hard-core, hence the name of this kind of softened interactions. This way, an original interaction such as a phenomenological Argonne v18 or a $\chi$EFT is transformed by a short-distance unitary transformation or a series of them (see [191] and references therein). These methods which allow a continuous change of resolution scale of the interaction and drive it towards a decoupled form are known as renormalization-group methods. The progressive decoupling can be done either by making trivially the interaction matrix elements go to zero above the running momentum cutoff ($V_{\text{lowk}}$ approach) or by driving the interaction towards a band-diagonal form when lowering the cutoff (similarity renormalization group, SRG, approach) [190, 191]. A relevant characteristic of these low-momentum interactions is their *universality* at the two-nucleon level since independently of the original interaction the evolved interactions are very similar, a property that is attributed to the common long-range pion physics of the initial potential [190].

For the VS-IMSRG calculations in Sec. 5.2 we have used the two nucleon interaction up to N$^3$LO (fourth order) of chiral perturbation theory by Entem and Machleidt (EM) [192] (known as EM 500 MeV, $\Lambda = 500$ MeV). The LECs have been fitted to reproduce nucleon-nucleon scattering data up to 290 MeV with a comparable $\chi^2$/datum to high-precision phenomenological potentials such as Argone v18. Other observables such as scattering lengths, effective ranges or deuteron properties (binding energy, radius, quadrupole moment) are in excellent agreement with the experiment. As for the three-nucleon interation we will use the N$^2$LO of Refs. [180, 193]. These NN and 3N interactions have been evolved using RG techniques to different resolution scales [194], but the one we have selected for our calculation is $\lambda(\Lambda) = 1.8(2.0)\text{fm}^{-1}$, known as EM1.8/2.0 interaction. It reproduces quite well ground states energies and nuclear drip lines for light- and medium-mass nuclei.

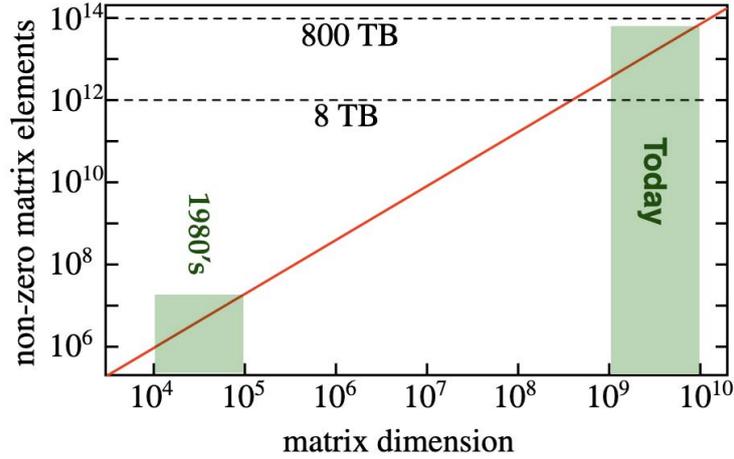

FIGURE 3.2: Approximate dependence of the number of non-zero matrix elements on the matrix dimension for NSM Hamiltonians. Examples of aggregate memory are shown, together with the values reachable in 1980s and in 2018. Figure taken from [195].

## 3.3    The Nuclear Shell Model

The major aim of quantum many-body theory is to solve accurately the stationary Schrö­dinger equation

$$H|\Phi_i^A\rangle = E_i^A|\Phi_i^A\rangle, \tag{3.3}$$

for a given Hamiltonian $H$, where $|\Phi_i^A\rangle$ ($E_i^A$) is the $i$ eigenstate (eigenvalue) of the $A$-body system. When the $A$-body quantum system is the atomic nucleus, the input Hamiltonian corresponds to a realistic interaction representing the strong interaction in terms of nu­cleonic degrees of freedom (neutrons and protons)

$$H = T + V_{2N} + V_{3N} + \ldots, \tag{3.4}$$

where $T$ denotes the intrinsic kinetic energy, $V_{2N}$ the two-nucleon potential, $V_{3N}$ the three-nucleon potential, and so on. In order to ensure a rapid convergence of the solution as well as a manageable basis dimension, an essential technical requirement is a decreasing cluster expansion hierarchy $|V_{2N}| \gg |V_{3N}| \gg \ldots$. Otherwise there would be a large num­ber of many-body matrix elements to be handled in the calculation. Then, the limitation comes mainly from the memory needed to store and diagonalize the non-zero matrix elements of the N-body potentials (see Fig. 3.2).

The $A$-body wave function $|\Phi_i^A\rangle$ depends on the coordinates, spin and isospin degrees of freedom of each nucleon. Therefore as the number of nucleons increases, the exact solution of Eq. (3.3) becomes more involved and the computational cost prohibitive for $A \geq 20$. There are two main approximations to tackle this problem [195], either to reduce the number of active nucleons and as a result the dimensions of the many-body Hilbert space, or while maintaining the full many-body space to simplify the complexity of the wave function.

The first approach is the *interacting shell model*, which we will describe in detail here as it is the many-body method mainly used in our study. During the thirties several measurements of nuclear observables such as the binding energy per nucleon, the proton and neutron separation energies or nuclear charge radii among others, predicted the occurrence of the so-called *magic numbers*, particular values for the number of neutrons or protons in nuclei for which these nuclear properties present discontinuities. In 1949 M. Goeppert-Mayer, and O. Haxel *et al.* [196, 197] introduced the nuclear shell model as a framework to describe all these phenomena derived from nuclear structure. They proposed a spherical mean field consisting of an isotropic harmonic oscillator plus a strongly attractive spin-orbit coupling potential. With the inclusion of the latter, the spectrum of single particle energies exhibits a shell structure that can explain the magic numbers observed.

The second approach is the basis of *energy density functional theory*. In either effective approach a consistent modified Hamiltonian has to be constructed so as to include the correlations with the excluded nucleons or included in the full many-body wave function. This is done in such a way that the effective interaction acting on the simplified wavefunction reproduces the observables of the complete many-body Hilbert space

$$H|\Phi_i^A\rangle = E_i^A|\Phi_i^A\rangle \longrightarrow H^{\text{eff}}|\Phi_i^{\text{eff}}\rangle = E_i|\Phi_i^{\text{eff}}\rangle. \tag{3.5}$$

In a first approximation one can think that nucleons move in a mean-field potential generated by the remaining $A - 1$ nucleons, so the problem of $A$ strongly interacting particles turns into a problem of $A$ non-interacting particles in an external mean field which is known as the *non-interacting shell model* or *independent particle model* [61]. But in reality the potential is more complex and includes two-body forces and in general many body terms (Eq. (3.4)). One can add and subtract a mean field potential

$$H = \left[\sum_{i=1}^{A}\frac{-\hbar^2}{2m_N}\nabla_i^2 + \sum_{i=1}^{A}V_{mf}(\boldsymbol{r}_i)\right] + \left[\sum_{i<j}^{A}V(\boldsymbol{r}_i,\boldsymbol{r}_j) - \sum_{i=1}^{A}V_{mf}(\boldsymbol{r}_i) + \dots\right],$$
$$H = H_0 + V_{\text{res}}, \tag{3.6}$$

where $\boldsymbol{r}_i$ are the coordinates of the *i*-nucleon, and separate the Hamiltonian into a noninteracting one body part $H_0$ and a residual many-body interaction $V_{\text{res}}$.

The mean field Hamiltonian is relatively easy to solve because we can treat it as separable $A$-body problem, and we only have to find the stationary singe-particle states or the single particle orbitals, $\phi_n(\boldsymbol{r}_i)$, for the one-nucleon Hamiltonian $h(\boldsymbol{r}_i)$

$$H_0 = \sum_{i=1}^{A} h(\boldsymbol{r}_i) \longrightarrow h(\boldsymbol{r}_i)\phi_{\alpha_i}(\boldsymbol{r}_i) = \varepsilon_{\alpha_i}\phi_{\alpha_i}(\boldsymbol{r}_i), \tag{3.7}$$

$$H_0\Phi_0(\boldsymbol{r}_1,\boldsymbol{r}_2,\ldots,\boldsymbol{r}_A) = E\Phi_0(\boldsymbol{r}_1,\boldsymbol{r}_2,\ldots,\boldsymbol{r}_A), \tag{3.8}$$

with

$$E = \sum_{i=1}^{A} \varepsilon_{\alpha_i}, \quad \Phi_0(\boldsymbol{r}_1,\boldsymbol{r}_2,\ldots,\boldsymbol{r}_A) = \mathcal{A}\left[\Pi_{i=1}^{Z}\phi_{\alpha_i}^{(\pi)}(\boldsymbol{r}_i)\right]\mathcal{A}\left[\Pi_{i=1}^{N}\phi_{\alpha_i}^{(\nu)}(\boldsymbol{r}_i)\right]. \tag{3.9}$$

Here $\varepsilon_{\alpha_i}$ are the single-particle energies and $\mathcal{A}$ is the antisymmetrizing operator. The $A$-body wave function can be factorized as a product of the proton ($\pi$) and neutron($\nu$) wave functions. A simple ansatz for the antisymmetrized many-body wave function is the direct product of proton and neutron antisymmetrized product states or Slater determinants,

$$\mathcal{A}_X = \frac{1}{\sqrt{X!}}\sum_{P\in S_X}\text{sign}(P)\Pi_{ij}P_{ij}, \quad X = Z, N \tag{3.10}$$

where $P_{ij}$ is the permutation operator of the pair $(i,j)$, $\text{sign}(P) = -1$ for an odd number of pair permutations and $S_X$ is the permutation group of $X$ elements. A common choice for the mean field potential is the harmonic oscillator or a Woods-Saxon potential plus a spin-orbit term, hence the eigenfunctions $\phi_{\alpha}^{(\pi,\nu)}(\boldsymbol{r})$ are states of the coupled *ls*-basis

$$\phi_{\alpha}(\boldsymbol{r}) = g_{nl}(r)\left[Y_l(\theta,\varphi)\chi_{\frac{1}{2}}\right]_{jm_j} \tag{3.11}$$

with $g_{nl}(r)$ the harmonic oscillator functions, $Y_{lm_l}(\theta,\varphi)$ are the spherical harmonics and $\chi_{\frac{1}{2}m_s}$ is a spin-$\frac{1}{2}$ spinor. In general, for a spherical potential the single-particle wavefunctions are labeled by their radial, orbital angular momentum, and total angular momentum quantum numbers, $n$, $l$, and $j$, respectively. This set of quantum numbers is labeled collectively here as $a \equiv (n,l,j)$. Each $j$ has $(2j+1)$ $m_j$ states, and the associated single-particle states are labeled by $\alpha = (a, m_j)$. Then, according to the Pauli exclusion principle, protons and neutrons fill these single particle states labeled as $nlj$, with $l$ represented in the spectroscopic notation $s$, $p$, $f$, $g$, $h$, $i$, etc for $l = 0,1,2,3,4,5,\ldots$, respectively. In a spherical plus a spin-orbit potential, one has the sequence of orbitals $0s_{1/2}$, $0p_{3/2}$, $0p_{1/2}$, $0d_{5/2}$, $0d_{3/2}$, $1s_{1/2}$, $0f_{7/2}$, $0f_{5/2}$, etc.

### 3.3.1   Interacting Shell Model

The non-interacting shell model is a simple formulation which explains some nuclear phenomena, but it is an approximated solution to Eq. (3.3) and the residual interaction in Eq. (3.6) is not usually small. Therefore, direct diagonalization of the full Hamiltonian in terms of the basis of the mean field is performed [198].

In this case the problem arises as we increase $A$, since the diagonalization in the full

Hilbert space becomes extremely costly or even unfeasible. Despite this, using the fact that there is a large energy separation between major shells or magic numbers, excitations across shells are highly suppressed. This allows one to separate the single particle states for the mean field potential into an inert *core*, a *valence* or *configuration* space and an *external* space. The inert core is composed by orbitals that are always full, while the external space is constituted by the orbitals that are always empty. It is in the valence space where the dynamics take place and we have particle-hole excitations.

The shell-model Hamiltonian ($H^{\text{eff}}$) in the configuration space or in the occupation number representation (see Appendix B for details) consists in a sum of one-body plus two-body interactions

$$H^{\text{eff}} = \sum_i \varepsilon_i c_i^\dagger c_i + \sum_{ijkl} \bar{v}_{ijkl} c_i^\dagger c_j^\dagger c_l c_k, \tag{3.12}$$

where $c_i^\dagger (c_i)$ are the the fermionic creation and annihilation operators associated with the single particle state $i$, respectively. Additionally, $\varepsilon_i$ is the energy of the single-particle state and $\bar{v}_{ijkl} = v_{ijkl} - v_{ijlk}$ is the antisymmetrized TBMEs. In these parameters are implicitly encoded the information of the external space. Although 3$N$ forces are neglected, in practice they can be effectively captured in the single-particle energies and TBMEs. As we have described in Sec. 3.2, there are two alternative paths to construct $H_{\text{eff}}$, either by using some Hilbert-space projection techniques which retain the link with the "microscopic" inter-nucleon forces or by fitting these parameters so as to reproduce available experimental data, as the binding energies and excitation energies. Usually, a combination of the two of them is followed.

The many-body wave functions are linear combinations of orthogonal Slater determinants $|\phi_i\rangle$ representing different configuration of particle-hole excitations of those single-particle orbitals

$$|\Phi_i\rangle = \sum_j c_{ij} |\phi_j\rangle \tag{3.13}$$

where the coefficients $c_{ij}$ are obtained from the diagonalization of the effective interaction in the valence space. With the exact diagonalization all correlations that are captured by the effective interaction go into the wave functions $|\Phi_i\rangle$. However, a thorough treatment of correlations restricts the dimension of the model space, with current maximal dimension of $10^{11} - 10^{12}$ [199] many-body configurations (Slater determinants).

### 3.3.2 *M*-scheme SM basis

In order to perform the shell-model calculations one has to chose a proper many-body basis. Since the Hamiltonian is rotationally invariant one can find eigenstates of the Hamiltonian which are simultaneous eigenstates of the total angular momentum ($J^2$) and its projection ($J_z$). The energy normally depends on $J$, so eigenstates with different $J$ are not degenerate, but for a given $J$ the energy does not depend on $M$. Depending on the choice of the basis, the Hamiltonian can be explicityly block-diagonal in $J$, and these different choices is what is known as *basis schemes*.

There are three main possible choices: the $M$-scheme, the $J$-coupled scheme, and the $JT$-coupled scheme. In the $M$-scheme framework, the Slater determinants have a definite total angular momentum projection $M$. $M$ and $T_z$ are good quantum numbers which means that all possible values of $J$ and $T$ are in the basis, therefore the dimensions of the Hamiltonian matrix is maximal. However, since $H$ commutes with $J^2$ and $J_z$, if we take all states of a given $M$ and diagonalize $H$, the eigenstates will have a good $J$. In this scheme the wave function of the nuclear state in Eq. (3.13) translates into

$$|JMTT_z\rangle = \sum_{i=1}^{D} c_i |i, MT_z\rangle, \qquad (3.14)$$

where $D$ is the number of the basis states allowed with a a specified M or *M-scheme dimension*. The coefficients $c_i$ are obtained by solving the Schrödinger equation in matrix form, which in the $M$-scheme is

$$\sum_{i=1}^{D} H_{ij}^{\text{eff}} c_j = E c_i, \qquad (3.15)$$

where $H_{ij}^{\text{eff}} = \langle i, MT_z | H^{\text{eff}} | j, MT_z \rangle$. Furthermore, each basis state is a simple tensor product of a proton and neutron Slater determinant as in Eq. (3.8). Since the $M$ quantum number is additive, we have $M = M_n + M_p$ and every proton Slater determinant with $M_p$ must be combined with a neutron Slater determinant $M_n = M - M_p$. If $N(Z)_v$ are the number of neutrons and protons in the valence space, respectively, the number of Slater determinants that can be built in the configuration space is

$$D = \binom{d_v}{N_v} \times \binom{d_v}{Z_v}, \qquad (3.16)$$

where $d_v$ is the dimension of the valence space $d_v = \sum_j (2j + 1)$ which contains orbitals denoted by $nl_j$, as defined before. Consider for example $^{26}$Si which has $Z = 14$ and $N = 12$. A proper configuration space would be the $sd$ shell, which includes the $1s_{1/2}$, $0d_{5/2}$ and $0d_{3/2}$ orbitals. In this case for both protons and neutrons we have $d_v = 12$, and $N_v = 4$ and $Z_v = 6$ so $D = 457380$. Although the basis wave functions $|i, MT_z\rangle$ do not have a definite total angular momentum $J$, since they are obtained upon diagonalization of the shell-model Hamiltonian which commutes with $J^2$, nuclear states of Eq. (3.14) automatically have good total angular momentum $J$. In practice, one typically chooses $M = 0$ for $A$ even nuclei, and $M = \frac{1}{2}$ for $A$ odd nuclei.

On the other hand, in the $J$ or $JT$-coupling schemes the basis in the $M$-scheme is divided in blocks of smaller dimension constructed using involved angular momentum algebra [200, 201]. In these cases the basis dimension is rather small compared with the *M*-scheme (see Table 3.1), but in the later the Hamiltonian matrix is sparse and easy to diagonalize using the *Lanczos algorithm* as we will see in Sec. 3.3.4. Nevertheless, the *J* coupled scheme has several advantages: the lower basis dimension together with a well defined angular momentum allows one to carry out large matrix operations (Lanczos iterations as we discuss below) without memory limitations [202].

### 3.3.3 Shell-model codes

Several shell-model codes have been developed over the years such as the hybrid (*M*-scheme and *J*-scheme) OXBASH and MSU-NSCL [203, 204], which evolved into Nu-ShellX@MSU [205]. The Strasbourg-Madrid nuclear-theory group developed ANTOINE (*M*-scheme) and NATHAN (*J*-scheme) codes [200, 202], which are the codes that have been used for the large-scale shell model computations in this thesis. Other codes for large-scale parallel shell-model calculations are MSHELL [206] and MSHELL64 [207].

With the growth of computing resources designed for massive parallel computing, several codes which exploit efficiently those resources have been developed such as within *M*-scheme MFDn [208–210], BIGSTICK [211, 212] and KSHELL [213], or within *J*-coupled scheme by MFDnJ [214]. Currently, these codes have an hybrid MPI (Message Passing Interface) and OpenMP (Open Multi-processing) parallelization. The MFDn(J) and BIGSTICK codes were developed within the UNEDF (Universal Nuclear Energy Density Functional) project [215], and were mainly created for no-core shell model (though BIGSTICK can be used in phenomenological shell-model) and can deal explicitly with three-body forces.

All these codes are written in FORTRAN, from Fortran 77 in ANTOINE to Fortran 90/95 for more recent codes, and the majority of them use the Lanczos algorithm to diagonalize the Hamiltonian. This algorithm is suitable for nuclear-structure calculations, where one is never interested in a complete set of eigensolutions but mainly in the ground or low-lying excited states.

| Nuclei | Configuration space | *M*-scheme dim | *J*-scheme dim |
|--------|---------------------|----------------|----------------|
| $^8$Be | *p*-shell: $0p_{3/2}, 0p_{1/2}$ | 51 | 9 |
| $^{32}$Si | *sd*-shell: $0d_{5/2}, 1s_{1/2}, 0d_{3/2}$ | 93710 | 3372 |
| $^{62}$Zn | *pf*-shell: $0f_{7/2}, 1p_{3/2}, 1p_{1/2}, 0f_{5/2}$ | $1.58 \times 10^9$ | $2.20 \times 10^7$ |
| $^{78}$Y | *pfg*-shell: $1p_{3/2}, 1p_{1/2}, 0f_{5/2}, 0g_{9/2}$ | $1.31 \times 10^{10}$ | $1.11 \times 10^8$ |
| $^{130}$Sm | *sdgh*-shell: $0g_{7/2}, 1d_{5/2}, 1d_{3/2}, 2s_{1/2}, 0h_{11/2}$ | $2.06 \times 10^{15}$ | $9.58 \times 10^{12}$ |

TABLE 3.1: *M*-scheme and *J*-scheme basis dimensions for several models spaces. The $M = 0$ and $J = 0^+$ subspaces are shown.

### 3.3.4 Lanczos method

For standard diagonalization methods the CPU time scales as $D^3$, with $D$ the dimension of the matrix, which makes them prohibitively long for large-scale shell model computations. This is why the Lanczos method [216] was introduced in shell-model calculations by the Glasgow group in 1977 [217]. Two main properties make the Lanczos method well suited for nuclear-structure many-body calculations: most of the time only the few lowest eigenvalues and eigenstates are needed and the Hamiltonian matrix is sparse.

The basic idea of the Lanczos algorithm is to construct iteratively an orthogonal basis (*Krylov subspace*) by repeated applications of the Hamiltonian matrix (or the matrix to be diagonalized), starting from an initial normalized vector, $|v_1\rangle$, called the *pivot state*. In this new basis the Hamiltonian is tridiagonal. The power of this algorithm roots in its simplicity together with its ease to find extremal eigenvalues of the matrix to diagonalize in the Krylov subspace, since those eigenvalues quickly converge to those of the full space.

The standard algorithm is as follows: from the initial normalized state $|v_1\rangle$ one generates after $k-1$-iterations a sequence of orthonormal vectors (Lanczos vectors) $\{|v_i\rangle\}_{i=1}^k$:

$$
\begin{aligned}
H|v_1\rangle &= \alpha_1|v_1\rangle + \beta_1|v_2\rangle \\
H|v_2\rangle &= \beta_1|v_1\rangle + \alpha_2|v_2\rangle + \beta_2|v_3\rangle \\
&\vdots \\
H|v_i\rangle &= \beta_{i-1}|v_{i-1}\rangle + \alpha_i|v_i\rangle + \beta_i|v_{i+1}\rangle \\
&\vdots \\
H|v_k\rangle &= \beta_{k-1}|v_{k-1}\rangle + \alpha_k|v_k\rangle,
\end{aligned}
\tag{3.17}
$$

with $\langle v_i|v_j\rangle = \delta_{ij}$. In each iteration, a new vector is generated by imposing orthogonality with the previous vector and normalization. From the orthonormality of the Lanczos vectors the Hamiltonian matrix is tridiagonal in this basis:

$$
H = \begin{pmatrix}
\alpha_1 & \beta_1 & & & \\
\beta_1 & \alpha_2 & \beta_2 & & \\
& \beta_2 & \alpha_3 & \ddots & \\
& & \ddots & \ddots & \beta_{k-1} \\
& & & \beta_{k-1} & \alpha_k
\end{pmatrix}.
\tag{3.18}
$$

In practice the recurrence steps for constructing the Lanczos vectors are

(1) $|u_i\rangle = H|v_i\rangle$

(2) $\alpha_i = \langle v_i | u_i \rangle$

(3) $|u_i\rangle = |u_i\rangle - \alpha_i |v_i\rangle$, for $i = 1$

(4) $|u_i\rangle = |u_i\rangle - \beta_{i-1}|v_{i-1}\rangle$, if $i > 1$

(5) $\beta_i = (\langle u_i | u_i \rangle)^{1/2}$

(6) $|v_{i+1}\rangle = \beta_i^{-1}|u_i\rangle$

The matrix is then diagonalized until a required criterium of convergence is reached. However, due to the limited floating-point machine precision there are rounding errors, and this could lead to a Lanczos vector $|v_{i+1}\rangle$ which could be not perfectly orthonormal to all the previous vectors. After many iterations this may cause lower energy states to reappear. For this reason a reorthogonalization could be needed.

The Lanczos algorithm can be employed as a projection method with $J^2$ to produce eigenvectors of good angular momentum $J$. If one uses vectors with good $J^2$ as a pivot of a Lanczos calculation, the successive Lanczos vectors will have up to rounding errors the same $J$ as the pivot, and convergence will be reached more efficiently. The same rounding problems could generate 'spurious' eigenvectors which do not have the proper $J$, and therefore a $J$ projection is advisable.

Although the computing power has increased considerably during the last years, other eigensolvers have been implemented to overcome an insufficient memory storage of the Lanczos vectors needed for convergence, such as for example of the *thick-restart Lanczos* [213] (where the Lanczos algorithm after some iterations is restarted and the number of Lanczos vectors is reduced to those with the lowest eigenvalues), the LOBPCG (Locally Optimal Block Preconditioned Gradient) or the filter diagonalization methods [218, 219].

## 3.4 Valence Space In-Medium Similarity Renormalization Group

The idea of decoupling energy scales introduced to construct interactions with amenable properties for the subsequent many-body methods presented in Sec. 3.2.3 can also be used to deal with the many-body method itself. In fact, the similarity renormalization group can be applied to decouple physics at different excitation energy scales of the nucleus and to transform the whole Hamiltonian into a block or band diagonal form. This non-perturbative *ab initio* method is known as in-medium similarity renormalization group (IMSRG) [220].

### 3.4.1 Similiarity renormalization group

The SRG was proposed independently by Glazek and Wilson [221] and Wegner [222]. The idea behind is to transform the original Hamiltonian into a block- or band-diagonal form by a continuous unitary transformation parametrized by a one-dimensional parameter $s$

$$H(s) = U(s)H(0)U^\dagger(s). \tag{3.19}$$

Since it is a unitary transformation the spectrum of the Hamiltonian is preserved. Mathematically, the transformed Hamiltonian can be written as a diagonal plus an off-diagonal term

$$H(s) = H^d(s) + H^{od}(s), \tag{3.20}$$

and find $U(s)$ such that

$$H(s) \xrightarrow{s \to \infty} H^d(s), \quad H^{od}(s) \xrightarrow{s \to \infty} 0. \tag{3.21}$$

In the valence space (VS) formulation of IMSRG, the valence space is decoupled from the excluded region, therefore $H^d = H^{VS}$ and $H^{od} = H^{\text{excl}}$. In order to solve for $H(s)$, we take the derivative of Eq. (3.19) with respect to $s$

$$\frac{dH(s)}{ds} = \frac{dU(s)}{ds}H(0)U^\dagger + U(s)H(0)\frac{dU^\dagger}{ds} = \frac{dU(s)}{ds}U^\dagger H(s) + H(s)U(s)\frac{dU^\dagger}{ds}. \tag{3.22}$$

Since $U^\dagger(s)U(s) = \mathbb{1}$ and defining the anti-hermitian operator

$$\eta(s) \equiv \frac{dU(s)}{ds}U^\dagger(s) = -\eta^\dagger(s), \tag{3.23}$$

we can rewrite Eq. (3.22) as

$$\frac{dH(s)}{ds} = [\eta(s), H(s)], \tag{3.24}$$

This equation is known as the *flow equation* for the Hamiltonian which describes the evolution of $H(s)$ under the action of the generator $\eta(s)$. From Eq. (3.23) we can write a differential equation for $U(s)$,

$$\frac{dU(s)}{ds} = \eta(s)U(s) \Longrightarrow U(s) = \mathcal{S}\exp\int_0^s ds'\eta(s'), \tag{3.25}$$

where $\mathcal{S}$ is the $s$-ordering operator, which reorders operators from left to right in descending order in $s'$. Equivalently, Eq. (3.25) can be written as a product of infinitesimal unitary transformations

$$U(s) = \lim_{n \to \infty} \Pi_{i=0}^{i=n} e^{\eta(s_i)\delta s_i}, \quad s_{i+1} = s_i + \delta s_i, \quad s = \sum_{i=0}^{i=n} s_i. \tag{3.26}$$

From these equations one can see that $\eta(s)$ is infinitesimally generating changes to the resolution scale encoded by the similarity transformation of Eq. (3.19). Hence the name of $s$ as a flow parameter. However, $U(s)$ can not be written as a single exponential and therefore the use of the simplifying Baker-Campbell-Hausdorff expansion is not applicable to simplify the expansion of the transformed Hamiltonian. Without any further approximation, the solution to the problem will require the storage of the generator at all flowing trajectory points which would make it numerically unrealistic.

There are several choices for the generator to drive the Hamiltonian to a block diagonal form [223]. The most common are the Wegner [222] and White [224] generators. The Wegner generator is defined as

$$\eta(s) \equiv \left[ H^d, H^{od} \right] + \text{h.c.}. \tag{3.27}$$

When introduced in the flow equation (3.24), and under the assumption that the transformation induced by $\eta$ suppresses $H^{od}$, the asymptotic scaling of the off-diagonal entries of the Hamiltonian is

$$H_{ij}^{od}(s) \simeq e^{-s(E_i - E_j)^2} H_{ij}^{od}(0), \quad \text{with} \quad E_i - E_j \equiv \langle i | H^d(s) | i \rangle - \langle j | H^d(s) | j \rangle. \tag{3.28}$$

On the other hand the White operator is defined by

$$\eta(s) \equiv \frac{H_{ij}^{od}}{E_i - E_j} + \text{h.c.}, \tag{3.29}$$

with an asymptotic scaling that goes as

$$H_{ij}^{od}(s) \simeq e^{-s} H_{ij}^{od}(0). \tag{3.30}$$

As it can be seen each generator produces a very different asymptotic behaviour. While the Wegner generator drives to zero more quickly matrix elements corresponding to states with larger energy difference, the White generator acts on all matrix simultaneously. Numerically it is more advantageous to use the White generator if there is no energy degenerancies that can make the operator divergent. For this reason it is the generator that will be used in the IMSRG numerical code employed in the calculations shown in Sec. 5.2.

A valuable characteristic of the SRG transformation is that at the same time that the Hamiltonian is evolved, other operators that represent observables of interest of the nuclear system such as electromagnetic or electroweak transitions may be evolved consistently with the same unitary transformation. If $O$ is the Hermitian operator that represents this observable, then the similarity transformation for it reads

$$O(s) = U(s)O(0)U^\dagger(s), \tag{3.31}$$

and its flow equation is

$$\frac{dO(s)}{ds} = [\eta(s), O(s)]. \tag{3.32}$$

In the case of the double magnetic dipole $\gamma\gamma$ decay studied in Sec. 5.2, the **M1** operator and its isoscalar and isovector components were evolved.

### 3.4.2   Reference state normal ordering-IMSRG(2)

The application of the SRG to nuclear forces was done in free space to construct 2N and 3N interactions with improved many-body convergence to use as inputs in ab initio calculations [191, 225]. In this framework $H(s)$ and $\eta(s)$ are normal-ordered with respect to the vacuum. There is an alternative which is to perform the SRG in the nuclear medium, the $A$-body system, and therefore we need a method which helps us to find $H(s)$ that can be used as a valence space Hamiltonian. In order to do this, it is needed to introduce an "In-Medium"(IM) normal ordering of operators with respect to a reference state $|\Phi\rangle$ (the nuclear core). The advantage of the in-medium normal ordering is that it captures many effects of induced many-body interactions, inherent to the SRG evolution, by means of lower order many-body interactions of the normal-ordered Hamiltonian. This powerful feature of the IMSRG simplifies computationally to find the solution to the many-body Schrödinger equation, and therefore extends the reach of this *ab initio* method to heavier mass nuclei.

Here we assume that $|\Phi\rangle$ represents a reference state (for example the Hartree-Fock ground state), so that it obeys

$$\langle\Phi|\{c_i^\dagger\dots c_j\}|\Phi\rangle = 0. \tag{3.33}$$

With the help of the Wick's Theorem, the second-quantized Hamiltonian with two- and three-body interactions can be written in terms of normal-ordered operators as [220]

$$H = E + \sum_{ij} f_{ij}\{c_i^\dagger c_j\} + \frac{1}{2!^2}\sum_{ijkl}\Gamma_{ijkl}\{c_i^\dagger c_j^\dagger c_k c_l\} + \frac{1}{3!^2}\sum_{ijklmn} W_{ijklmn}\{c_i^\dagger c_j^\dagger c_k^\dagger c_l c_m c_n\}, \tag{3.34}$$

where the coefficients of the normal ordered Hamiltonian are

$$E = \sum_{ii} T_{ii} n_i + \frac{1}{2} \sum_{ij} V_{ij}^{(2)} n_i n_j + \frac{1}{6} \sum_{ijk} V_{ij}^{(3)} n_i n_j n_k, \tag{3.35}$$

$$f_{ij} = T_{ij} + \sum_k V_{ikjk}^{(2)} n_k + \frac{1}{2} \sum_{kl} V_{ikjl}^{(3)} n_k n_l, \tag{3.36}$$

$$\Gamma_{ijkl} = V_{ijkl}^{(2)} + \frac{1}{4} \sum_m V_{ijmklm}^{(3)} n_m, \tag{3.37}$$

$$W_{ijklmn} = V_{ijklmn}^{(3)}. \tag{3.38}$$

In these expressions $n_i$ is the occupation number in the reference state $n_i \equiv \langle \Phi | c_i^\dagger c_i | \Phi \rangle$, hence the sums run only over states that are occupied in the reference state. From Eqs. (3.35)-(3.38) one can see that the zero-, one-, and two-body parts of the Hamiltonian incorporate in-medium contributions from the free-space three-nucleon interaction $V^{(3)}$.

This is an important result, since truncating the IM-SRG flow equations to normal-ordered two-body operators, which is known as IMSRG(2) truncation scheme, will capture the three-nucleon physics through Eqs. (3.35)-(3.37). Currently, this is the standard truncation for nuclear structure applications.

It has been shown that the omission of normal-ordered three-body part for low-momentum interactions gives rise to a deviation in the ground state and excited state energies of light and medium-mass nuclei of 1-2%, a result that supports the truncation of IMSRG(2). Furthermore, it provides an effective way to account for three-nucleon forces effects in many-body calculations without the involved computational cost of treating directly three-body operators. Also it scales polynomially with the size of the single-particle basis $D$ ($\mathcal{O}(D^6)$) [226].

### 3.4.3 Magnus formulation

In order to solve numerically the evolution Equations (3.22) and (3.32), direct integration of this system of ordinary differential equations demands large memory, limiting the computation to lower dimensional model spaces. Additionally, the truncation of the unitary matrix under the $s$-ordering inevitably causes non-unitarity. These non-unitary resolution errors break the equivalence between the evolved and initial Hamiltonian or other observables we want to evolve consistently. A possible solution to this problem comes from the "Magnus formulation" of the IMSRG [227].

The idea behind this formulation is to express the unitary matrix as

$$U(s) = e^{\Omega(s)}, \quad \Omega^\dagger(s) = -\Omega(s), \quad \text{and} \quad \Omega(0) = 0. \tag{3.39}$$

This allows one to write the flow equation for $\Omega$ as

$$\frac{d\Omega}{ds} = \sum_{k=0}^{\infty} \frac{B_k}{k!} \mathrm{ad}_{\Omega}^k[\eta(s)] = \eta(s) + \frac{1}{2!}[\Omega, \eta(s)] + \frac{1}{3!}[\Omega, [\Omega, \eta(s)]] + \dots, \quad (3.40)$$

where $B_k$ are the Bernoulli numbers, and

$$\mathrm{ad}_{\Omega}^0 = \eta, \quad \mathrm{ad}_{\Omega}^k = [\Omega, \mathrm{ad}_{\Omega}^{k-1}(\eta)]. \quad (3.41)$$

From the condition $\Omega^{\dagger}(s) = -\Omega(s)$ and using the Baker-Cambell-Hausdorff formula

$$H(s) = e^{\Omega(s)} H e^{-\Omega(s)} = \sum_{k=0}^{\infty} \frac{1}{k!} \mathrm{ad}_{\Omega}^k[H] = H + [\Omega, H] + \frac{1}{2}[\Omega, [\Omega, H]] + \dots \quad (3.42)$$

Therefore the differential equations in Eqs. (3.22) or (3.32) are simplified to a series of nested commutators for the Hamiltonian or the consistently evolved operator. Additionally, one has an ordinary differential equation for $\Omega(s)$ that can be solved using a first order Euler step method without any lost of accuracy and saving considerable memory. The series of nested commutators in Eqs. (3.42) and (3.40) are evaluated recursively until some criteria of convergence is reached.

As in the IMSRG(2), the Magnus(2) formulation truncates $\eta(s)$ and $\Omega(s)$ together with their commutator at the two-body level. The advantage of the Magnus expansion relies on the fact that even making this truncation unitarity is not lost. This formulation has been implemented in the VS-IMSRG code employed in the results shown in Sec. 5.2.

### 3.4.4 VS-IMSRG Model space

As we have seen in Sec. 3.3, to construct the nuclear wave-function a typical choice for the single particle wave-functions is the harmonic oscillator basis. Since these wave-functions are separated into a core and a valence space, one must specify how many single particle excitations are included in the calculations.

The number of single-particle basis states in a calculation is given by $N_{\max}$ or $e_{\max}$, defined by $e_{\max} = (2n + l)_{\max}$, with the radial quantum number $n$ and angular momentum $l$. When $3N$ forces are included it is needed to impose additional truncation on $E_{3\max} = (e_{\max}^{(1)} + e_{\max}^{(2)} + e_{\max}^{(3)})_{\max}$. Usually the value of these parameters is increased until a convergence is reached for the expectation value of the observable under study.

# Chapter 4

# Double-Gamma decay

Electromagnetic transitions in nuclei provide a powerful tool to test nuclear structure properties as well as to give useful information on the dynamics of the relevant degrees of freedom that enter into play at a given energy. Furthermore, the electromagnetic nature of the interaction allows a clear interpretation of the experimental data. The interest in going to processes in which two photons are involved is because new aspects of the nuclear probe can be revealed by two photon decay and scattering experiments.

The first theoretical study of a two-photon ($\gamma\gamma$) process was made by Maria Göeppert-Mayer in 1929 [228, 229] within atoms, but it was not until the sixties and seventies with the growing interest in photonuclear physics that the theoretical $\gamma\gamma$ transitions where considered in nuclei [230–233]. The use of electromagnetic decays from isobaric analog states to connect electromagnetic with electroweak transitions has been done since quite long ago [234]. In this early work, the measurement of the electric dipole $\gamma$ decay from the isobaric analog state (IAS) of $^{141}$Pr is used to obtain the first forbidden $\beta$ decay NME for $^{141}$Ce. More recently, some works provide a detailed analysis based in the spin-isospin common nature of the strong, weak and electromagnetic interactions oriented to learn about the spin-isospin structure of the excitations and transitions [235] or focused more on the neutrino-nuclear responses relevant for astrophysics and laboratory searches of neutrino properties [146].

In Sec. 2.5 we have seen that a recent approach to improve the nuclear description of $0\nu\beta\beta$ decay is through the search of correlated nuclear observables whose experimental measurement is more accessible than $0\nu\beta\beta$ decay. Developments in nuclear spectroscopy make particularly interesting to investigate the possible correlation between $\gamma\gamma$ and $0\nu\beta\beta$ decay.

This chapter is organized as follows: first in Sect.4.1 we derive the theoretical formalism of two-photon decay by nuclei, explaining the main approximations made in this kind of photo-nuclear processes. In Sect. 4.2 we focus on the study of the double magnetic dipole NME from the double isobaric analog state (DIAS) of the $\beta\beta$ initial nucleus.

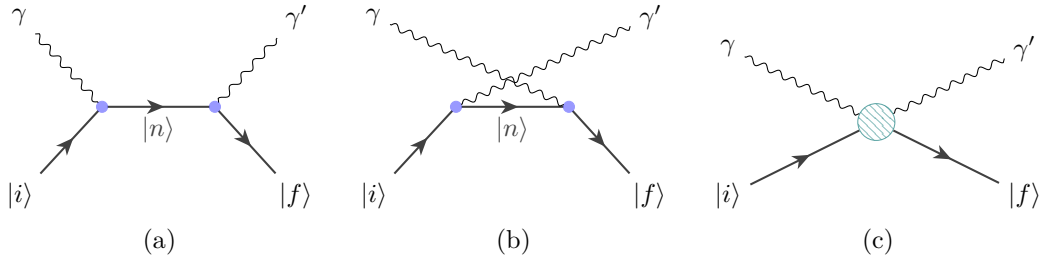

FIGURE 4.1: Feynman diagrams contributing to $\gamma\gamma$ decay:(a) direct and (b) crossed two-current amplitude, (c) two-photon contribution.

In Sec. 4.3 we give an overview of past experimental measurements of double gamma decay. Finally, in Sec. 4.4 the first steps towards a measurement of $\gamma\gamma(M1M1)$ from DIAS are sketched, describing the most relevant theoretical calculations needed for a future experimental set up.

## 4.1 The formalism for two-photon decay

Double gamma decay is a second order process in quantum electrodynamics where a quantum system in an excited state emits simultaneously two photons whose energy sum matches the transition energy and whose individual spectra are continuous. In the case of atomic nuclei it has been studied many times in the literature [230–233] with particular emphasis in $0^+ \rightarrow 0^+$ transitions because of its most favourable experimental measurement prospects. Based on the results of Ref. [233] the authors of Ref. [10] further develop the energy and angular correlations in this decay mode.

In this section we describe the main ingredients behind the formal framework for the $\gamma\gamma$-decay process

$$N_i(P_i) \longrightarrow N_f(P_f) + \gamma_\lambda(k) + \gamma_{\lambda'}(k'),  \tag{4.1}$$

where $N_i(P_i)$, $N_f(P_f)$ is a short notation for the nucleus being in the states $|i\boldsymbol{P}_i\rangle$ and $|f\boldsymbol{P}_f\rangle$ with four-momentum $P_i^\mu = (E_i, \boldsymbol{P}_i)$ and $P_f^\mu = (E_f, \boldsymbol{P}_f)$ respectively. The two photons have four-momenta $k^\mu = (k_0, \boldsymbol{k})$, $k'^\mu = (k'_0, \boldsymbol{k}')$ and helicities $\lambda$ and $\lambda'$. Figure 4.1 shows the Feynman diagrams contributing to this decay up to second order in perturbation theory.

### 4.1.1 Perturbation analysis for electromagnetic interaction in nuclei

The Hamiltonian describing the photo-nuclear process is given by the sum of the uncoupled nuclear ($H_n$) and radiation ($H_\gamma$) Hamiltonians and the interaction ($H_{\text{int}}$) between them

$$H = H_{\text{n}} + H_\gamma + H_{\text{int}}.  \tag{4.2}$$

Hence, nuclear wave functions are solutions of the Schrödinger equation with a Hamiltonian which is the sum of the individual nucleons kinetic energies along with NN and 3N interaction potentials as described in Sec. 3.3. Since the energies of photons $(10 - 40\,\text{MeV})$ are small compared with the rest energy of the nucleus $(\sim A938\,\text{MeV})$[1] this is a good approximation. The overall center of mass motion can be separated from the intrinsic part, therefore the nuclear eigenvalue problem is described by

$$H_{\text{n}}|n, \boldsymbol{P}_n\rangle = E_n|n, \boldsymbol{P}_n\rangle, \tag{4.3}$$

$$E_n = \frac{\boldsymbol{P}_n^2}{2Am} + w_n, \tag{4.4}$$

where $|n, \boldsymbol{P}_n\rangle = |\xi_n J_n M_n\rangle \otimes |\boldsymbol{P}_n\rangle$. $|\boldsymbol{P}_n\rangle$ are simply plane waves, while $|\xi_n J_n M_n\rangle$ are the eigenstates of the intrinsic Hamiltonian characterized by the nuclear angular momentum $J_n$ and its projection $M_n$ for the state $|n\rangle$. $\xi_n$ parametrizes all the other quantum numbers needed to describe the nuclear eigenstate.

The interaction of the electromagnetic field with the nucleus has to be Lorentz and gauge invariant. Furthermore, it should be linear and quadratic in the electromagnetic potential in order to describe the two photon decay at first and second order in perturbation theory. Therefore, the general form for the interaction is [233]

$$H_{\text{int}} = \int d^4x J^\mu(x) A_\mu(x) + \frac{1}{2} \int d^4x d^4y A_\mu(x) B^{\mu\nu}(x,y) A_\nu(y) \tag{4.5}$$

where $J_\mu(x)$ is the nuclear current operator and $A^\mu(x)$ is the electromagnetic field operator. $B_{\mu\nu}(x,y)$ is the so called "seagull" or two-photon operator. They can be obtained from the functional derivatives of the full Hamiltonian $H(A)$

$$J^\mu(x) = \left.\frac{\delta H(A_\mu)}{\delta A_\mu}\right|_{A_\mu \equiv 0}, \tag{4.6}$$

and

$$B^{\mu\nu}(x,y) = \left.\frac{\delta^2 H(A_\mu)}{\delta A_\mu(x)\delta A_\nu(y)}\right|_{A_\mu \equiv 0}. \tag{4.7}$$

Gauge invariance sets restrictions in the form of the current operator through the continuity equation and also links both the two-photon and the current operators. However, to derive the general expression for the two-photon decay amplitude it is not needed the explicit form of the current and two-photon operators. We will write them explicitly once we have to compute the nuclear matrix elements.

---

[1] Or for example the average nucleon velocity for $A = 10$ is of the order of $\sim 0.2c$.

We will use ordinary time-dependent perturbation theory in the interaction representation with second-quantized photon fields, that is, the nuclear observables evolve according to the unperturbed nuclear Hamiltonian

$$J^\mu(x) = e^{iH_n x_0} J^\mu(\boldsymbol{x}) e^{-iH_n x_0}, \tag{4.8}$$

$$B^{\mu\nu}(x,y) = e^{iH_n x_0} B^{\mu\nu}(\boldsymbol{x}, \boldsymbol{y}) e^{-iH_n x_0} \delta(x_0 - y_0). \tag{4.9}$$

The scattering matrix for this process is given by

$$\mathcal{S} = T\left[ \exp\left( -i \int_{-\infty}^{\infty} H_{\text{int}}(x) d^4x \right) \right], \tag{4.10}$$

and the terms that contribute to the process in Eq. (4.1) to the lowest order are

$$\mathcal{S}^{(1)} = -\frac{i}{2} \int d^4x\, d^4y\, T[B_{\mu\nu}(x,y) A^\mu(x) A^\nu(y)], \tag{4.11}$$

$$\mathcal{S}^{(2)} = -\frac{1}{2} \int d^4x\, d^4y\, T[J_\mu(x) A^\mu(x) J_\mu(y) A^\mu(y)]. \tag{4.12}$$

Photon and fermion fields operators commute, therefore time ordered products can be written as

$$\mathcal{S}^{(1)} = -\frac{i}{2} \int d^4x\, d^4y\, B_{\mu\nu}(x,y) T[A^\mu(x) A^\nu(y)], \tag{4.13}$$

$$\mathcal{S}^{(2)} = -\frac{1}{2} \int d^4x\, d^4y\, T[J_\mu(x) J_\mu(y)] T[A^\mu(x) A^\mu(y)], \tag{4.14}$$

and the transition amplitude is given by $\mathcal{M} = \langle f | \mathcal{T} | i \rangle$, where the transition matrix is $\mathcal{S} = \mathbb{I} + i\mathcal{T}$. In our case the initial and final nuclear states are

$$|i\rangle = |\xi_i J_i M_i, \boldsymbol{P}_i\rangle \otimes |0\rangle_\gamma, \tag{4.15}$$

$$|f\rangle = |\xi_f J_f M_f, \boldsymbol{P}_f\rangle \otimes |\boldsymbol{k}\varepsilon^{(\lambda)}; \boldsymbol{k}'\varepsilon^{(\lambda')}\rangle, \tag{4.16}$$

being

$$|\boldsymbol{k}\varepsilon^{(\lambda)}; \boldsymbol{k}'\varepsilon^{(\lambda')}\rangle = a_{\boldsymbol{k},\lambda}^\dagger a_{\boldsymbol{k}',\lambda'}^\dagger |0\rangle_\gamma. \tag{4.17}$$

The operators $a_{\boldsymbol{k},\lambda}^\dagger$ and $a_{\boldsymbol{k},\lambda}$ are the creation and annihilation operators for photons with the wave vector $\boldsymbol{k}$ and circular polarization $\lambda = \pm 1$, and satisfy the commutation relations

$$[a_{\boldsymbol{k}\lambda}, a_{\boldsymbol{k}'\lambda'}^\dagger] = -(2\pi)^3 2k_0 \delta^{(3)}(\boldsymbol{k} - \boldsymbol{k}') \eta^{\lambda\lambda'}. \tag{4.18}$$

The rest of the commutators vanish. For real photons the scalar and longitudinal states do not appear so $-\eta^{\lambda\lambda'} = \delta_{\lambda\lambda'}$. The photon field can be expanded in terms of the creation and annihilation operators as

$$A_\mu(x) = \sum_\lambda \int \frac{d^3\boldsymbol{k}}{(2\pi)^3} \frac{1}{2k_0} \{\varepsilon_{\mu\lambda}(k) a_{\boldsymbol{k}\lambda} e^{-ikx} + \varepsilon^*_{\mu\lambda}(k) a^\dagger_{\boldsymbol{k}\lambda} e^{ikx}\}, \tag{4.19}$$

where $\varepsilon_{\mu\lambda}(k)$ is the polarization vector. Since $[A_\mu(x), A_\nu(y)] = 0$, the time ordered product is irrelevant

$$\langle 0_\gamma | a_{\boldsymbol{k}\lambda} a_{\boldsymbol{k}'\lambda'} T[A_\mu(x), A_\nu(y)] | 0_\gamma \rangle = \langle 0_\gamma | a_{\boldsymbol{k}\lambda} a_{\boldsymbol{k}'\lambda'} A_\mu(x) A_\nu(y) | 0_\gamma \rangle, \tag{4.20}$$

hence one finds

$$\langle 0_\gamma | a_{k\lambda_1} a_{k'\lambda_2} A_\mu(x), A_\nu(y) | 0_\gamma \rangle = \varepsilon^*_{\mu\lambda}(k) \varepsilon^*_{\nu\lambda'}(k') e^{i(kx+k'y)} + \varepsilon^*_{\mu\lambda'}(k') \varepsilon^*_{\nu\lambda}(k) e^{i(k'x+ky)}. \tag{4.21}$$

Introducing Eq. (4.21) into Eq. (4.13) and Eq. (4.14) and taking into account the symmetry properties of the integrals (note that from Eq. (4.37) the two-photon operator has the symmetry $B_{\mu\nu}(x, y) = B_{\nu\mu}(y, x)$), the transition amplitudes are

$$\mathcal{M}^{(i)}_{\text{fi}}(k'\lambda', k\lambda) = i l^{\mu\nu}(k'\lambda', k\lambda) \int d^4x d^4y e^{i(k'x+ky)} N^{(i)}_{\mu\nu}(x, y), \tag{4.22}$$

with

$$l^{\mu\nu}(k'\lambda', k\lambda) = \varepsilon^{\mu*}_{\lambda'}(k') \varepsilon^{\nu*}_\lambda(k), \tag{4.23}$$

and

$$N^{(1)}_{\mu\nu}(x, y) = i\langle \xi_f J_f M_f, \boldsymbol{P}_f | B_{\mu\nu}(x, y) | \xi_i J_i M_i, \boldsymbol{P}_i \rangle, \tag{4.24}$$

$$N^{(2)}_{\mu\nu}(x, y) = \langle \xi_f J_f M_f, \boldsymbol{P}_f | T[J_\mu(x) J_\nu(y)] | \xi_i J_i M_i, \boldsymbol{P}_i \rangle. \tag{4.25}$$

If we introduce the integral representation of the Heaviside distribution that appears in the time ordering operator, as well as the closure relation between the two current operators the transition amplitudes are

$$\mathcal{M}^{(i)}_{\text{fi}}(k'\lambda', k\lambda) = (2\pi)\delta(k_0 + k'_0 + P_f - P_i) l^{\mu\nu}(k\lambda, k'\lambda') \int d^3x d^3y e^{-i(\boldsymbol{k}'\cdot\boldsymbol{x}+\boldsymbol{k}\cdot\boldsymbol{y})} N^{(i)}_{\mu\nu}(\boldsymbol{x}, \boldsymbol{y}), \tag{4.26}$$

with $N^{(1)}_{\mu\nu}(\boldsymbol{x}, \boldsymbol{y}) = -B_{\mu\nu}(\boldsymbol{x}, \boldsymbol{y})$ and

$$N^{(2)}_{\mu\nu}(\boldsymbol{x}, \boldsymbol{y}) = \sum_n \left[ \frac{\langle \xi_f J_f M_f, \boldsymbol{P}_f | J_\mu(\boldsymbol{x}) | \xi_n J_f M_n, \boldsymbol{P}_n \rangle \langle \xi_n J_f M_n, \boldsymbol{P}_n | J_\nu(\boldsymbol{y}) | \xi_i J_i M_i, \boldsymbol{P}_i \rangle}{E_n + k_0 - E_i - i\epsilon} \right. $$
$$\left. + \frac{\langle \xi_f J_f M_f, \boldsymbol{P}_f | J_\mu(\boldsymbol{y}) | \xi_n J_f M_n, \boldsymbol{P}_n \rangle \langle \xi_n J_f M_n, \boldsymbol{P}_n | J_\nu(\boldsymbol{x}) | \xi_i J_i M_i, \boldsymbol{P}_i \rangle}{E_n + k'_0 - E_i - i\epsilon} \right]. \tag{4.27}$$

The centre of mass contribution to the transition amplitude can be omitted since we have the freedom to chose as the coordinate system the one that coincides with the centre of mass and assume that the nucleus is initially at rest. Moreover, it can also be assumed that the nucleus in the final state is at rest since recoils effects $\sim O(k_0/Am_n)$ are negligible. This allows one to replace $|n, \boldsymbol{P}_n\rangle$ by the intrinsic state $\rightarrow |\xi_n J_n M_n\rangle$ in Eq. (4.26). It is convenient to define the transition operator as

$$T_{\text{fi}}^{(i)}(k'\lambda', k\lambda) = \int d^3x d^3y \, l^{\mu\nu}(k'\lambda', k\lambda) e^{-i(\boldsymbol{k}'\cdot\boldsymbol{x} + \boldsymbol{k}\cdot\boldsymbol{y})} N_{\mu\nu}^{(i)}(\boldsymbol{x}, \boldsymbol{y}), \qquad (4.28)$$

where

$$N_{\mu\nu}^{(2)}(\boldsymbol{x}, \boldsymbol{y}) = J_\mu(\boldsymbol{x}) \frac{1}{H_{\text{nucl}} - E_i + k_0 - i\epsilon} J_\nu(\boldsymbol{y}) + J_\nu(\boldsymbol{y}) \frac{1}{H_{\text{nucl}} - E_i + k_0' - i\epsilon} J_\mu(\boldsymbol{x}). \qquad (4.29)$$

The good angular momentum of the quantum states makes appropriate the introduction of the multipolar expansion so the electromagnetic transition operator can be expanded as a superposition of irreducible tensor operators of definite angular momentum, as we will describe in Sec. 4.1.3.

### 4.1.2   Nuclear current and two-photon operators in the non-relativistic limit

In the simplest picture of nuclei, they are modeled as a system of non-relativistic point particles. Within this approach the electromagnetic current operator is just the sum of the individual current operators associated to each proton and neutron. This description corresponds to what we have already introduced in Sec. 2.2.1 as the impulse approximation. However, a more complete description needs to include the missing subnuclear degrees of freedom.

As we have seen the current operator can be derived from the full Hamiltonian and it is restricted by gauge invariance through the continuity equation

$$\boldsymbol{\nabla} \cdot \boldsymbol{j} + i[H, \rho(x)] = 0, \qquad (4.30)$$

which also links the current with the nuclear interaction.

The nucleon-nucleon interaction can be described at long internucleon distances by pion exchange, and at medium and short distances by heavier meson exchange or contact interactions. Moreover, inside nuclei there are other phenomena not associated with meson exchange but with excitations of the nucleon (isobar configurations), where the $\Delta$-isobar is considered the dominant contribution [236]. In this framework effective two- and many-body operators acting on nucleon's generalized coordinates arise from eliminating subnucleon degrees of freedom. Therefore the nuclear electromagnetic current and two-photon operator can be expressed as a sum of one-, two-, and many-body operators

$$\rho(\boldsymbol{q}) = \sum_i \rho_i^{(1)}(\boldsymbol{q}) + \sum_{i<j} \rho_{ij}^{(2)}(\boldsymbol{q}) + \ldots, \tag{4.31}$$

$$\boldsymbol{j}(\boldsymbol{q}) = \sum_i \boldsymbol{j}_i^{(1)}(\boldsymbol{q}) + \sum_{i<j} \boldsymbol{j}_{ij}^{(2)}(\boldsymbol{q}) + \ldots, \tag{4.32}$$

$$B_{lm}(\boldsymbol{q},\boldsymbol{q}') = \sum_i B_{lm,i}^{(1)}(\boldsymbol{q},\boldsymbol{q}') + \sum_{i<j} B_{lm,ij}^{(2)}(\boldsymbol{q},\boldsymbol{q}') + \ldots. \tag{4.33}$$

In our work we only consider one-body operators and restrict to the non-relativistic limit, which is a good approximation for the range of photon energies in the nuclear transitions studied in our work. Considering only terms up to $1/m$, with $m$ the nucleon mass, the non-relativistic expansion of the one-body current and two-photon operators are

$$\rho_{NR}^{(1)}(\boldsymbol{q}) = \sum_i e_i e^{i\boldsymbol{q}\cdot\boldsymbol{r}_i}, \quad | \quad \rho_{NR}^{(1)}(\boldsymbol{x}) = \sum_i e_i \delta(\boldsymbol{x}-\boldsymbol{r}_i) \tag{4.34}$$

$$\boldsymbol{j}_{NR}^{(1)}(\boldsymbol{q}) = \sum_i \frac{e_i}{2m}\{\boldsymbol{p}_i, e^{i\boldsymbol{q}\cdot\boldsymbol{r}_i}\} - \sum_i \frac{i\mu_i}{2m}\boldsymbol{q}\times\boldsymbol{\sigma}_i e^{i\boldsymbol{q}\cdot\boldsymbol{r}_i}, \tag{4.35}$$

$$\boldsymbol{j}_{NR}^{(1)}(\boldsymbol{x}) = \sum_i \frac{e_i}{2m}\{\boldsymbol{p}_i, \delta(\boldsymbol{x}-\boldsymbol{r_i})\} + \sum_i \frac{i\mu_i}{2m}\boldsymbol{\nabla}_i\times\boldsymbol{\sigma}_i\delta(\boldsymbol{x}-\boldsymbol{r_i}) \tag{4.36}$$

$$B_{lm}^{(1)}(\boldsymbol{q},\boldsymbol{q}') = -\sum_i \frac{e_i^2}{m}e^{i(\boldsymbol{q}+\boldsymbol{q}')\cdot\boldsymbol{r}_i}\delta_{lm} \quad | \quad B_{lm}^{(1)}(\boldsymbol{x},\boldsymbol{y}) = -\sum_i \frac{e_i^2}{m}\delta_{lm}\delta(\boldsymbol{x}-\boldsymbol{r_i})\delta(\boldsymbol{y}-\boldsymbol{r_i}). \tag{4.37}$$

Here the charge and magnetic moments of the $i$-nucleon are defined by

$$e_i \equiv \frac{e}{2}(1+t_3(i)), \tag{4.38}$$

$$\mu_i \equiv \mu_p \frac{1}{2}(1+t_3(i)) + \mu_n \frac{1}{2}(1-t_3(i)), \tag{4.39}$$

where $\mu_{p,n}$ are the proton and neutron magnetic moments, and $t_3$ is the third component of the nucleon isospin operator (see App. B for details). The magnetic dipole operator is $\boldsymbol{\mu}_{n,p} = \mu_N(g_l^{(n,p)}\boldsymbol{l} + g_s^{(n,p)}\boldsymbol{s})$, with the nuclear magneton $\mu_N = e\hbar/2m_pc$ and orbital and spin g-factors $g_l^{(n)} = 0$, $g_l^{(p)} = 1$, $g_s^{(n)} = -3.826$ and $g_s^{(p)} = 5.586$. It is implicit here that both the proton and the neutron do not have internal structure. As in the case of the axial current, here the internal structure of nucleons is parametrized by the form factors $F_1(Q^2)$ and $F_2(Q^2)$, known as Dirac and Pauli form factors respectively, which enter in the covariant definition of the single-nucleon current by

$$j^\mu = \bar{u}(\boldsymbol{p}',\sigma')\left[F_1(q^2)\gamma^\mu + F_2(q^2)\frac{i\sigma^{\mu\nu}q_\nu}{2m}\right]u(\boldsymbol{p},\sigma), \quad q = p'-p. \tag{4.40}$$

The isospin structure of the form factors takes the form [188] $F_i = (F_i^S + F_i^V t_3)/2$, hence the charge and magnetic moment defined in Eqs. (4.38) and (4.39) are generalized to

$$e_i \equiv \frac{1}{2} \left[ G_{E,i}^S(q^2) + G_{E,i}^V(q^2) t_3(i) \right], \tag{4.41}$$

$$\mu_i \equiv \frac{1}{2} \left[ G_{M,i}^S(q^2) + G_{M,i}^V(q^2) t_3(i) \right], \tag{4.42}$$

with

$$G_E(q^2) = F_1(q^2) - \frac{q^2}{4m^2} F_2(q^2), \quad G_M(q^2) = F_1(q^2) + F_2(q^2). \tag{4.43}$$

At low momentum transfer they take the values $G_E^S(0) = G_E^V(0) = 1$, $G_M^S(0) = \mu_p + \mu_n$ and $G_M^V(0) = \mu_p - \mu_n$, so we recover the definitions in Eq. (4.38) and Eq. (4.39).

### 4.1.3   Multipole decomposition

Due to the fact that nuclear states have a well defined total angular momentum it is useful to decompose the charge and current operators in terms of operators which transfer a definite angular momentum. The multipole expansion is given by [10]

$$\varepsilon_\lambda^\mu(k) e^{i\boldsymbol{k}\cdot\boldsymbol{x}} = -(2\pi)^{\frac{1}{2}} \sum_{\substack{L,M \\ S=0,1}} \hat{L} \lambda^S A_{L,M}^\mu(S, k_0, \boldsymbol{x}) D_{M\lambda}^L(R), \tag{4.44}$$

where for $S = 0, 1$ we have the electric $A_{L,M}^\mu(E, k_0, \boldsymbol{x})$ and magnetic multipoles $A_{L,M}^\mu(M, k_0, \boldsymbol{x})$, respectively. $D_{M\lambda}^L(R)$ are the Wigner D-matrices, with $R$ representing the rotation of the quantization axis into the direction of the wave vector $\boldsymbol{k}$, and $\hat{L} \equiv (2L+1)^{1/2}$.

Since there is some freedom in the selection of the vector potential, the authors in Ref. [10] choose the Landau gauge because within this gauge the multipole expansion is simplified in the physics regime where the photon wavelength is much longer than the dimensions of the nuclear probe ($kx \ll 1$), known as *long wave approximation*. In this gauge the scalar and vector components of the electric multipole of order $L$ are

$$A_{LM}^0(E, k_0, \boldsymbol{x}) = i^L \left( \frac{L+1}{L} \right)^{1/2} j_L(kx) Y_{LM}(\hat{\boldsymbol{x}}), \tag{4.45}$$

$$\boldsymbol{A}_{LM}(E, k_0, \boldsymbol{x}) = \frac{i^{L-1}}{k_0} \left[ \frac{\boldsymbol{\nabla} \times (\boldsymbol{x} \times \boldsymbol{\nabla})}{(L(L+1))^{1/2}} + \left( \frac{L+1}{L} \right)^{1/2} \boldsymbol{\nabla} \right] j_L(kx) Y_{LM}(\hat{\boldsymbol{x}}), \tag{4.46}$$

where $j_L(kx)$ is the spherical Bessel function. While, the scalar and vector components of the magnetic multipole of order $L$ are

$$A_{LM}^0(M, k_0, \boldsymbol{x}) = 0, \tag{4.47}$$

$$\boldsymbol{A}_{LM}(M, k_0, \boldsymbol{x}) = \frac{i^{L-1}}{(L(L+1))^{1/2}} (\boldsymbol{x} \times \boldsymbol{\nabla}) j_L(kx) Y_{LM}(\hat{\boldsymbol{x}}). \tag{4.48}$$

In the long-wave limit, the spherical Bessel function can be approximated by $j_L(kx) \simeq (kx)^L/(2L+1)!!$, which makes $\boldsymbol{A}_{LM}(E, k_0, \boldsymbol{x}) \simeq 0$.

### 4.1.4 Generalized nuclear polarizabilities

Introducing the multipolar expansion defined in Eq.(4.44) into $l^{\mu\nu}(k'\lambda', k\lambda)$ we obtain

$$l^{\mu\nu}(k'\lambda', k\lambda)e^{-i(\boldsymbol{k'}\cdot\boldsymbol{x}+\boldsymbol{k}\cdot\boldsymbol{y})} = 2\pi \sum_{\substack{L,L',M,M' \\ S,S'}} \zeta \hat{L}\hat{L}'\lambda^S \lambda'^{S'} A^{\mu}_{L',-M'}(S', k'_0, \boldsymbol{x}) A^{\nu}_{L,-M}(S, k_0, \boldsymbol{y})$$
$$\times D^{L*}_{M\lambda}(R) D^{L'*}_{M'\lambda'}(R'), \qquad (4.49)$$

where the phase $\zeta = (-1)^{L+L'+M+M'+\delta_{\mu0}+\delta_{\nu0}}$ comes from the complex conjugate of $A^{\mu}_{L,M}(S, k_0, \boldsymbol{x})$. As in Eq. (4.44) $R'$ is the rotation matrix of the quatization axis into the direction of the wave vector $\boldsymbol{k'}$. Taking into account that the direct product of two tensors can be expanded as a sum of irreducible tensors

$$A^{\nu}_{L',-M'} A^{\mu}_{L,-M} = \sum_J (-1)^{J-M_J} \hat{J} \begin{pmatrix} L & L' & J \\ M & M' & M_J \end{pmatrix} \left\{ A^{\nu}_L \otimes A^{\mu}_{L'} \right\}_{J,M_J}, \qquad (4.50)$$

where the 3$j$ symbol is defined from the Clebsch-Gordan $(LML'M'|JM_J)$ coefficient by

$$\begin{pmatrix} L & L' & J \\ M & M' & M_J \end{pmatrix} \equiv (-1)^{L-L'-M_J} \hat{J}^{-1} (LML'M'|J-M_J). \qquad (4.51)$$

The Wigner D-matrices satisfy the relation $D^{J*}_{MM'}(R) = (-1)^{M'-M} D^J_{-M,-M'}(R)$, together with the symmetry properties of the 3$j$ symbol, we can rewrite Eq. (4.49)

$$l^{\mu\nu}(k'\lambda', k\lambda)e^{-i(\boldsymbol{k'}\cdot\boldsymbol{x}+\boldsymbol{k}\cdot\boldsymbol{y})} = (2\pi) \sum_{\substack{J,L,L',M,M' \\ S,S'}} (-1)^{\delta_{\mu0}+\delta_{\nu0}+J+L+L'-M_J} \hat{L}\hat{L}'\hat{J}\lambda^S \lambda'^{S'} \begin{pmatrix} L & L' & J \\ M & M' & -M_J \end{pmatrix}$$
$$\times \left\{ A^{\mu}_{L'}(S', k'_0, \boldsymbol{x}) \otimes A^{\nu}_L(S, k_0, \boldsymbol{y}) \right\}_{J,M_J} D^L_{M-\lambda}(R) D^{L'}_{M'-\lambda'}(R'). \qquad (4.52)$$

If we introduce this expression into Eq. (4.28) we arrive at [10]

$$T^{(i)}_{\mathrm{fi}}(k'\lambda', k\lambda) = \sum_{\substack{J,L,L',M,M' \\ S,S'}} (-1)^{L+L'-M_J} \hat{J}^2 \lambda^S \lambda'^{S'} \begin{pmatrix} L & L' & J \\ M & M' & -M_J \end{pmatrix}$$
$$\times P^{LL'(i)}_{JM_J} D^L_{M-\lambda}(R) D^{L'}_{M'-\lambda'}(R'), \qquad (4.53)$$

where $P_J^{L'L(i)}$ is the *generalized polarizability* operator

$$P_{JM_J}^{L'L(i)} = (-1)^J (2\pi) \frac{\hat{L}\hat{L}'}{\hat{J}} \int d^3x d^3y (-1)^{\delta_{\mu 0}+\delta_{\nu 0}} \left\{ A_{L'}^\mu(S',k_0',\boldsymbol{x}) \otimes A_L^\nu(S,k_0,\boldsymbol{y}) \right\}_{J,M_J} N_{\mu\nu}^{(i)}(\boldsymbol{x},\boldsymbol{y}),$$
(4.54)

which is an irreducible tensor operator of rank $J$ and is the part that contains the dynamical information of the transition amplitude. Since the transition amplitude is the matrix element of Eq. (4.53), applying the Wigner-Eckart theorem we get

$$T_{\mathrm{fi}}^{(i)}(k'\lambda',k\lambda) = (-1)^{J_f - M_i} \sum_{\substack{J,L,L',M,M' \\ S,S'}} (-1)^{L+L'} \hat{J}^2 \lambda^S \lambda'^{S'} \begin{pmatrix} L & L' & J \\ M & M' & -M_J \end{pmatrix}$$

$$\times \begin{pmatrix} J_f & J & J_i \\ -M_f & M_J & M_i \end{pmatrix} \langle J_f || P_J^{LL'(i)} || J_i \rangle D_{M-\lambda}^L(R) D_{M'-\lambda'}^{L'}(R'). \quad (4.55)$$

Here, $\langle J_f || P_J^{LL'(i)} || J_i \rangle$ is the reduced matrix element of the generalized polarizability operator $P_J^{LL'(i)}$, related to the matrix element $\langle J_f M_f | P_{JM_J}^{L'L(i)} | J_i M_i \rangle$ by the Wigner-Eckart Theorem [61]. We do not use a different symbol to differentiate between operators and their matrix elements, but it is clear from the text and expressions.

The two-photon nuclear polarizability is

$$P_J^{L'L(1)} = (-1)^{J+1} (2\pi) \frac{\hat{L}\hat{L}'}{\hat{J}} \langle J_f || \int d^3x d^3y (-1)^{\delta_{\mu 0}+\delta_{\nu 0}} \left[ A_{L'}^\mu(S',k_0',\boldsymbol{x}) \otimes A_L^\nu(S,k_0,\boldsymbol{y}) \right]_J B_{\mu\nu}(\boldsymbol{x},\boldsymbol{y}) || J_i \rangle,$$
(4.56)

while the two-current polarizability reads

$$P_J^{L'L(2)} = (-1)^J (2\pi) \frac{\hat{L}\hat{L}'}{\hat{J}} \sum_{n,J_n} \left[ \begin{Bmatrix} L & L' & J \\ J_i & J_f & J_n \end{Bmatrix} \frac{\langle J_f || O_{L'}(S',k_0') || J_n \rangle \langle J_n || O_L(S,k_0) || J_i \rangle}{E_n - E_i + k_0} + \right.$$

$$\left. + (-1)^{J-L-L'} \begin{Bmatrix} L' & L & J \\ J_i & J_f & J_n \end{Bmatrix} \frac{\langle J_f || O_L(S,k_0) || J_n \rangle \langle J_n || O_{L'}(S',k_0') || J_i \rangle}{E_n - E_i + k_0'} \right],$$
(4.57)

where the summation in $n$ goes over all possible intermediate states compatible with the multipolarity of the operators $O_{LM}(S,k_0)$, which correspond to the electric and magnetic multipole operators

$$O_{LM}(S,k_0) = \int d^3\boldsymbol{x} (-1)^{\delta_{\mu 0}} J_\mu(\boldsymbol{x}) A_{LM}^\mu(S,k_0,\boldsymbol{x}). \quad (4.58)$$

To go further, the two-photon and current operators can be written in terms of one-body nucleon operators as defined in Eqs. (4.34)-(4.37). In the case of the two-photon polarizability we have

$$
P_{JM_J}^{L'L(1)} = (-1)^{J+1} (2\pi) \frac{\hat{L}\hat{L}'}{\hat{J}} \int d^3x \, d^3y \left[ A_{L'}^i(S', k_0', \boldsymbol{x}) \otimes A_L^j(S, k_0, \boldsymbol{y}) \right]_{JM_J} \left[ \frac{\delta_{ij}}{m} \sum_{i=1}^{A} e_i^2 \delta(\boldsymbol{x} - \boldsymbol{x}_i) \delta(\boldsymbol{y} - \boldsymbol{x}_i) \right],
$$

$$
= (-1)^{J+1} (2\pi) \sum_{i=1}^{A} \frac{e_i^2}{m} \left[ \boldsymbol{A}_L(M, k_0, \boldsymbol{x}_i) \otimes \boldsymbol{A}_{L'}(M, k_0', \boldsymbol{x}_i) \right]_{JM} \quad nonumber
$$

$$
= -i^{L+L'} \sqrt{\pi} \frac{\hat{L}^3 \hat{L}'^3}{\hat{J}} \begin{pmatrix} L' & L & J \\ 0 & 0 & 0 \end{pmatrix} \begin{Bmatrix} L' & L & J \\ L & L' & 1 \end{Bmatrix} \sum_{i=1}^{A} \frac{e_i^2}{m} j_L(k_0 x_i) j_{L'}(k_0' x_i) Y_{JM_J}(\hat{\boldsymbol{x}}_i), \tag{4.59}
$$

$$
\simeq -\sqrt{\pi} i^{L+L'} \frac{k_0^L k_0'^L \hat{L}^2 \hat{L}'^2}{\hat{J} [(2L+1)!!(2L'+1)!!]^{1/2}} \begin{pmatrix} L' & L & J \\ 0 & 0 & 0 \end{pmatrix} \begin{Bmatrix} L' & L & J \\ L & L' & 1 \end{Bmatrix} \sum_{i=1}^{A} \frac{e_i^2}{m} x_i^{L+L'} Y_{JM_J}(\hat{\boldsymbol{x}}_i). \tag{4.60}
$$

In the last line the long-wave approximation has been used. Since only the spatial components of the two-photon operator play a role in the non-relativistic limit, and we have the freedom in the redefinition of the vector potential so as make that only the magnetic multipoles contribute in the long-wave approximation to the two-photon polarizability, then we have

$$
P_{JM_J}^{LL'(1)} = P_{JM_J}^{LL'(1)}(ML, ML'). \tag{4.61}
$$

The case of the two-current polarizability is more general because the current operator has both the charge and current density components and each of them will contribute to the electric and magnetic multipoles respectively. Furthermore, the multipolarity of the two photons depends on the nuclear states involved as imposed by angular momentum selection rules and it can be experimentally identified by analysing the angular distribution of the emitted photons. More important is the fact that not all multipoles contribute with the same strength to the transition but there is a hierarchy, a feature that can be either an advantage or a drawback depending on the physics process we want to study. In general the two-current polarizability is given by

$$
P_{JM_J}^{LL'(2)} = P_{JM_J}^{LL'(2)}(EL, EL') + P_{JM_J}^{LL'(2)}(ML, ML') + P_{JM_J}^{LL'(2)}(EL, ML') + P_{JM_J}^{LL'(2)}(ML, EL'), \tag{4.62}
$$

but the physics process we will want to study will restrict the terms that enter in the decay rate.

Nonetheless, for completeness we will give the expression of the electric and magnetic multipoles. In the long-range limit only the charge density enters in the electric multipole operator and no knowledge of the nuclear current is necessary (Siegert theorem)

$$O_{LM}(E, k_0) = -\frac{i^L k_0^L}{(2L+1)!!}\sqrt{\frac{L+1}{L}}\sum_{i=1}^{A}\int d^3x\, e_i \delta^{(3)}(\boldsymbol{r} - \boldsymbol{x}_i)r^L Y_{LM}(\hat{\boldsymbol{x}})$$

$$= \alpha_L(E, k_0)\sum_{i=1}^{A}e_i x_i^L Y_{LM}(\hat{\boldsymbol{x}}_i),\tag{4.63}$$

$$\alpha_L(E, k_0) = -\frac{i^L k_0^L}{(2L+1)!!}\sqrt{\frac{L+1}{L}}.\tag{4.64}$$

Likewise the magnetic multipole operator is

$$O_{LM}(M, k_0) = -\frac{i^L k_0^L}{(2L+1)!!\sqrt{L(L+1)}}\sum_{i=1}^{A}\int d^3x\Big(\frac{e_i}{2m}\{\boldsymbol{p}_i, \delta(\boldsymbol{x} - \boldsymbol{x_i})\}+$$

$$+\sum_i \frac{i\mu_i}{2m}\boldsymbol{\nabla}_i \times \boldsymbol{\sigma}_i \delta(\boldsymbol{x} - \boldsymbol{x_i}))\cdot\Big((\boldsymbol{x}\times\boldsymbol{\nabla})x^L Y_{LM}(\hat{\boldsymbol{x}})\Big)$$

$$= \frac{\mu_N}{\hbar c}\alpha_L(M, k_0)\left[\sum_{i=1}^{A}\left(\frac{2}{L+1}g_l^{(i)}\boldsymbol{l}_i + g_s^{(i)}\boldsymbol{s}_i\right)\cdot\boldsymbol{\nabla}_i\left(x_i^L Y_{LM}(\hat{\boldsymbol{x}}_i)\right)\right],\tag{4.65}$$

with $\alpha_L(M, k_0) = i^{-1}\alpha_L(E, k_0)$.

In order to calculate the nuclear matrix elements of the electromagnetic transition operators, we have used the occupation number representation as described in App. C. In this framework the NMEs of a one-body tensor operator is decomposed into the product of the single-particle matrix element and the one-body transition density matrix. The analyti—cal part is encoded in the reduced single-particle matrix element, which we defined below for the electric and magnetic transition operators. Moreover, as we will discuss in Ch. 5, we used the nuclear many-body codes to extract numerically the transition densities matrices.

The reduced single-particle matrix element of the electric multipole operator in Eq. (4.63) in the harmonic oscillator basis defined in Ch. 3 is

$$\langle a||\boldsymbol{O}_L^i(E, k_0)||b\rangle = \alpha_L(E, k_0)\langle a||e_i r_i^L \boldsymbol{Y}_L(\hat{\boldsymbol{x}}_i)||b\rangle$$

$$= \alpha_L(E, k_0)\mathcal{R}_{ab}^L (-1)^{j_b - \frac{1}{2} + L}\frac{1 + (-1)^{l_a + l_b + L}}{2}\frac{\hat{j}_a \hat{j}_b \hat{L}}{\sqrt{4\pi}}\begin{pmatrix} j_a & j_b & L \\ \frac{1}{2} & -\frac{1}{2} & 0 \end{pmatrix},\tag{4.66}$$

with the radial integral $\mathcal{R}_{ab}^L$

$$\mathcal{R}_{ab}^L = \int_0^\infty g_{n_a l_a}(r)r^L g_{n_b l_b}r^2 dr.\tag{4.67}$$

In turn the reduced single-particle matrix element of the magnetic multipole operator in Eq. (4.65) is

$$
\begin{aligned}
\langle a||\boldsymbol{O}_L^i(M,k_0)||b\rangle =& \alpha_L(M,k_0)\frac{\mu_N}{\hbar c}\left[\frac{2}{L+1}g_l^{(i)}\langle a||\boldsymbol{l}_i\cdot\boldsymbol{\nabla}_i\left(x_i^L\boldsymbol{Y}_L(\hat{\boldsymbol{x}}_i)\right)||b\rangle\right.\\
&\left.+g_s^{(i)}\langle a||\boldsymbol{s}_i\cdot\boldsymbol{\nabla}_i\left(x_i^L\boldsymbol{Y}_L(\hat{\boldsymbol{x}}_i)\right)||b\rangle\right]\\
=& \alpha_L(M,k_0)\frac{\hat{j}_a\hat{j}_b\hat{L}}{\sqrt{4\pi}}\frac{1-(-1)^{l_a+l_b+L}}{2}\mathcal{R}_{ab}^{(L-1)}\begin{pmatrix}j_a & j_b & L\\ \frac{1}{2} & -\frac{1}{2} & 0\end{pmatrix}(L-\kappa)\\
&\times\left[g_l^{(i)}\left(1+\frac{\kappa_{ab}}{L+1}\right)-\frac{1}{2}g_s^{(i)}\right],
\end{aligned}
\tag{4.68}
$$

where $\kappa := (-1)^{l_a+j_a+\frac{1}{2}}\left(j_a+\frac{1}{2}\right)+(-1)^{l_a+j_b+L-\frac{1}{2}}\left(j_b+\frac{1}{2}\right)$.

### 4.1.5 Double-gamma decay selection rules and multipole hierarchy

Parity conservation in electromagnetic processes sets the selection rule

$$
\pi_i\pi_f = (-1)^{L+S+L'+S'}.
\tag{4.69}
$$

Therefore pure electric $\hat{\rho}_{JM_J}^{LL'(2)}(EL,EL')$ and pure magnetic $\hat{\rho}_{JM_J}^{LL'(1,2)}(ML,ML')$ polarizabilities connect states with parity change $\pi_i\pi_f = (-1)^{L+L'}$, while mixed polarizabilities connect states with the opposite parity change $\pi_i\pi_f = (-1)^{L+L'+1}$. This means that pure and mixed polarizabilities are not entangled.

On the other hand, angular momentum selection rules come from the tensor structure of the transition operator. The coupling of the angular momentum involved must satisfy the relations

$$
\Delta(J_iJ_fJ) \implies |J_i - J_f| \le J \le J_i + J_f,
\tag{4.70}
$$

$$
\Delta(LL'J) \implies |L - L'| \le J \le L + L',
\tag{4.71}
$$

$$
\Delta(J_iJ_nL') \implies |J_i - J_n| \le L' \le J_i + J_n,
\tag{4.72}
$$

$$
\Delta(J_fJ_nL) \implies |J_f - J_n| \le L \le J_f + J_n,
\tag{4.73}
$$

$$
\Delta(J_iJ_nL) \implies |J_i - J_n| \le L \le J_i + J_n,
\tag{4.74}
$$

$$
\Delta(J_fJ_nL') \implies |J_f - J_n| \le L' \le J_f + J_n,
\tag{4.75}
$$

in order for the transition amplitude not to vanish.

On the other hand, electromagnetic transitions follow a hierarchy, with electromagnetic multipoles of higher order being less important [61]. Also when comparing electric and magnetic multipoles of the same order electric multipoles have higher transition probabilities.

An order of magnitude estimation can be derived by assuming that the radial wave function $g_{nl}(r)$ is constant inside the nucleus and zero outside (see equations (6.60) and (6.61) in Ref. [61]). If we introduce this approximation into Eq. (4.55) for the two-current amplitude we have that magnetic multipoles are $\mathcal{O}(10^{-2})$ smaller than electric multipoles of the same order for transitions of the same energy, as it happen for single gamma decay. Therefore, $E1E1$, $M1M1$, and $M1E1$ are the dominant transitions. However, as we will discuss in Sec. 5.3.1 electric dipole transitions are difficult to calculate with the NSM, and our motivation has focused in the study of the relation between $\gamma\gamma(\text{M1M1})$ and $0\nu\beta\beta$-decay NMEs.

### 4.1.6 Transition probability

The differential decay rate is given by [10]

$$d\Gamma = (4\pi)^2 \sum_f \frac{d^3\boldsymbol{P}_f}{(2\pi)^3} \frac{d^3\boldsymbol{k}'}{(2\pi)^3} \frac{d^3\boldsymbol{k}}{(2\pi)^3} \frac{(2\pi)^4 \delta^4(P_f - P_i + k_0 + k_0')}{4k_0 k_0'} |\overline{T}_{\text{fi}}|^2, \qquad (4.76)$$

where the sum extends to all possible final states and the bar means an average over the initial states. The above equation is valid only if initial and final states are pure states. This is in general not the case and the transition probability has to be described in terms of the density matrices for the initial and final states. In this case we have density matrices both for initial and final nuclear states ($\rho^{i,f}$) and for the two photons ($\tau, \tau'$). Therefore

$$\sum_f |\overline{T}_{\text{fi}}|^2 = \sum_m T_{M_f M_i}^{\lambda'\lambda} \rho_{M_i M_i'}^i T_{M_f' M_i'}^{\mu'\mu*} \rho_{M_f M_f'}^f \tau_{\mu\lambda} \tau'_{\mu'\lambda'}, \qquad (4.77)$$

with $m \equiv (M_i, M_i', M_f, M_f', \lambda, \mu, \lambda', \mu')$. In the case of unoriented nuclei density matrices are

$$\rho_{M_i M_i'}^i \equiv \langle J_i M_i | \rho^i | J_i M_i' \rangle = \frac{1}{2J_i + 1} \delta_{M_i M_i'}, \quad \rho_{M_f M_f'}^f \equiv \langle J_f M_f | \rho^f | J_f M_f' \rangle = \delta_{M_f M_f'}. \qquad (4.78)$$

The above matrices represent the situation in which an average over initial states and a sum over final states is made, which characterizes the case where the initial states are randomly oriented and the final states are unobserved. Furthermore, in the case where only the direction of the emitted photons is determined experimentally and no information of the polarization is obtained, then the photon density matrices are

$$\tau_{\lambda\mu}^i \equiv \langle \lambda | \rho^i | \mu \rangle = \delta_{\lambda\mu}, \quad \tau_{\lambda'\mu'}^{\prime i} \equiv \langle \lambda' | \rho^i | \mu' \rangle = \delta_{\lambda'\mu'}, \qquad (4.79)$$

here again, summation over final states is considered.

Using the expansion of $T_{\mathrm{fi}}^{\lambda'\lambda}$ given by Eq. (4.55) in terms of irreducible tensors and omitting the superscript expansion we obtain

$$
\begin{aligned}
T_{M_f M_i}^{\lambda'\lambda} T_{M'_f M'_i}^{\mu'\mu*} =& (-1)^{M_i + M'_i} \sum_{\substack{J,L,L',M,M' \\ S,S'}} \sum_{\substack{J',K,K',N,N' \\ Q,Q'}} (-1)^{L+L'+K+K'} \hat{J}^2 \hat{J}'^2 \lambda^S \lambda'^{S'} \mu^Q \mu'^{Q'} \\
& \times \begin{pmatrix} L & L' & J \\ M & M' & -M_J \end{pmatrix} \begin{pmatrix} K & K' & J' \\ N & N' & -M'_J \end{pmatrix} \begin{pmatrix} J_f & J & J_i \\ -M_f & M_J & M_i \end{pmatrix} \begin{pmatrix} J_f & J' & J_i \\ -M'_f & M'_J & M'_i \end{pmatrix} \\
& \times P_J^{LL'} P_{J'}^{KK'*} D_{M-\lambda}^L(R) D_{M'-\lambda'}^{L'}(R') D_{N-\mu}^{K*}(R) D_{N'-\mu'}^{K'*}(R').
\end{aligned}
\tag{4.80}
$$

Therefore introducing the above expression into Eq. (4.77) and performing the sums in $m$ one gets

$$
\begin{aligned}
\sum_f |\overline{T}_{\mathrm{fi}}|^2 =& \hat{J}_i^{-2} \sum_{\substack{M_i,M_f \\ \lambda\lambda'}} \sum_{\substack{J,L,L',M,M' \\ S,S'}} \sum_{\substack{J',K,K',N,N' \\ Q,Q'}} (-1)^{L+L'+K+K'} \hat{J}^2 \hat{J}'^2 \lambda^{S+Q} \lambda'^{S'+Q'} \\
& \times \begin{pmatrix} L & L' & J \\ M & M' & -M_J \end{pmatrix} \begin{pmatrix} K & K' & J' \\ N & N' & -M'_J \end{pmatrix} \begin{pmatrix} J_f & J & J_i \\ -M_f & M_J & M_i \end{pmatrix} \begin{pmatrix} J_f & J' & J_i \\ -M_f & M'_J & M_i \end{pmatrix} \\
& \times P_J^{LL'} P_{J'}^{KK'*} D_{M-\lambda}^L(R) D_{M'-\lambda'}^{L'}(R') D_{N-\lambda}^{K*}(R) D_{N'-\lambda'}^{K'*}(R').
\end{aligned}
\tag{4.81}
$$

Combining the Clebsch-Gordan series expansion of the product of two $D$ matrices and their complex transformation properties, and using sums of $3j$ symbols over all the angular projection quantum numbers we arrive at

$$
\begin{aligned}
\sum_f |\overline{T}_{\mathrm{fi}}|^2 =& \hat{J}_i^{-2} \sum_{\lambda,\lambda'} \sum_{\substack{J,L,L' \\ S,S'}} \sum_{\substack{J',K,K' \\ Q,Q'}} \sum_\alpha (-1)^{\lambda'+\lambda} \hat{J}^2 \lambda^{S+Q} \lambda'^{S'+Q'} (-1)^{L'+K'+J-\Lambda} \begin{Bmatrix} L & K & \alpha \\ K' & L' & J \end{Bmatrix} \\
& \times (L - \lambda K \lambda | \alpha 0)(L' - \lambda' K' \lambda' | \alpha 0) P_J^{LL'} P_{J'}^{KK'*} D_{\Lambda 0}^\alpha(R) D_{\Lambda 0}^\alpha(R').
\end{aligned}
\tag{4.82}
$$

Due to the symmetry of the initial state it is always possible to select our coordinate system in such a way that the quantization axis coincides with $\mathbf{k}'$. Then $R' = (0,0,0)$ and $D_{\Lambda 0}^\alpha(0) = \delta_{\Lambda,0}$ and $D_{00}^\alpha(R) = P_\alpha(x)$ ($x \equiv \cos\theta_{12}$). Expanding the Clebsch-Gordan coefficients in terms of $3j$ symbols and defining

$$
\begin{aligned}
g_J^{LL'KK'} =& \sum_{\lambda\lambda'} \hat{\alpha}^2 (-1)^{\lambda'+\lambda} (-1)^{L-K+J-\Lambda} \lambda^{S+Q} \lambda'^{S'+Q'} \begin{pmatrix} L & K & \alpha \\ -\lambda & \lambda & 0 \end{pmatrix} \begin{pmatrix} L' & K' & \alpha \\ -\lambda' & \lambda' & 0 \end{pmatrix} \begin{Bmatrix} L & K & \alpha \\ K' & L' & J \end{Bmatrix}, \\
=& \hat{\alpha}^2 (-1)^{L-K+J-\Lambda} \begin{pmatrix} L & K & \alpha \\ -1 & 1 & 0 \end{pmatrix} \begin{pmatrix} L' & K' & \alpha \\ -1 & 1 & 0 \end{pmatrix} \begin{Bmatrix} L & K & \alpha \\ K' & L' & J \end{Bmatrix} \\
& \times \left[ 1 + (-1)^{S+Q+L+K+\alpha} \right] \left[ 1 + (-1)^{S'+Q'+L'+K'+\alpha} \right],
\end{aligned}
\tag{4.83}
$$

finally we arrive to

$$
\sum_f |\overline{T}_{\mathrm{fi}}|^2 = \hat{J}_i^{-2} \sum_{\substack{J,L,L' \\ S,S'}} \sum_{\substack{J',K,K' \\ Q,Q'}} \sum_\alpha \hat{J}^2 P_J^{LL'} P_{J'}^{KK'*} g_J^{LL'KK'} P_\alpha(x).
\tag{4.84}
$$

The general expression for the differential decay rate for an experiment that only measures the direction of the two photons and with unoriented and unobserved nuclei is

$$\frac{d\Gamma}{dx\,dk_0} = \frac{k_0 k_0'}{\pi} \sum_f |\overline{T}_{\text{fi}}|^2, \quad k_0 + k_0' = E_i - E_f \equiv Q_{\gamma\gamma}. \tag{4.85}$$

## 4.2 Double magnetic dipole nuclear matrix element from double isobaric analog states

The main motivation for our study of $\gamma\gamma$ decay is to investigate the possible correlation with $0\nu\beta\beta$ decay. If the nuclear structure aspects of the two transition operators follow a clear relationship when studied across a wide range of nuclei and effective interactions, a measurement of one operator could be used to make a prediction of the range of expected values of the other within a certain confidence level.

The use of electromagnetic decays from isobaric analog states to connect electromagnetic with electroweak transitions has been done since quite long ago [234]. In this early work, the measurement of the electric dipole $\gamma$ decay from the isobaric analog state (IAS) of $^{141}$Pr is used to obtain the first forbidden $\beta$ decay NME for $^{141}$Ce. More recently, some works provide a detailed analysis based in the spin-isospin common nature of the strong, weak and electromagnetic interactions oriented to learn about the spin-isospin structure of the excitations and transitions [235] or focused more on the neutrino-nuclear responses relevant for astrophysics and laboratory searches of neutrino properties [146]. The reason of choosing an electromagnetic observable as a possible observable related to $0\nu\beta\beta$ is because its measurement allows a clear determination of the NMEs once a suitable experimental set up is designed. A reasonable starting point would be to choose an operator with a similar spin dependence as the $\beta\beta$ operator, so double magnetic dipole decay seems the best candidate. Moreover, since the nuclear structure dependence of the NME comes from the transition density matrices, a closer spin-isospin structure of both transition densities can be expected to translate into an enhanced correspondence between both NMEs. To achieve this goal isospin symmetry guarantees a strong similarity between $\gamma\gamma$ and $\beta\beta$ decays if we choose the decay of the double isobaric analogue state (DIAS) of the initial $\beta\beta$-decay state as represented in Fig. 4.2. Then the $\gamma\gamma$ and $\beta\beta$ decays comprise the nuclei

$$\gamma\gamma: \qquad {}_{Z}^{A}Y_{N}^{*} \rightarrow {}_{Z}^{A}Y_{N} + 2\gamma \tag{4.86}$$

$$0\nu\beta\beta: \qquad {}_{Z-2}^{A}X_{N+2} \rightarrow {}_{Z}^{A}Y_{N} + 2e^{-} \tag{4.87}$$

with $N, Z$ the neutron and proton number. ${}_{Z}^{A}Y_{N}^{*}$ represents the excited $\beta\beta$ final nucleus in the DIAS. This is a state with isospin $T = T_z + 2$, with $T_z$ the isospin third component of

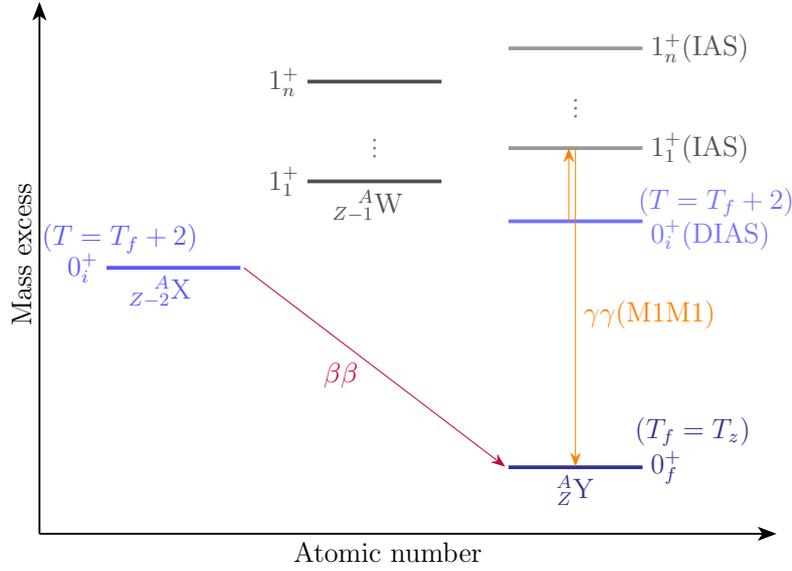

FIGURE 4.2: Nuclear states involved in $\gamma\gamma(M1M1)$ and $\beta\beta$ decays. In light blue are shown the initial $\beta\beta$ state, $0_i^+$, and its double isobaric analog state $0_i^+$(DIAS), the initial $\gamma\gamma$ state. In dark blue the common final $\beta\beta$ and $\gamma\gamma$ state $0_f^+$. In gray are shown all the intermediate $1_n^+$ isobaric analog states, which connect $0_i^+$(DIAS) and $0_f^+$ through the isovector component of the magnetic dipole operator $\boldsymbol{M1}$.

the $\beta\beta$ final nucleus, $T_z = (N - Z)/2$. To summarize, $\gamma\gamma$ decay involves

$$|0_i^+\rangle_{\gamma\gamma} \equiv |0_i^+\rangle_{\beta\beta}(\text{DIAS}) = \frac{T^- T^-}{K^{1/2}} |0_i^+\rangle_{\beta\beta}, \tag{4.88}$$

$$|0_f^+\rangle_{\gamma\gamma} \equiv |0_f^+\rangle_{\beta\beta}, \tag{4.89}$$

with $K$ a normalization constant and $T^- = \sum_i^A t_i^-$ the nuclear isospin lowering operator, which only changes $T_z$.

In a $0^+ \to 0^+$ $\gamma\gamma$ decay, angular momentum selection rules Eq. (4.70) impose $J = 0$, which in turn requires that $L = L'$ as given by Eq. (4.71). Since both multipoles must have the same order, from parity conservation one has that $S = S'$, so that both multipoles have either electric or magnetic character, that is in $P_0^{LL(2)}$ mixed terms like $P_0^{LL(2)}(ML, EL)$ and $P_0^{LL(2)}(EL, ML)$ are forbidden. Like in $\gamma$ decay there are no $E0E0$ or $M0M0$ transitions, because for $L = 0$, $M0$ vanishes (linked with the absence of magnetic monopoles) while $E0$ is a constant which can not connect two different nuclear states. This is the reason why $0^+ \longrightarrow 0^+$ $\gamma$ decay is forbidden. However, in the same way as in $\gamma$ decay $E0$ transitions are possible via *internal conversion* or *internal pair production*. However, experimentally they can be separated quite easily [10].

The leading contribution to the nuclear $0^+_{\text{DIAS}} \longrightarrow 0^+_{\text{GS}}$ $\gamma\gamma$ decay comes from $E1E1$ and $M1M1$ multipole transitions, but only the magnetic dipole operator, $\mathbf{M1}$, defined as

$$\mathbf{M1} = \mu_N \sqrt{\frac{3}{4\pi}} \sum_{i=1}^{A} (g_i^l \mathbf{l}_i + g_i^s \mathbf{s}_i), \tag{4.90}$$

has similar spin properties as the Gamow-Teller spin-isospin operator ($\boldsymbol{\sigma} \cdot \boldsymbol{\tau}$), which is the dominant contribution to $0\nu\beta\beta$ decay. All the parameters in Eq. (4.90), have been introduced in Sec. 4.1.2, where the electromagnetic charge and current operators were defined. Moreover, $\boldsymbol{l}$ is the angular-momentum operator and $\mathbf{s} = \frac{1}{2}\boldsymbol{\sigma}$ is the spin operator.

Although the electric dipole NME can be dominant, the angular distributions of $\gamma\gamma(E1E1)$, $\gamma\gamma(M1M1)$ transitions and of their interference are different and in principle they could be separated experimentally. On the other hand, the $\mathbf{M1}$ operator connects states with the same parity and angular momentum change of one unit, so the complete set in Eq. (4.57) are all states with spin-parity $J^{\phi} = 1^+$. Furthermore, in order to connect $|0^+_i\rangle_{\gamma\gamma}$ and $|0^+_f\rangle_{\gamma\gamma}$ these intermediate states must be the isobaric analog states of the intermediate nucleus with isospin $T = T_z + 1$, as depicted in Fig. 4.2, which assumes no isospin mixing.

As shown in Sec. 4.1.4 the dynamics of the transition operator is codified in the generalized polarizabilities and in the case of $\gamma\gamma(M1M1)$ decays it is given by

$$P_0^{11(2)}(M1M1, k_0, k'_0) = \frac{4\pi}{3\sqrt{3}} k_0 k'_0 \sum_n \left[ \frac{\langle 0^+_f ||\mathbf{M1}|| 1^+_n \rangle \langle 1^+_n ||\mathbf{M1}|| 0^+_i \rangle}{E_n - E_i + k'_0} + \frac{\langle 0^+_f ||\mathbf{M1}|| 1^+_n \rangle \langle 1^+_n ||\mathbf{M1}|| 0^+_i \rangle}{E_n - E_i + k_0} \right], \tag{4.91}$$

which depends on the two photon energies. However, since $k_0 + k'_0 = E_i - E_f$ one can rewrite the equation above as

$$P_0^{11(2)}(M1M1, k_0, k'_0) = \frac{2\pi}{3\sqrt{3}} k_0 k'_0 \sum_n \frac{\langle 0^+_f ||\mathbf{M1}|| 1^+_n \rangle \langle 1^+_n ||\mathbf{M1}|| 0^+_i \rangle}{\varepsilon_n \left(1 - \frac{\Delta\varepsilon^2}{2\varepsilon_n^2}\right)}, \tag{4.92}$$

where $\varepsilon_n = E_n - (E_i + E_f)/2$ and $\Delta\varepsilon = k_0 - k'_0$. In order to make the $\gamma\gamma$ NMEs independent of the photon energies we will restrict to the case where both photons share equally the available energy in the transition, $k_0 = k'_0 = Q_{\gamma\gamma}/2$. Since the transition probability is symmetric under $k_0 \leftrightarrow k'_0$ it is reasonable to start analysing this situation, which in turn is the most probable. In Sec. 4.3 we will discuss how such requierement can be realized experimentally. In our study of the correlation with the $0\nu\beta\beta$ NME ($M^{0\nu}$) we just need to calculate the $M^{\gamma\gamma}(M1M1)$ NME

$$M^{\gamma\gamma}(M1M1) = \sum_n \frac{1}{\varepsilon_n} \langle 0^+_f ||\mathbf{M1}|| 1^+_n \rangle \langle 1^+_n ||\mathbf{M1}|| 0^+_i \rangle. \tag{4.93}$$

The condition $k_0 \approx k'_0$ is needed because the $0^+_{\text{DIAS}}$ tends to be high-energy states and as we will show in Sec. 5 dominant dipole transitions between initial and final states usually have approximately an energy similar to $Q_{\gamma\gamma}$.

On the other hand, the first contribution to the two-photon polarizability $P_0^{11(1)}(ML, ML)$ is $P_0^{11(1)}(M1, M1)$ which at leading order corresponds to an one-body operator of Eq. (4.60), that can not connect $0_{DIAS}^+$ with $0_{GS}^+$ if they are states of good isospin. Although it is expected that these transitions probabilities are small, we computed their associated nuclear matrix element for different nuclei and effective interactions and we have verified that they are negligible.

## 4.3 Experimental measurements of double gamma decay

Experimental measurements of $\gamma\gamma$ decay of nuclear states are very rare and until quite recently limited only to the double magic nuclei $^{16}$O [237, 238], $^{40}$Ca [239] and $^{90}$Zr [239]. However, its atomic counterpart was studied quite long ago [240] mainly because the advantage of detecting less energetic photons emitted by electronic transitions (photoelectric absorption vs Compton scattering), and because of the advantageous population of the adequate excited state by the use of laser technology.

In general, the first excited state of nuclei have spin greater than $0^+$ and as a consequence first order $\gamma$ decay is allowed. However, for a few nuclei this state has $J^\pi = 0^+$ and $\gamma$ decay is strictly forbidden only competing with internal conversion and pair formation that can be separated experimentally. This translates into a branching ratio of $\Gamma_{\gamma\gamma}/\Gamma_{tot} \sim 10^{-4}$ for $0_2^+ \to 0_1^+$ $\gamma\gamma$ decay [10].

Only a few years ago, with the improvements in crystal detectors' energy resolution and timing properties, it was possible to overcome experimental challenges and detect $\gamma\gamma$ decay in competition with $\gamma$ decay [241] in $^{137}$Ba for the first time. This measurement is usually referred to as *competitive $\gamma\gamma$* or *$\gamma\gamma/\gamma$ decay*. The main difficulty faced by these kind of experiments is the unavoidable background of the much more likely $\gamma$ decay, with branching ratios of $\Gamma_{\gamma\gamma}/\Gamma_{tot} \sim 10^{-6}$. $\gamma$ decay, apart of being $10^6$ times more likely, can easily suffer Compton scattering leaving part of its energy in one detector of the experimental setup and the rest in other detector as sketched in Fig. 4.3. Although there were previous searches of the competitive $\gamma\gamma$ decay achieving good results with NaI [242], NaI(Tl) [243] and HPGe [244] detectors, none of them could report a positive signal with enough statistical significance for a discovery.

In Ref. [241] they measured $\gamma\gamma$ decay from the nuclear excited state $J^\pi = 11/2^-$ to the ground state of $^{137}$Ba with $J^\pi = 3/2^+$, populating the initial state through $\beta^-$ decays of $^{137}$Cs. The experiment used a planar equal relative angle configuration of five LaBr$_3$ : Ce detectors providing angular distribution data points of $72°$ and $144°$. A lead shielding between each detector was arranged in order to suppress the unwanted Compton scattering of 661.66keV. The collaboration reported a $\gamma\gamma/\gamma$ decay signal with a statistical significance of $5.5\sigma$ (standard desviations) determining a branching ratio of $\Gamma_{\gamma\gamma}/\Gamma_\gamma = 2.05 \times 10^{-6}$. From the observed angular correlations for the two data points at $72°$ and $144°$ together with the energy spectrum of individual photons at $72°$, the *off diagonal* (or mixed) polarizabilities for the quadrupole-quadrupole transition, $\alpha_{M2E2} = 33.9(2.9)$, and octuple-dipole transition $\alpha_{E3M1} = 10.1(4.2)$, were obtained from the fit. These data can be then be used to benchmark nuclear models used to compute these observables.

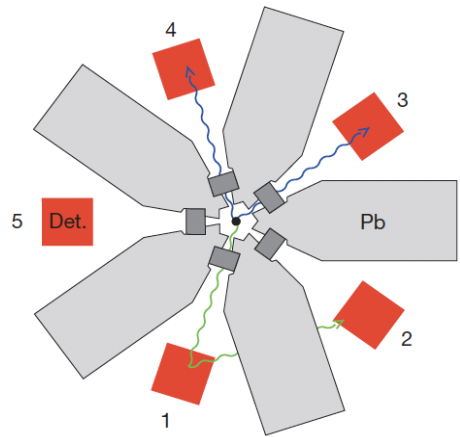

FIGURE 4.3: Experimental set up scheme of the experimental set up used in Ref. [241] for the measurement of the competitive $\gamma\gamma/\gamma$-decay in $^{137}$Ba, together with a representation of signal(blue) and background events(green). Figure taken from Ref. [241].

As it is common in experimental sciences an independent confirmation of this kind of breakthroughs is needed. With a similar experimental set up, the collaboration in Ref. [245] used the ELI Gamma Above Neutron Threshold (ELIGANT) detector system at the Extreme Light Infrastructure–Nuclear Physics (ELI-NP) facility to study this rare decay. They confirmed the existence of the competitive $\gamma\gamma$ decay with a statistical significance of $8.7\sigma$ and measured a branching ratio of $\Gamma_{\gamma\gamma}/\Gamma_\gamma = 2.62 \times 10^{-6}$. The detector set up was optimized to obtain a clear signal over a wide angular range with eleven $LaBr_3 : Ce$ and $CeBr_3$ detectectors. The $CeBr_3$ detectors were included to characterize and separate any possible natural radioactivity coming from lanthanum. As in Ref. [241], the distance from the source and the lead shielding were designed to separate true coincident events from multiple Compton scattering (maximize signal-to-noise ratio). However, their results showed an important $E3M1$ contribution to the $\gamma\gamma$ decay in contradiction with the initial results reported in Ref. [241].

While theoretical calculations presented in Ref. [245] using energy-density-functional theory together with quasiparticle-phonon model and Monte Carlo shell model are consistent with the best fit $\alpha_{E3M1}$ values, the calculated $\alpha_{M2E2}$ is not well described by data. Altogether, the results shown in Refs. [241, 245] demonstrate the potential of $\gamma\gamma$ decay to extract nuclear structure properties of atomic nuclei, and motivate current models to improve their predictions as well as encourage further experimental measurements.

## 4.4 Steps towards $\gamma\gamma(M1M1)$ measurements from DIAS

The lack of experimental measurements of $\gamma\gamma$ decay in atomic nuclei reflects the difficulties that these kind of experiments face. However, their clean connection with fundamental physics makes them a powerful tool to test nuclear models. They are not the only testing ground to investigate the electromagnetic behaviour of nuclei, other processes such as electron scattering or photo-nuclear reactions has been used to test nuclear degrees of freedom [236, 246].

The main interest of this work is to analyse the connection between the electroweak and electromagnetic sectors inside the nuclear domain, particularly by $0\nu\beta\beta$ and $\gamma\gamma(M1M1)$ nuclear decays. If there is a strong relation between both observables, the measurement of the possibly more accessible $\gamma\gamma(M1M1)$ nuclear decay not only would be a major breakthrough for nuclear spectroscopy but also would help in the prediction of $M^{0\nu}$ with an uncertainty that is based on systematic nuclear calculations and also based on an experimental measurement.

The first steps towards $\gamma\gamma(M1M1)$ nuclear decay measurement from DIAS are underway [153]. The aim of this pioneering proposal is to set the basis for a future program to measure the $\gamma\gamma$ decay first from the DIAS in $^{48}$Ti, but with the aim to extend it to other $\beta\beta$-final nuclei. $^{48}$Ti is a suitable candidate because it is the final nucleus of the $0\nu\beta\beta$ decay of $^{48}$Ca and because the $0^+_{DIAS}$ state at 17.379 MeV excitation energy is already known experimentally [247]. Since it is a high energy state, it is particle unbound, so Q-value systematics would allow nucleon emission. However, if isospin symmetry were exact its particle decay branch would be forbidden. Nonetheless, since isospin symmetry is approximate, particle emission from $0^+_{DIAS}$ is possible although strongly suppressed, making the $\gamma$-branch small but not negligible. Therefore, the first goal is to measure the $\gamma$- and particle widths of the $0^+_{DIAS}(T=4)$ state. In this respect, we have predicted the particle and single gamma widths with the NSM for the $0^+_{DIAS}$ in $^{48}$Ti and other interesting candidates in connection with $\beta\beta$ decay. These calculations are presented in Sec. 5.3.

Experimental access to DIAS is difficult mainly because they are high-lying nuclear states in a densely populated region, and therefore populating the state with the proper spin and isospin represents a challenge. Therefore, the difficulty of populating such states is what makes these measurements so scarce in the literature. So far, their interest had been the study of the quadratic form of the isobaric mass multiplet equation and deviations of it. For example for the isobaric $T=2$ multiplet, there are six complete $0^+_{T=2}$ quintets: $A = 8, 20, 24, 28, 32$ and $36$ for which the $0^+_{DIAS}$ in $^8$Be, $^{20}$Ne, $^{24}$Mg, $^{28}$Si, $^{32}$S and

$^{36}$Ar has been measured [248, 249]. States with $T = T_z + 2$ are usually populated by two-neutron pickup reactions $(p,t)$ or ($^3$He,n). The reaction $^{50}$Ti(p,t)$^{48}$Ti could be used to populate the $0^+_{T=4}$ state in $^{48}$Ti but it requires protons above 40 MeV and its predicted cross section is expected to be around 200 nb. Analog two-nucleon transfer reactions as (p,t) or ($^3$He,n) are under study by the Laboratori Nazionali di Legnaro (LNL) group [250] to populate $0^+$ DIAS such as $^{50}$Ti($^{24}$Mg,$^{26}$Mg)$^{48}$Ti and $^{46}$Ca($^{24}$Mg,$^{22}$Ne)$^{48}$Ti, respectively.

Previous to any measurement of a rare decay such as $\gamma\gamma$ decay with competing decaying processes, one must perform some testing simulations of the experimental set up beforehand. In Ref. [251] the first study of AGATA [252] (Advanced GAmma Tracking Array) capabilities to measure the competitive $\gamma\gamma/\gamma$ decay was presented. Despite the fact that previous attempts to measure $\gamma\gamma$ decay with HPGe were unsuccessful, mainly because their lower time resolution need to discriminate background events, technical advances have combined position sensitive Ge crystals with tracking capabilities while preserving their high-energy resolution. The AGATA spectrometer implements this new $\gamma$-detection concept with segmented Ge crystals, opening a new path to the measurement of competitive $\gamma\gamma/\gamma$ decay with a completely different experimental technique as used in previous positive results [241, 245]. Although this first feasibility study reveals some experimental challenges for $\gamma\gamma/\gamma$ decay measurements, it shows the key points in which the detector should be improved: tracking algorithm, increased position resolution, increased solid angle coverage.

Currently, a preliminary feasibility study of $\gamma\gamma$(M1M1) decay measurement from DIAS with LaBr$_3$ scintillator detectors using a spherical layout, instead of a planar one as in Refs. [241, 245], is underway [250, 253]. The study aims to look for ways to maximize the $\gamma\gamma$ detection efficiency to compensate the small branching ratios for this process $\Gamma_{\gamma\gamma}/\Gamma \sim 10^{-8} - 10^{-9}$ calculated in this thesis as presented in Sec. 5.3.2. Additionally, it aims to investigate ways to discriminate $\gamma\gamma$ from competing transitions using the good timing resolution of LaBr$_3$ detectors together with angular and energy measurements. In order to assess the performance of the experimental set-up and to optimize physical cuts on acquired data, it is important to take as starting point in the simulation a reliable prediction of the branching ratios of $\gamma\gamma$ decay with competing processes such as proton emission, $\gamma$ decay and internal pair conversion. This has been one of our contributions to the preliminary feasibility study of the measurement of $\gamma\gamma$M1M1-decay. However, a complete theoretical characterization is still needed and it is left for future work.

# Chapter 5

# Results

The possibility to answer fundamental questions about the nature of neutrinos and the symmetries that govern their mass generation and the matter-antimatter asymmetry in the universe has drawn the attention to $0\nu\beta\beta$ decay as a unique process to shed light to these long-standing puzzles. However, the NMEs of $0\nu\beta\beta$-decay are poorly known, and what is worst they enter quadratically into $0\nu\beta\beta$ decay half-life, $T_{1/2}^{0\nu}$. Throughout the years an intense theoretical effort has been dedicated to improve the many-body methods employed to make predictions as well as to understand different aspects of this hypothetical and very rare process. With the improvements in both areas, theoretical quantification of the error associated to the NMEs is starting to become feasible.

On the experimental side, nuclear structure measurements [138] or muon capture [254] are useful tools for constraining nuclear models, but they are not directly related with $0\nu\beta\beta$ decay. Particular interest has been placed in nuclear reactions such as two-nucleon transfer [142, 255] or double charge exchange transitions [256], since they give a valuable information on $0\nu\beta\beta$ decay in the same sense as charge-exchange reactions do for $\beta$ decay.

The good correlation found [147] in the NSM between $0\nu\beta\beta$ decay and double Gamow-Teller (DGT) transitions, encouraged the measurement of the later as well as the search for new correlations with observables possibly more accessible from the experimental point of view. The results presented in Ref. [147] from energy-density functional theory [257] follow the same correlation. This triggered an intense study of the $0\nu\beta\beta$-DGT correlation within other nuclear models such as the NSM with variational Monte Carlo, projected Hartree-Fock-Bogoliubov theory, no-core shell model, the in-medium generator coordinate method, and VS-IMSRG and QRPA as discussed in detail in Ref. [150].

The main goal of our study has been to begin with the analysis of the possible relation between second-order electromagnetic transitions and $0\nu\beta\beta$ decay motivated by the isospin symmetry and the similarity between M1M1 and $0\nu\beta\beta$ operators [12]. The good correlation found using the NSM in Ref. [12] for the transition $0^+_{DIAS} \rightarrow 0^+_{gs}$, encouraged our study of $\gamma\gamma$ decay within the VS-IMSRG, which is still ongoing. Additionally, the authors of Ref. [150] also look into this correlation within the QRPA method.

Although electromagnetic transitions are very promising as we will discuss at the end of this chapter, both DGT or M1M1 proposed transitions have not been measured yet. A major advancement was performed recently when we investigated the correlation with $2\nu\beta\beta$ [13], finding a particularly good correlation. Together with the experimental measurement of $2\nu\beta\beta$ decay, the theoretical correlation has been used for the first time to predict $0\nu\beta\beta$ NMEs with theoretical uncertainties.

Overall, this chapter has as a main goal to deepen into the details of the correlation study between $0\nu\beta\beta$-decay and $\gamma\gamma$-decay NMEs. The NSM results and major consequences of the $0\nu\beta\beta$-$\gamma\gamma$ correlation are described in Sec. 5.1. In Sec. 5.2 we introduce the preliminary study of $\gamma\gamma$-M1M1 decay from $0^+_{DIAS}$ and the role of the isospin mixing within the VS-IMSRG method. Afterwards, in Sec. 5.3 we present the first theoretical predictions of the competing processes of $\gamma\gamma$ decay $0^+_{DIAS} \rightarrow 0^+_{gs}$. In Sec. 5.4 we summarize the results of the correlation between $0\nu\beta\beta$ and $2\nu\beta\beta$ NMEs, performed within the NSM and the QRPA in Ref. [13]. Finally, in Sec. 5.5 the main results for the decay of $^{136}$Xe to the first excited state of $^{136}$Ba, published in Ref. [14] are discussed.

## 5.1 Shell model analysis of $\gamma\gamma(M1M1)$ decay and its correlation with $0\nu\beta\beta$ decay

As we have seen in Sec. 4.2 the dynamics of the $\gamma\gamma$ transition operator is encoded in the generalized polarizability, which for the case of $\gamma\gamma(M1M1)$ decay is given by Eq. (4.2). Since $P_0^{11(2)}(M1M1, k_0, k'_0)$ depends on the energies of the two emitted photons, to simplify the problem we restrict our analysis to the case where the two photons share their energy, $k_0 = k'_0 = (E_i - E_f)/2 =: Q/2$, which in turn is the most probable situation. Therefore in our study of the correlation we just need to calculate the NME presented in Eq. (4.93):

$$M^{\gamma\gamma}(M1M1) = \sum_n \frac{\langle 0^+_f || \mathbf{M1} || 1^+_n \rangle \langle 1^+_n || \mathbf{M1} || 0^+_i \rangle}{E_n - \frac{1}{2}(E_i + E_f)}, \qquad (5.1)$$

with the magnetic dipole operator defined in Eq. (4.90).

Since we are interested in the $\gamma\gamma$ connection with $\beta\beta$ decay, we have seen that isospin symmetry ensures the closest similarity between both processes if the states involved are those in Eqs. (4.88) and (4.89)

$$|0^+_i\rangle_{\gamma\gamma} \equiv |0^+_i\rangle_{\beta\beta}(\text{DIAS}) = \frac{T^- T^-}{K^{1/2}} |0^+_i\rangle_{\beta\beta}, \qquad (5.2)$$

$$|0^+_f\rangle_{\gamma\gamma} \equiv |0^+_f\rangle_{\beta\beta}. \qquad (5.3)$$

Equation (5.2) assumes that there is no isospin mixing. Furthermore, in all the $\gamma\gamma$ transitions selected for the correlation analysis, the connection between $|0_i^+\rangle_{\gamma\gamma}$ and $|0_f^+\rangle_{\gamma\gamma}$ is done by the isovector component of the magnetic dipole **M1** operator to the isobaric analog states $1_n^+$ with $T_n = T_f + 1$. The cases for which the inital or final $\beta\beta$ nuclei have $N = Z$ were analysed although not included in the $\gamma\gamma$-$\beta\beta$ correlation study since they have much higher $M^{\gamma\gamma}$ NMEs lying outside the correlation. For this nuclei each of the orbital, spin and spin-orbit components of the numerator of $M^{\gamma\gamma}$ has a much higher NMEs than the rest of the nuclei, and therefore they are outside the correlation. The condition $k_0 \simeq k_0'$ is needed because for transitions between $0^+(DIAS)$ and the ground state, $E_n - E_i \gg E_i$ and therefore Eq. (5.1) is not a good approximation of the actual $\gamma\gamma(M1M1)$ NME. A deeper explanation about this approximation will be given in Sec. 5.3.

### 5.1.1   Analysis of the NSM calculation of $\gamma\gamma(M1M1)$ NMEs

Once all the theoretical details are summarized and clear the next step is to perform the many body calculations. This has been done using the shell model codes ANTOINE [200, 202] and NATHAN [202].

In order to construct the NMEs, electromagnetic and weak transition subroutines implemented in ANTOINE and NATHAN were modified to compute one-body transition densities, $\langle \xi_f J_f ||[c_a^\dagger \tilde{c}_b]_1|| \xi_i J_i \rangle$ as defined in Eq. (C.2) of App. C. In this way they could be used to determine NMEs of any tensor operator with the same irreducible nature.

First, we calculate the initial $\beta\beta$ state $|0_i^+\rangle_{\beta\beta}$, that is, the ground state of the $\beta\beta$ parent nucleus, and rotate it in isospin space to obtain $|0_i^+\rangle_{\gamma\gamma}$ as in Eq. (5.2). Afterwards, we compute the ground state of the $\gamma\gamma$ emitting nucleus which corresponds to the $\beta\beta$ decay final nucleus as in Eq. (5.3). Finally, to obtain the set of intermediate states $\{1_n^+\}$ we perform a Lanczos strength function calculation. The Lanczos strength function method is based on the arbitrariness of selecting the pivot in the Lanczos algorithm to construct the strength distribution of any operator, which essentially gives its spectral decomposition. If we choose the isovector component of the magnetic dipole operator acting on the ground state as a pivot, $|v_1\rangle = \mathbf{M1}_{\text{IV}}|0_f^+\rangle$ and normalize it, we can expand it in the eigenbasis as

$$|u_1\rangle = \frac{\mathbf{M1}_{\text{IV}}|0_f^+\rangle}{\langle v_1|v_1\rangle} = \sum_n a_n |n\rangle, \qquad (5.4)$$

where the coefficients $a_n$ are the amplitude of the pivot in the $n$th eigenstate. At a given iteration $k$, the Lanczos vector is obtained by ortogonalization of $H^k|u_1\rangle$ to all the previous Lanczos vectors $|i\rangle$, $i < k$. Therefore, after the diagonalization of the tridiagonal $k \times k$ matrix, if $k$ is less than the dimension of the space, we get an approximate set of eigenstates with the proper angular-momentum and isospin quantum numbers, $|1_n^+, T_n = T_f + 1\rangle$. Additionally, the Lanczos strength function method ensures that only a few intermediate states are needed in order to get a good estimation of the $M^{\gamma\gamma}(M1M1)$ NME, and a quick convergence is reached as it is shown in Fig. 5.1. This convergence pattern is similar in other effective interactions usded.

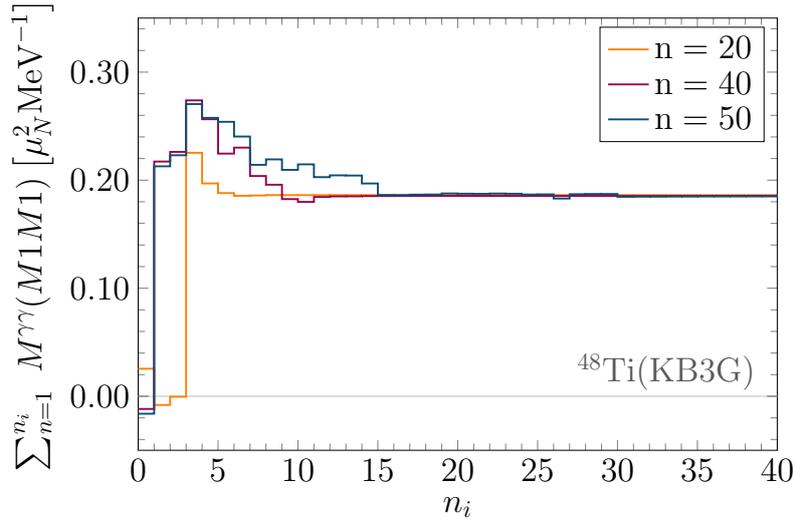

FIGURE 5.1: Cumulative sum of the $M^{\gamma\gamma}(M1M1)$ NME defined in Eq. (5.1), as a function of the number of intermediate states $n$ included using the Lanczos strength function. The plot shows the calculations of $M^{\gamma\gamma}(M1M1)$ for $^{48}$Ti using the KB3G effective interaction. The sets with total number of intermediate states $N = 20, 40$ and $50$ are shown.

Furthermore, we define the $\hat{M}^{\gamma\gamma}(M1M1)$ as the NME of the numerator in Eq. (5.1)

$$\hat{M}^{\gamma\gamma}(M1M1) = \sum_n \langle 0_f^+ ||\mathbf{M1}|| 1_n^+ \rangle \langle 1_n^+ ||\mathbf{M1}|| 0_i^+ \rangle, \tag{5.5}$$

$$= \langle 0_f^+ ||\mathbf{M1M1}|| 0_i^+ \rangle. \tag{5.6}$$

The $\hat{M}^{\gamma\gamma}(M1M1)$ NME allows one to compute its exact value, that is the value one would get if the infinite sum could be done, and that it has been calculated from the overlap of the vectors $\mathbf{M1}|0_i^+\rangle$ and $\mathbf{M1}|0_f^+\rangle$. Figure 5.2 shows the results for the cumulative sum of $\hat{M}^{\gamma\gamma}(M1M1)$ as a function of the number of intermediate states for $^{48}$Ti with the KB3G effective interaction (left panel) and for $^{72}$Zn with the GCN2850 effective interaction (right panel). Figure 5.2 also shows the exact value of this operator, by an horizontal dot-dashed line. Therefore, one can conclude that only a few intermediate states are needed to have a good approximation of the exact NME value. The minimum number of intermediate states for which this is achieved within $\sim 1\%$ is considered as a good criteria of convergence and used afterwards in the computation of $M^{\gamma\gamma}(M1M1)$ NMEs for which an exact value can not be computed in practice.

The energy denominator in $M^{\gamma\gamma}(M1M1)$ is evaluated using experimental energies when data is available [258, 259]. We have studied the effect of such energy corrections for $^{48}$Ti, because it is the only nucleus among all the nuclei studied for which the experimental energies $E_f$, $E_i(DIAS)$ and also the energy of the isobaric analog state $J^\pi = 6^+$ with $T = T_f + 1$ are known. Taking the latter, together with the calculated energy difference between the $6^+$ and $1^+$ states with $T = T_z + 1$, we correct the energy of the intermediate states $E_n$. With this experimental correction we find that $M^{\gamma\gamma}(M1M1)$ only deviates from the result obtained with calculated energies by 0.2%. Usually, the energy of DIAS or $T = T_z + 1$ states is unknown and since isospin-breaking effects cancel in $\varepsilon_n$ to a very good approximation [260], we have used the experimental data on states of the same isospin multiplet in neighboring nuclei: the $\beta\beta$ parent to fix $E_i$, and the $\beta\beta$ intermediate nucleus, when available, for $E_1$. Using these experimental energies the results for the NME $M^{\gamma\gamma}(M1M1)$ are modified less than 5% when compared to NME $M^{\gamma\gamma}(M1M1)$ calculated with the computed energies.

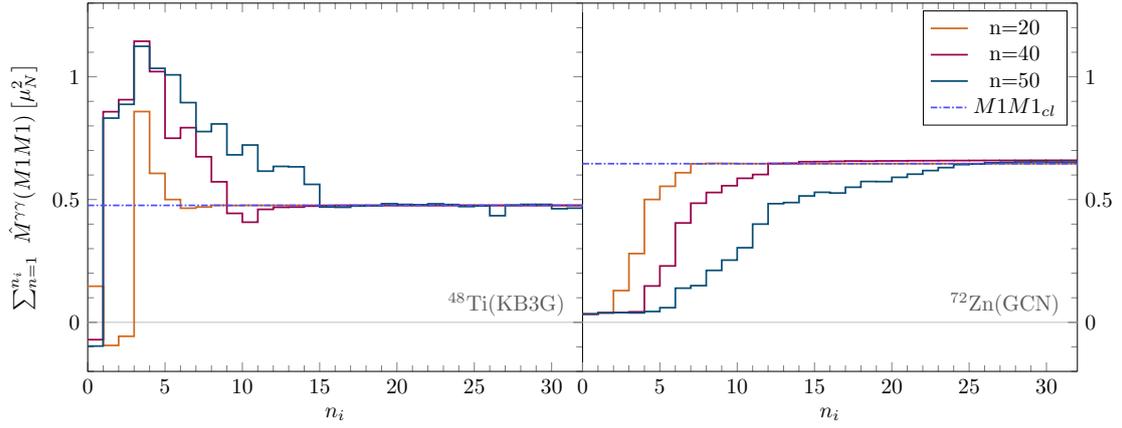

FIGURE 5.2: Cumulative sum of the $\tilde{M}^{\gamma\gamma}(M1M1)$ NME as a function of the number of intermediate states $n_i$ included. The plot shows the calculations done for $^{48}$Ti with the KB3G effective interaction (left panel) and in $^{72}$Zn with the GCN2850 effective interaction (right panel). The sets with total number of intermediate states $n = 20, 40$ and $50$ are shown, together with the closure value $\langle 0_f^+ || \mathbf{M1M1} || 0_i^+ \rangle$ displayed as a blue dot-dashed line.

After introducing the energy correction, we can evaluate $M^{\gamma\gamma}(M1M1)$ as a function of the energy of the intermediate states as shown in Fig. 5.3. In this figure the $M^{\gamma\gamma}(M1M1)$ NMEs are shown for three nuclei $^{48}$Ti, $^{82}$Se and $^{128}$Te in each of the configurations spaces used.

We obtain converged results around 1% after $20 - 100$ iterations in the Lanczos strength function method. A characteristic feature of the $M^{\gamma\gamma}(M1M1)$ NME illustrated in Fig. 5.3 is that intermediate states above 15 MeV do not contribute to the NME and there are only a few states which dominate the transition. A less evident feature that will have important consequences in our correlation study is the fact that the states that contribute most in lighter nuclei have lower energies than the ones that dominate in heavier nuclei, being the overall ratio of around 0.5. As we will see later, there are also slight differences

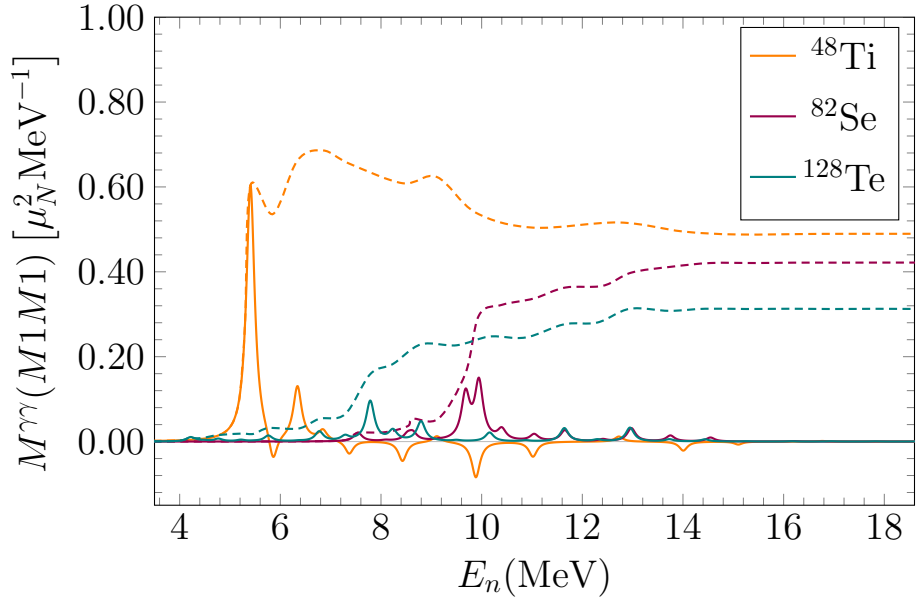

FIGURE 5.3: Individual states contribution (solid lines) and cumulative (dashed lines) values of the $M^{\gamma\gamma}(M1M1)$ NME as a function of the excitation energy of the intermediate states $E_n$. The plot shows the results of the $0^+_{DIAS} \longrightarrow 0^+_{gs}$ for $^{48}$Ti, $^{82}$Se and $^{128}$Te nuclei using the effective interactions KB3G, GCN2850 and QX respectively. The data has been smoothed by a superposition of 0.1 MeV width Lorentzians.

between different effective interactions.

Finally, in order to compare weak and electromagnetic decays we need to take into account that both processes are described by tensor operators of the same rank but different isospin projections. While the $0\nu\beta\beta$ operator changes $N$ and $Z$ by two units ($|\Delta T_z| = 2$), the $\gamma\gamma$ decay operator does not change $T_z$. Therefore, we limit to the comparison of the isospin-reduced NMEs, that is multiplying $M^{\gamma\gamma}(M1M1)$ NME by a factor that is the ratio of Clebsch-Gordan coefficients coming from aplying the Wigner-Eckart theorem to the isospin space, $\alpha = \frac{1}{2}\sqrt{(2+T_f)(3+2T_f)}$.

### 5.1.2   Correlation between $\gamma\gamma$-M1M1 and $0\nu\beta\beta$ decay NME

The results in Fig. 5.4 show the calculations performed for a wide range of $46 \leq A \leq 136$ nuclei within the nuclear shell model. They cover three different configuration spaces spanned by the following harmonic oscillator single-particle orbitals, for both protons and neutrons, and with isospin-symmetric interactions

- $40 f_{7/2}$, $1p_{3/2}$, $0f_{5/2}$ and $1p_{1/2}$ ($pf$ configuration space): $^{46-58}$Ti, $^{50-58}$Cr and $^{54-60}$Fe with the KB3G [169] and GXPF1B [170] effective interactions.

- $1p_{3/2}$, $0f_{5/2}$, $1p_{1/2}$ and $0g_{9/2}$ ($pfg$ configuration space): $^{72-76}$Zn $^{74-80}$Ge, $^{76-82}$Se, $^{82,84}$Kr with the GCN2850 [171], JUN45 [172] and JJ4BB [173] interactions.

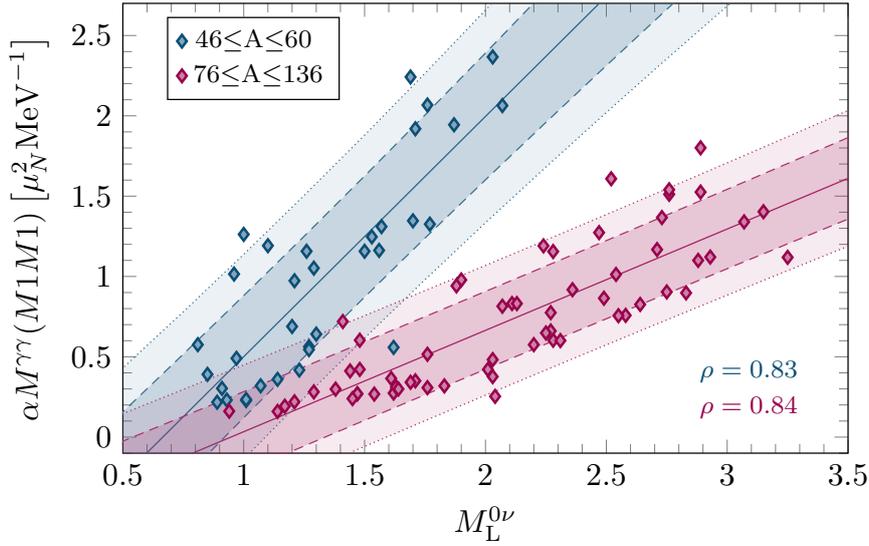

FIGURE 5.4: Standard $0\nu\beta\beta$ NMEs ($M_L^{0\nu}$), from Ref. [147] vs double-magnetic dipole ($M^{\gamma\gamma}(M1M1)$) NMEs obtained with different interactions for each configuration space. The $\alpha$ factor comes from comparing reduced matrix elements in isospin space. Solid lines correspond to the best linear fit while dashed and dotted lines represent the prediction bands at 68(90)% confidence level, respectively.

- $1d_{5/2}$, $0g_{7/2}$, $2s_{1/2}$, $1d_{3/2}$ and $0h_{11/2}$ (*sdgh* space): $^{124-132}$Te, $^{130-134}$Xe and $^{134,136}$Ba with the GCN5082 [171] and QX [174] interactions.

They had been computed using the shell model codes ANTOINE [200, 202] and NATHAN [202]. The $0\nu\beta\beta$ NMEs, calculated with the same configuration spaces and nuclear interactions, are taken from Ref. [147].

Figure 5.4 illustrates the relation between the standard $0\nu\beta\beta$-decay and double-magnetic dipole $M^{\gamma\gamma}(M1M1)$ NMEs. It shows a very good linear correlation both for the $pf$-shell nuclei including nineteen nuclei containing titanium, chromium and iron isotopes. Also, Fig. 5.4 presents a very good correlation for *pfg*- together with *sdgh*-shell nuclei. The data include NMEs for twenty five nuclei including zinc, germanium, selenium, krypton, tellurium, xenon and barium isotopes. These $M^{\gamma\gamma}(M1M1)$ NMEs have been calculated using the free-particle or *bare* orbital and spin *g*-factors, but we observe nearly the same correlation when we use the effective *g*-factors that are in slightly better agreement with the experimental magnetic dipole moments and transitions, taking $g_i^{s,\mathrm{eff}} = 0.9 g_i^s$, $g_p^{l,\mathrm{eff}} = g_p^l + 0.1$, $g_n^{l,\mathrm{eff}} = g_n^l - 0.1$ in the $pf$ shell [261], and $g_i^{s,\mathrm{eff}} = 0.7 g_i^s$ for *pfg* nuclei [172]. An important feature is that the correlation is independent of the effective interaction used.

A linear regression analysis to the function $M_L^{0\nu} = a + b M^{\gamma\gamma}$ gives best-fit parameters $a = 0.872, b = 0.459$ for $46 \leq A \leq 60$, and $a = 1.29, b = 1.11$ for $72 \leq A \leq 136$, while the correlation coefficients are $\rho = 0.83$ and $\rho = 0.84$, respectively. These linear fits are shown in Fig. 5.4 by solid lines, together with the 68% and 90% confidence level (CL) prediction bands (dashed and dotted lines respectively). As we will see in Sec. 5.1.3, prediction bands could be used along with a hypothetical measurement of $\gamma\gamma$-M1M1 to obtain $0\nu\beta\beta$-decay NMEs for the particular nuclei that undergo the electromagnetic transition from the $0_{\text{DIAS}}^{+}$ state to the ground state.

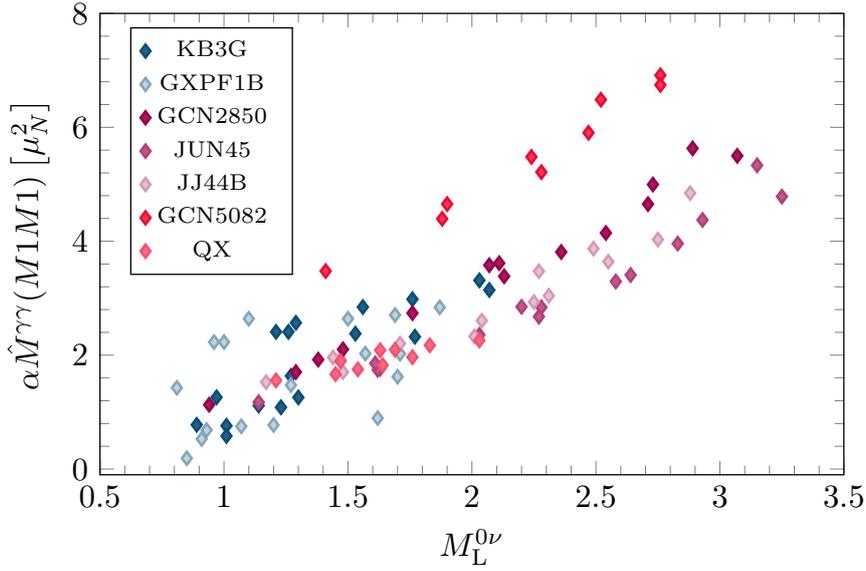

FIGURE 5.5: Standard $0\nu\beta\beta$ NME, $M_L^{0\nu}$, from Ref. [147]) vs the numerator of double-magnetic dipole operator, $\hat{M}^{\gamma\gamma}(M1M1)$, obtained with different interactions for each configuration space.

Figure 5.4 shows that the linear correlation is different for lighter than heavier nuclei, and this is due to the energy denominator in $M^{\gamma\gamma}(M1M1)$. This is clearly seen when only the numerator of the $M^{\gamma\gamma}(M1M1)$ operator, $\hat{M}^{\gamma\gamma}$, is plotted as we can see in Fig. 5.5. If one takes the energy denominator of the states that contribute most to $\hat{M}^{\gamma\gamma}(M1M1)$ and computes the mean for all the nuclei that follow the same correlation, that is $pf$ and $pfg$-$sdgh$ model spaces respectively, the ratio between these two mean values is $\overline{DE}_{n_d}(pfg + gds)/\overline{DE}_{n_d}(pf) = 2.33$, while the ratio of slopes in the best fit in Fig. 5.4 is 2.41. This effect can be seen in Fig. 5.3 for that sample of nuclei, since it is a systematic feature that the intermediate states that contribute most to $M^{\gamma\gamma}(M1M1)$ in $pf$-nuclei lie at lower energies compared to those with $A \geq 72$. It is also remarkable that the correlations have a weak dependence on the energy denominator, otherwise the connection between the two operators could have worsened.

In Fig. 5.5 $M^{\gamma\gamma}(M1M1)$ NMEs for heavier nuclei in the $sdhg$ space computed with the GCN5082 interaction lie regularly in the upper region of the correlation, in contrast with the values obtained with the QX interaction. This is due to two effects. One is that the NMEs of $\hat{M}^{\gamma\gamma}(M1M1)$ are systematically higher for GCN5082 as shown in Fig. 5.6. The origin of this effect is the dominant components of the **M1** matrix element in the $h_{\frac{11}{2}}$-channel for the states that most contribute to the M1M1. Considering only this matrix element we find that $\hat{M}^{\gamma\gamma}(M1M1)[h_{\frac{11}{2}}h_{\frac{11}{2}}]$ matrix elements in GCN5082 is 2.27 times higher than in QX, for $^{132}$Te. The second effect is the energy denominator, which is smaller for dominant states calculated with the GCN5082 interaction compared with the QX interaction, with ratio 1.44. A naive multiplication of both factors gives a value of 3.21 which is quite close to

$$M^{\gamma\gamma}(M1M1)[h_{\frac{11}{2}}h_{\frac{11}{2}};\text{GCN5082}]/M^{\gamma\gamma}(M1M1)[h_{\frac{11}{2}}h_{\frac{11}{2}};\text{QX}] = 3.28. \quad (5.7)$$

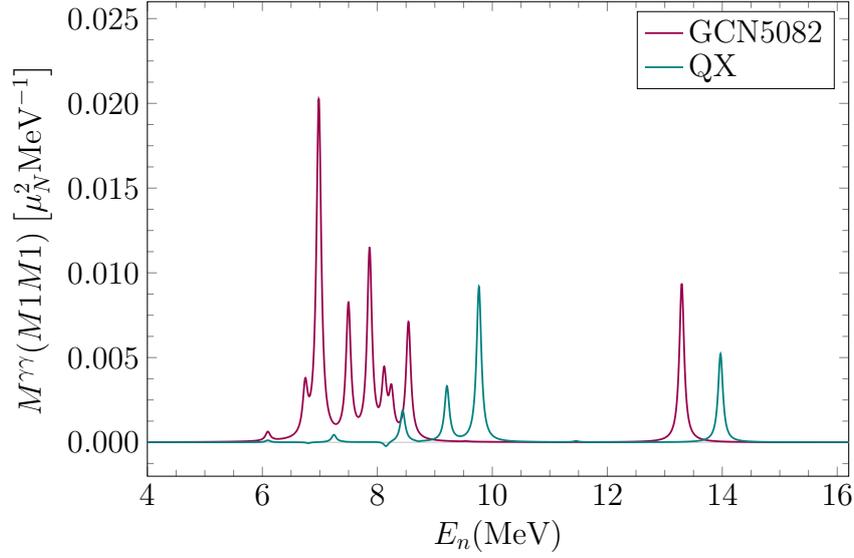

FIGURE 5.6: Contributions to $\hat{M}^{\gamma\gamma}(M1M1)$ NMEs as a function of the energy of the intermediate state for $^{132}$Te calculated with the GCN5082 and the QX interactions.

Likewise, a similar effect applies also to Xe and Ba.

The origin of the correlation can be better understood decomposing the $\hat{M}^{\gamma\gamma}(M1M1)$ NME into its spin, orbital and interference components

$$\hat{M}^{\gamma\gamma} = \hat{M}_{ss'}^{\gamma\gamma} + \hat{M}_{ll'}^{\gamma\gamma} + \hat{M}_{ls}^{\gamma\gamma}. \quad (5.8)$$

Figure 5.7 shows the three parts of a few nuclei within the different model spaces used. For some nuclei, like $^{72}$Zn, the spin part dominates. Consequently as $M_{ss'}^{\gamma\gamma}$ is proportional to the double Gamow-Teller operator a very good correlation is expected for these nuclei [147]. However, for heavy nuclei like $^{134}$Xe or $^{136}$Ba in *sdgh* shell, the orbital part $M_{ll'}^{\gamma\gamma}$ dominates. On the other hand the interference term $M_{ls}^{\gamma\gamma}$ is relatively smaller compared with the other components and can be of opposite sign. Beside this, the fact that the spin and orbital contributions to $\gamma\gamma$ decay systematically have the same sign preventing cancellations explains why the correlation is present even if the orbital part is dominant.

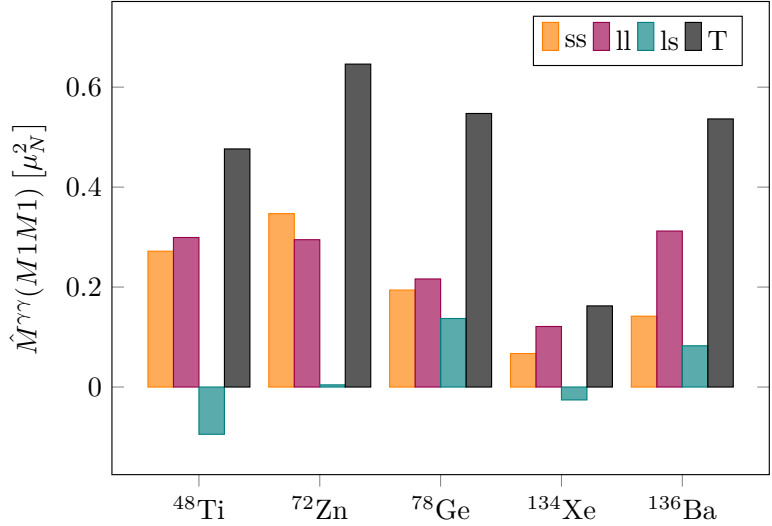

FIGURE 5.7: Different contributions to the numerator NME $\hat{M}^{\gamma\gamma}$ for several nuclei: total (T), spin $\hat{M}_{ss}^{\gamma\gamma}$ (ss), orbital $\hat{M}_{ll}^{\gamma\gamma}$ (ll) and interference $\hat{M}_{ls}^{\gamma\gamma}$ (ls) terms.

Additionally, we can gain further insight on the correlation between $\gamma\gamma$ and $0\nu\beta\beta$ NMEs decomposing each component (spin, orbital and interference) of the numerator $\hat{M}^{\gamma\gamma}$ in terms of the two-nucleon total angular momenta $\mathcal{J}$ for the nucleons involved in the transition. Figure 5.8 shows that $\hat{M}^{\gamma\gamma}$ is dominated by $\mathcal{J}=0$ pairs as it happens for $0\nu\beta\beta$ NMEs [262, 263], and partial cancellations with the higher components $\mathcal{J}>0$ make at the end a rather small NME. These cancellations are a common effect in all the components, but they are more remarkable for $M_{ss}^{\gamma\gamma}$ as one could have anticipated due to the spin-isospin SU(4) symmetry of the isovector spin operator [84, 264]. The leading contribution of $\mathcal{J}=0$ indicates that pairs coupled to $\mathcal{S}=\mathcal{L}=0$ are the most important for $\gamma\gamma$ DIAS to ground state transitions, hence $s_1 s_2 = (\mathcal{S}^2 - 3/2)/2 < 0$, and in the same way $l_1 l_2 < 0$. This, together with the fact that the spin and orbital isovector $g$-factors have the same sign, explains that the hierarchy in Fig. 5.8 leads to the absence of cancellations resulting in the $\gamma\gamma$ correlation with $0\nu\beta\beta$ decay.

Finally, the correlation between $\gamma\gamma(M1M1)$ DIAS to ground state transition and $0\nu\beta\beta$ decay has been recently studied within the framework of the QRPA in Ref. [150]. In this study, a good linear correlation is observed when several isoscalar proton-neutron pairing strengths ($g_{pp}^{T=0}$) used in usual $\beta$ and $\beta\beta$ decay calculations are considered. Figure 5.9 shows the comparison between the correlations found in the NSM and QRPA. Although both methods find a good correlation they differ: the slope of the linear fit in the NSM is about a factor two larger than the one obtained in the QRPA, while the later correlation is slightly shifted to higher values of $M^{0\nu}$. Moreover, the authors also find a good correlation between DGT and $0\nu\beta\beta$-decay NMEs when a range of $g_{pp}^{T=0}$ values is taken into account, contrary to the results reported for QRPA in Fig. 4 of Ref. [147]. They suggest that the discrepancy between the QRPA and other many-body methods could be due to the different contribution of $1^+$ multipoles in $0\nu\beta\beta$-decay NME. Also, they discuss that the QRPA correlation could change if deformation is included in the QRPA.

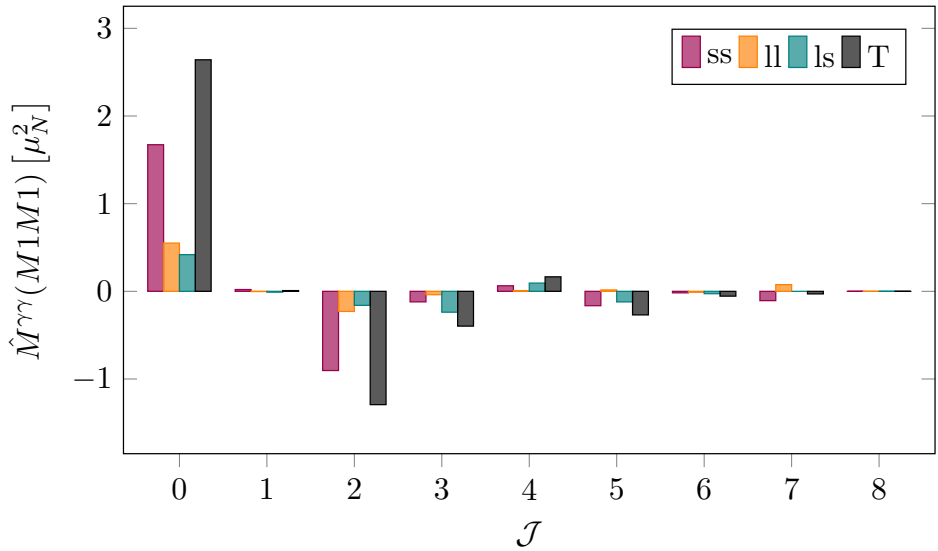

FIGURE 5.8: Decomposition of the $^{136}$Ba numerator NME $\hat{M}^{\gamma\gamma}$, in terms of the two-nucleon angular momenta $\mathcal{J}$: total (T), spin $\hat{M}_{ss}^{\gamma\gamma}$ (ss), orbital $\hat{M}_{ll}^{\gamma\gamma}$ (ll) and interference $\hat{M}_{ls}^{\gamma\gamma}$ (ls) parts.

### 5.1.3  A path towards uncertainty quantification of $0\nu\beta\beta$-decay NMEs in the NSM

The connection between $0\nu\beta\beta$ decay and $\gamma\gamma$ M1M1 transitions from DIAS to the ground state allows one to quantify the uncertainty to $0\nu\beta\beta$-decay NMEs from a hypothetical measurement of this $\gamma\gamma$ transition. This uncertainty and mean value of $0\nu\beta\beta$-decay NMEs predicted from the best linear fit to $M^{0\nu} = a + bM^{\gamma\gamma}$ rely on systematic calculations from several effective interactions across a wide range of the nuclear chart. This kind of studies constitute an alternative tool to improve our understanding of $0\nu\beta\beta$-decay NMEs guided by the experimental constrains.

To illustrate how this works we will suppose that $\gamma\gamma(M1M1)$ for transitions from the DIAS to the ground state have been measured with an uncertainty of $\pm15\%$ as in other second order $\gamma\gamma$ transitions [241, 245]. On the other hand, since we have the freedom to choose the value of what would be the $\gamma\gamma(M1M1)$ NME experimentally inferred, we will take the $M^{\gamma\gamma}(M1M1)$ that leads to $M_L^{0\nu}$ centered among the current predictions found in the literature [62].

Figure 5.10 compares the current spread in the standard $0\nu\beta\beta$-decay NMEs [62] represented by shaded bands, with their values and uncertainty derived from the correlation and prediction band at 90% CL from Fig. 5.4, represented by points with error bars. In addition, we also show in Fig. 5.10 the NSM results in the literature as crosses. There is a good agreement between predicted values from correlation and NMEs from other methods. The NMEs obtained from the NSM correlation are compatible with previous NSM values reported [62].

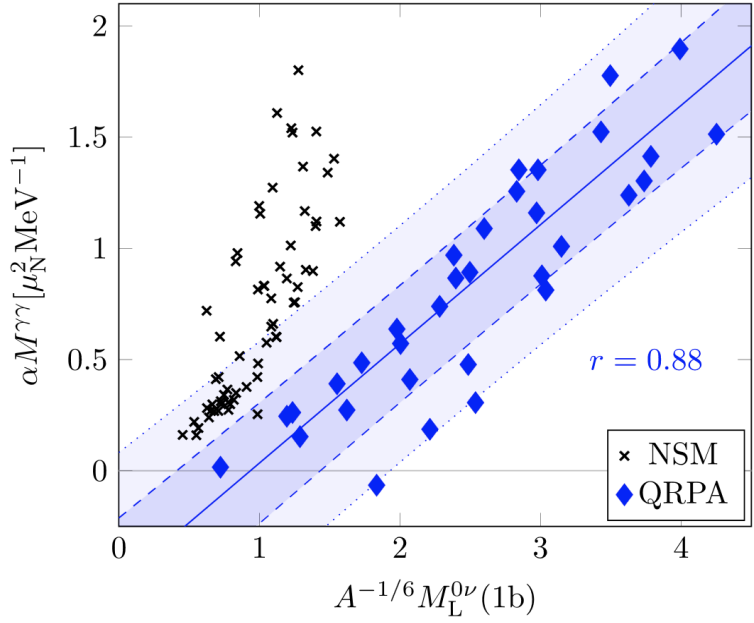

FIGURE 5.9: Comparison of the $0\nu\beta\beta$-decay NME (scaled by $A^{-1/6}$) vs $\gamma\gamma(M1M1)$ decay NME, obtained with the QRPA and NSM. The NSM results are from Ref. [12], and the QRPA ones from Ref. [150]. Blue lines show the best linear fit (solid) and 68% and 95% CL prediction bands (dashed and dotted) for the QRPA results. Figure taken from [150].

Therefore, based on the correlation obtained from tens of nuclei and different interactions, and assuming a hypothetical value for the experimental $M^{\gamma\gamma}(M1M1)$ with an error, we could estimate $M_L^{0\nu}$ with an uncertainty which improves constrains with respect to the spread in the combined many-body results. More importantly, this represents a NSM uncertainty obtained from systematic calculations for tens of nuclei using several interactions, which is more reliable than a collection of results, as previously estimated in the literature.

### 5.1.4 Implications to the sensitivity analysis of current and future $0\nu\beta\beta$ experiments

A reduction of the current spread in $M_L^{0\nu}$ values would have important consequences for $0\nu\beta\beta$ experiments and the limits set by them. Let us assume by now an ideal $0\nu\beta\beta$ experiment with no background. Then the number of expected signal events as a function

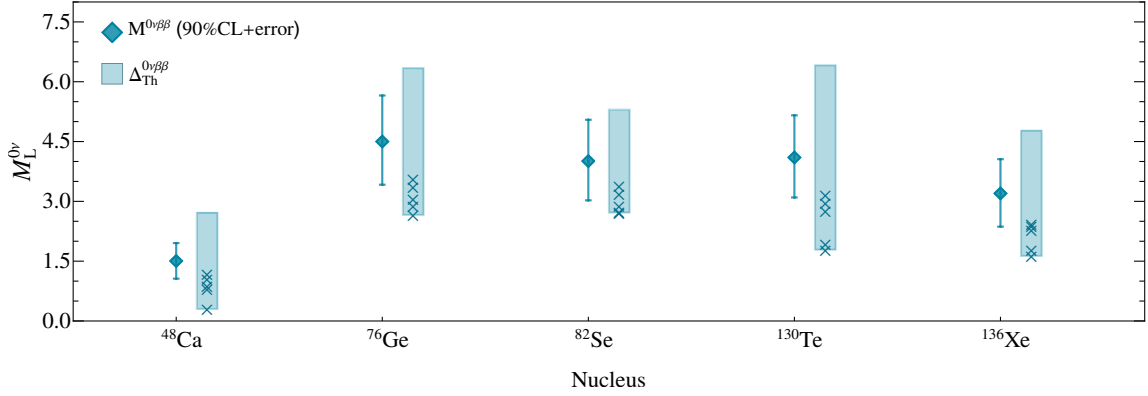

FIGURE 5.10: Data points with error bars represent the standard $0\nu\beta\beta$-decay NMEs ($M_L^{0\nu}$) obtained from correlation in Fig. 5.4 and predictions bands at 90% CL assuming an error in the experimental meassurement of $\pm15\%$, and its current spread ($\Delta_{\text{Th}}^{0\nu\beta\beta}$) reported in litrature [62]. Crosses represent NSM $M_L^{0\nu}$, also taken from Ref. [62].

of the half-life is given by

$$N = \log 2 \frac{\varepsilon N_A}{W} \frac{Mt}{T_{1/2}^{0\nu}}, \qquad (5.9)$$

where $N_A$ is the Avogadro number, $W$ is the atomic weight of the $\beta\beta$ isotope, $\varepsilon$ is the signal detection efficiency, $M$ is the source mass and $t$ the experiment running time. When combined with Eq. (2.40) we find that the sensitivity to $m_{\beta\beta}$ is [132]

$$\mathcal{S}(m_{\beta\beta}) = \mathcal{K} \sqrt{\frac{\overline{N}}{Mt}}, \qquad (5.10)$$

where

$$\mathcal{K} = \left( \frac{m_e^2 W}{\log 2 N_A G_{01} g_A^2 M_{0\nu}^2} \right)^{1/2}, \qquad (5.11)$$

and $\overline{N}$ is the average upper limit on the number of events expected in the experiment under no-signal hypothesis. Here, we will follow the definition of sensitivity of an experiment as in [132], that is, as the *average upper limit one would get from an ensemble of experiments with the expected background and no true signal*. For an ideal experiment the expected background is $b = 0$.

Figure 5.11 shows the sensitivity at 90% CL of a $^{136}$Xe experiment as a function of the exposure for the value $M^{0\nu}$ predicted from the best fit to the correlation. There are three bands from lighter to darker colors representing the sensitivity at 90% CL obtained from the minimum and maximum $M^{0\nu}$ values reported in the literature [62] ($\Delta_{\text{Th}}^{0\nu\beta\beta}$ lighter blue band), and equivalently for the minimum and maximum values estimated from 68% CL($\Delta_{\text{Est+Err}}^{0\nu\beta\beta}$ darker blue band) and 90% CL ($\Delta_{\text{Est+Err}}^{0\nu\beta\beta}$ medium blue band) prediction bands showed in Fig. 5.4.

In Fig. 5.11 it is also shown the minimum effective mass parameter corresponding to the inverted mass ordering $(m_{\beta\beta}^{\min})_{\text{IO}} = 18.4 \pm 1.3\text{eV}$[265]. An important consequence of Fig. 5.11 is the quantitative reduction in the exposure of any experiment to completely cover the inverted ordering region when comparing the maximal crossing points of these bands with $m_{\beta\beta}^{\min,\text{IO}}$ line. In the case of $^{136}$Xe, from Fig. 5.11 the derived reduction in the exposure is 9641(9065)Kg · yr for 68%(90%) CL prediction bands.

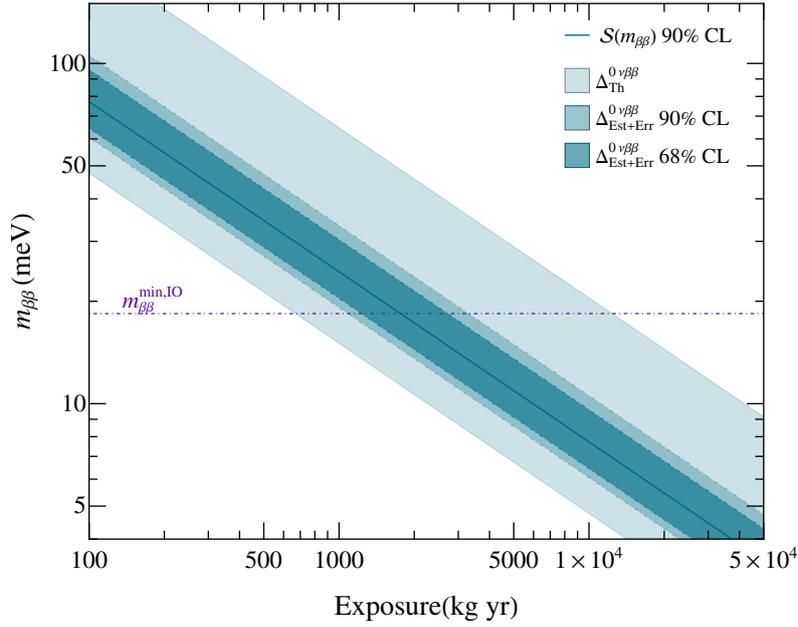

FIGURE 5.11: Sensitivity of an ideal $^{136}$Xe experiment at 90% CL as a function of the exposure (solid line), and the bands inferred from the minimum and maximum values of the $M^{0\nu}$ NME reported in literature($\Delta_{\text{Th}}^{0\nu\beta\beta}$ light blue band), and equivalently for the minimum and maximum values estimated from 68% CL($\Delta_{\text{Est+Err}}^{0\nu\beta\beta}$ dark blue band) and 90% CL ($\Delta_{\text{Est+Err}}^{0\nu\beta\beta}$ medium blue band) prediction bands.

In the case of an experiment with Poisson-distributed background of mean $b$, the unified approach method presented in Refs. [127, 132] allows one to write the average upper limit or sensitivity by

$$\mathcal{S}(b) = \sum_{n=0}^{\infty} \text{Po}(n|b)\mathcal{U}(n|b), \tag{5.12}$$

where $\text{Po}(n|b)$ is the Poisson-distributed background of mean $b$

$$\text{Po}(n|b) = \frac{b^n}{n!}e^{-b}, \tag{5.13}$$

and $\mathcal{U}(n|b)$ is a function that yields the unified approach upper limit at a particular CL for a particular observation $n$ and background level $b$. The values for $\mathcal{U}(n|b)$ are tabulated for several CL [266]. We will use the key parameters of a real Xe high-pressure time projection chamber of one ton [128]. The resulting sensitivity is shown in Fig. 5.12. From this data one could find that this experiment can reach a half-life sensitivity of $1.4 \times 10^{27}$yr in 5 years taking data.

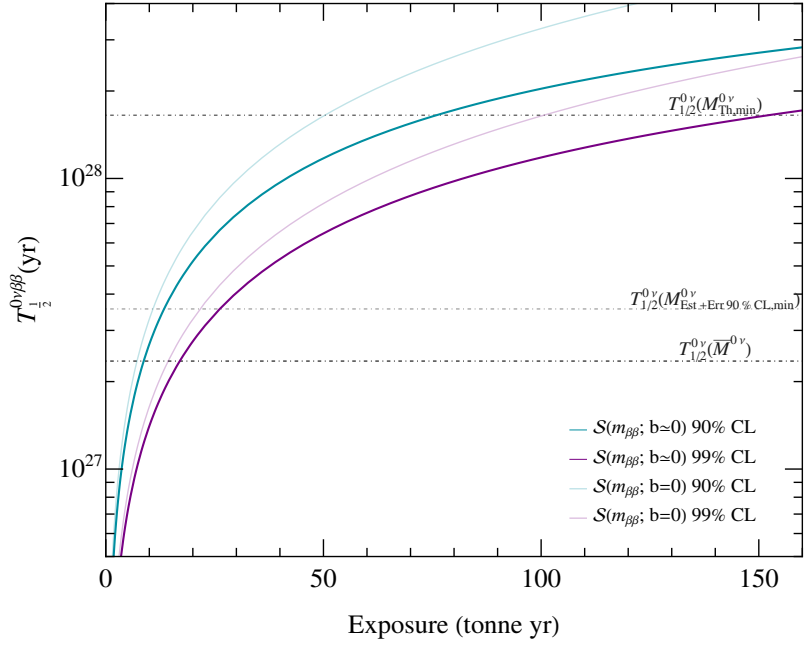

FIGURE 5.12: Sensitivity to $^{136}$Xe $0\nu\beta\beta$ decay half-life of a low-background experiment ($b \simeq 0$) at 90%(99%) CL in dark blue(purple) color as a function of the exposure. In light blue(purple) color it is shown the sensitivity to $0\nu\beta\beta$ decay half-life of an ideal experiment at 90%(99%) CL. The horizontal lines represent the calculated half-life values for $(m_{\beta\beta}^{\min})_{\text{IO}}$ and for the minimum value obtained from the correlation, $T_{1/2}^{0\nu}(M_{\text{Est+Err,90% CL,min}}^{0\nu})$, the minimum value of the literature NMEs, $T_{1/2}^{0\nu}(M_{\text{Th,min}}^{0\nu})$, and the mean value of the literature NMEs, $T_{1/2}^{0\nu}(\overline{M}_{\text{Th}}^{0\nu})$.

Figure 5.12 includes three horizontal lines which represent the half-life for $(m_{\beta\beta}^{\min})_{\text{IO}}$ for the central values of all the literature values $0\nu\beta\beta$ NME, $T_{1/2}^{0\nu}(\overline{M}_{\text{Th}}^{0\nu})$, the smallest NMEs among literature values, $T_{1/2}^{0\nu}(M_{\text{Th,min}}^{0\nu})$, and from the minimum NME obtained from the prediction band at 90% CL in Fig. 5.4, $T_{1/2}^{0\nu}(M_{\text{Est+Err,90% CL,min}}^{0\nu})$. In Fig. 5.12 it is also shown in blue (purple) lines the sensitivity to the $^{136}$Xe $0\nu\beta\beta$ decay half-life of a low-background experiment ($b \simeq 0$) in dark color at 90% (99%) CL as a function of exposure. The lighter colors represent exactly the same as before but for an ideal experiment ($b = 0$).

The main result derived from Figure 5.12 is the considerable reduction in the exposure necessary to reach the bottom of the inverted mass ordering when one assumes the minimum NME derived from the correlation, $(M_{\text{Est+Err90% CL}}^{0\nu})_{\min}$, and the mean value of the literature NMEs $(M_{\text{Th}}^{0\nu})_{\min}$.

Apart from the considerable reduction in the amount of material and time that one would get from a measurement of $\gamma\gamma$-M1M1 decay, the range of $m_{\beta\beta}$ from current lower limits in $T_{1/2}^{0\nu}$ is also improved, as it is shown in Figure 5.13 for the KamLAND-Zen lastest results [122]. The lightest shaded orange areas in Fig. 5.13 correspond to the uncertainty in the NMEs from current calculations for $^{136}$Xe, while the darker ones to the uncertainty derived from the estimated NMEs from the correlation in Fig. 5.4 obtained from 68%(90%) CL prediction bands for $^{136}$Xe.

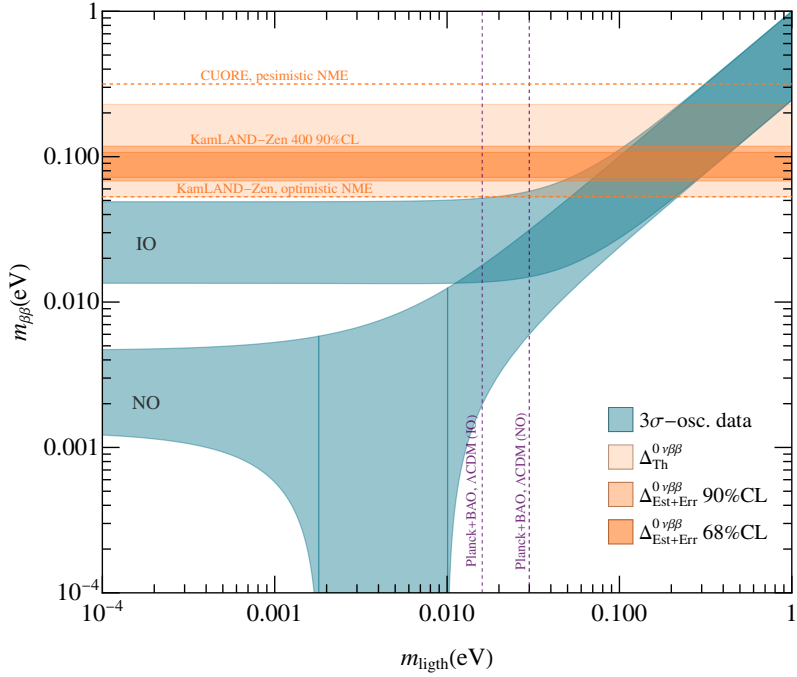

FIGURE 5.13: Effective Majorana mass $m_{\beta\beta}$ of the electron neutrino as a function of the mass $m_{\text{light}}$ of the lightest neutrino eigenstate for normal (NO) and inverterd mass ordering (IO). Filled blue regions include the uncertainty related to the CP-violation phases. To extract $m_{\beta\beta}$ ranges from experimental limits to $T^{0\nu}_{1/2}$, it is necessary the use of NMEs. The horizontal dashed lines show 90% CL current upper limits from $0\nu\beta\beta$ searches. It is also shown the tightest and loosest limits among those reported in the literature [62], namely the most stringent from KamLAND-Zen (labeled "KamLAND-Zen, optimistic NME"), and the less stringent from CUORE (labeled "CUORE, pesimistic NME"). Orange lightest shaded areas correspond to the uncertainty in the NMEs from current calculations for $^{136}$Xe, and the darker orange shaded areas the uncertainty derived from the estimated NMEs from the correlation in Fig. 5.4 obtained from 68%(90%) CL prediction bands for $^{136}$Xe. The vertical dashed line corresponds to 95% upper limits on $\sum m_\nu$ from cosmological observations, translated into upper limits on $m_{\text{light}}$ using the information from oscillation experiments [26].

## 5.2 VS-IMSRG $\gamma\gamma$(M1M1) decay study and the role of isospin mixing

In this section we will review the main results of the first study of the $\gamma\gamma$-M1M1 transitions within the VS-IMSRG. However, this study is still ongoing and further work is needed before moving into the correlation $\gamma\gamma$-M1M1 and $0\nu\beta\beta$-decay NMEs. All the results have been obtained using the chiral EFT interaction NN+3N EM1.8/2.0 [190, 192, 194].

To perform the IM-SRG evolution of the Hamiltonian and operators we will use `imsrg++`,

a software developed by R. Stroberg at TRIUMF [267]. Then the valence-space Hamiltonian is diagonalized, and the NMEs are calculated using the shell-model code KSHELL [11] and a python library developed by T. Miyagi [268]. We have used the IMSRG(2) approximation truncating the operators at the normal-ordered two-body level, using the Magnus formulation to derive the unitary transformation to decouple the valence-space Hamiltonian as described in Sec. 3.4. Furthermore, the harmonic-ocillator basis has been chosen with a frequency parameter $\hbar w = 16$ MeV. The single-particle space has been truncated to states with $e = 2n + l \leq e_{\max}$, and the 3N matrix elements have been truncated by $e_1 + e_2 + e_3 \leq E_{3\max}$ with $E_{3\max} = 24$.

First, we have performed a convergence analysis for the double magnetic dipole operator $\hat{M}^{\gamma\gamma}(M1M1)$, calculating the NMEs from $0^+_{\mathrm{DIAS}}$ to $0^+_{\mathrm{GS}}$ using as the Lanczos strength function to get the intermediate $1^+_n$ states. This has been done for several even-even nuclei in the $sd$-shell: $^{20-28}$Ne, $^{22-26,30}$Mg, $^{24-32}$Si, $^{32-34}$S and $^{34-36}$Ar. As discussed in Sec. 5.1, $\hat{M}^{\gamma\gamma}(M1M1)$ can be calculated assuming the sum over all intermediate states guiding us not only in the numerical convergence of the NMEs but also to benchmark them against its exact value.

| State | Nuclei | 1.8/2.0 EM | USDA | USDB | Exp. |
|---|---|---|---|---|---|
| $0^+_{\mathrm{DIAS}}$ | $^{20}$Ne | 16.60 | 16.96 | 16.84 | 16.73 |
| $1^+_{\mathrm{IAS}}$ | $^{20}$Ne | 10.47 | 11.19 | 11.17 | 11.26 |
| $0^+_{\mathrm{DIAS}}$ | $^{24}$Mg | 16.22 | 15.35 | 15.38 | 15.44 |
| $1^+_{\mathrm{IAS}}$ | $^{24}$Mg | 11.00 | 9.85 | 9.94 | 9.97 |
| $0^+_{\mathrm{DIAS}}$ | $^{28}$Si | 16.48 | 15.52 | 15.36 | 15.23 |
| $0^+_{\mathrm{DIAS}}$ | $^{32}$S | 13.16 | 11.91 | 11.93 | 12.05 |
| $1^+_{\mathrm{IAS}}$ | $^{32}$S | 8.58 | 6.78 | 6.84 | 7.00 |
| $0^+_{\mathrm{DIAS}}$ | $^{36}$Ar | 12.97 | 10.88 | 10.86 | 10.85 |

TABLE 5.1: Comparison between predicted energy values for the VS-IMSRG(2) obtained using the EM 1.8/2.0 interaction and truncation $e_{\max} = 12$, NSM predicted values obtained using the USDA, USDB [168] interactions, and experimental data. Energy units are in MeV.

On the other hand for nuclei in the $sd$-shell there are more data available for the energies of the isobaric analog states. In Table 5.1 we show the calculated energies for the double isobaric analog state, $0^+_{DIAS}$, and the isobaric analog state $1^+_{\mathrm{IAS}}$ calculated with both the VS-IMSRG(2) with the EM1.8/2.0 interaction and with NSM using the USDA, and USDB interactions. Table 5.1 also shows the experimental excitation energies of these states. Although there is a better agreement between calculations with USDA and USDB interactions and data, being relative errors around (1-3)%, the relative errors of the predicted values within VS-IMSRG(2) using the EM 1.8/2.0 interaction are below or around 10% except for the excitation energy of $0^+_{DIAS}$ of $^{36}$Ar where the error is much higher and around 20%.

Figure 5.14 shows an example of good convergence of the VS-IMSRG $\hat{M}^{\gamma\gamma}(M1M1)$ obtained with very few intermediate states as in $^{20}$Ne (left panel), while for $^{30}$Mg (right panel) the convergence in the VS-IMSRG is achieved for higher number of intermediate states. When one looks at the $M^{\gamma\gamma}(M1M1)$ convergence, it can worsen slightly as seen in Fig. 5.15. In general, we observe a better convergence for nuclei with isospin $T = 0$, and a worse one if the isospin is higher. This might be in part because the higher the isospin the higher lying is the DIAS state within the $0^+$ spectrum and also higher its isospin mixing with $T_{gs}$ and $T_{gs} + 1$ sates due to a high density of states. Since in the chiral NN+3N EM1.8/2.0 interaction used in the VS-IMSRG calculations isospin symmetry is slightly broken, it is worth to go deeper in order to understand the impact of isospin breaking better.

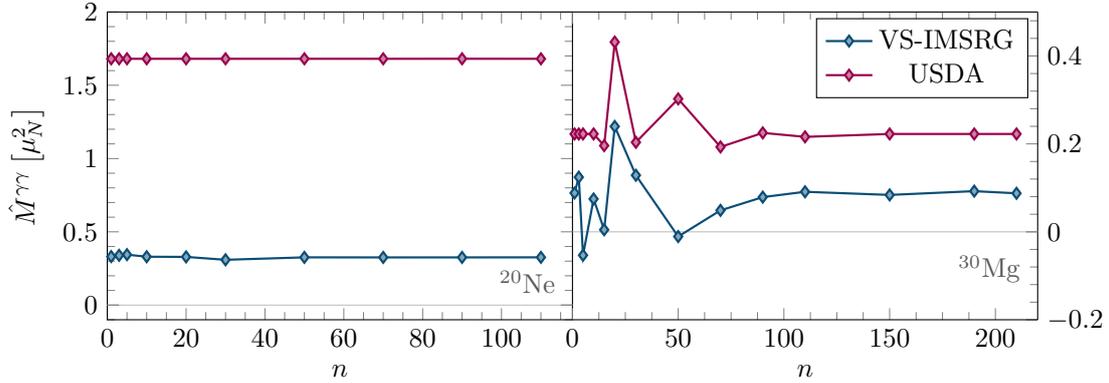

FIGURE 5.14: Convergence of $\hat{M}^{\gamma\gamma}(M1M1)$ (blue data points) as a function of the number $n$ of intermediate states $1^+_n$ for $^{20}$Ne (left panel) and $^{30}$Mg (right panel) calculated with the VS-IMSRG method. Each point represents $\hat{M}^{\gamma\gamma}(M1M1)$ for that number of intermediate states. For comparison, $\hat{M}^{\gamma\gamma}(M1M1)$ calculated with the NSM with the USDA interaction are also shown (red data points). Dashed lines represent exact closure values for each model.

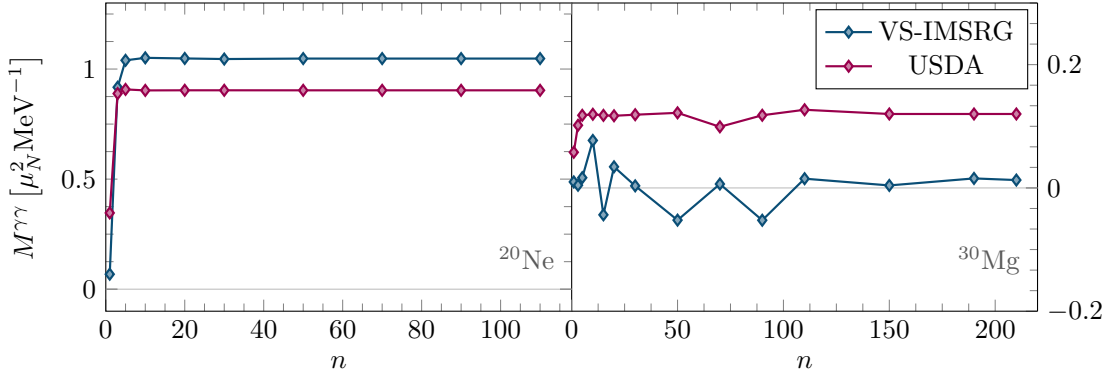

FIGURE 5.15: Convergence of $M^{\gamma\gamma}(M1M1)$ (blue data points) as a function of the number $n$ of intermediate states $1_n^+$ for $^{20}$Ne (left panel) and $^{30}$Mg (right panel) calculated with the VS-IMSRG method. Each point represents $M^{\gamma\gamma}(M1M1)$ for that number of intermediate states. For comparison, $M^{\gamma\gamma}(M1M1)$ calculated with the NSM with the USDA interaction are also shown (red data points).

To study the isospin mixing of the $0^+$ DIAS, we will assume that it can be mixed with $0^+$ states with isospin $T = T_f + 1$ and $T = T_f$, being the major component $T = T_f + 2$. Therefore, we can write the double isobaric analog state as the superposition

$$|0^+_{\text{DIAS}}\rangle = \alpha_{T_f+2}|0^+, T_f + 2\rangle + \alpha_{T_f+1}|0^+, T_f + 1\rangle + \alpha_{T_f}|0^+, T_f\rangle. \quad (5.14)$$

The coefficients $\alpha$ are the isospin amplitudes and they are a result of the isospin-violating terms in the nuclear Hamiltonian. The main source of isospin mixing comes from the Coulomb interaction, but other terms as the charge-dependence nuclear force or the neutron-proton mass difference can contribute. To obtain these coefficients we have computed $|0^+, T_f + i\rangle$ states and calculated the overlap with the $|0^+_{\text{DIAS}}\rangle$ state.

Figure 5.16 illustrates the isospin amplitudes of the $|0^+_{\text{DIAS}}\rangle$ state defined in Eq. (5.14) for each of the nuclei where we calculated the $M^{\gamma\gamma}(M1M1)$ NME. In this figure we see that in general for nuclei with $T = 0$, the isospin mixing is lower than for nuclei with higher isospin, but we find some exceptions, like for example $^{22}$Ne, $^{22}$Mg or $^{30,32}$Si. To go further we also computed the overlap between $|0^+_{\text{DIAS}}\rangle$ calculated with the EM1.8/2.0 and with the phenomenological interaction USDA, and we have checked that states with more isospin mixing are related with a lower overlap between the EM1.8/2.0 and USDA states.

In Figure 5.17 are shown the $\hat{M}^{\gamma\gamma}(M1M1)$ NMEs obtained with the VS-IMSRG(2) and with the NSM. The NSM NMEs calculated with the USDA and USDB interactions are represented as blue bands. On the other hand, this figure shows three different VS-IMSRG NMEs for each nuclei. The complete VS-IMSRG(2) NMEs are shown in dark green diamonds, the same NMEs but without the two-body component in light green diamonds, and VS-IMSRG NME without the consistent evolution of the transition operator discussed in Sec. 3.4.1 in cyan triangles. From this figure we can see that the effect of the two-body component in the SRG-evolved M1 operator is small, while the effect of the evolution is in some nuclei considerably higher, specially in nuclei with $T = 0$. Figure 5.17 also shows the relation between the degree of isospin mixing with the overlap of IAS calculated with isospin-conserving and isospin-breaking interactions.

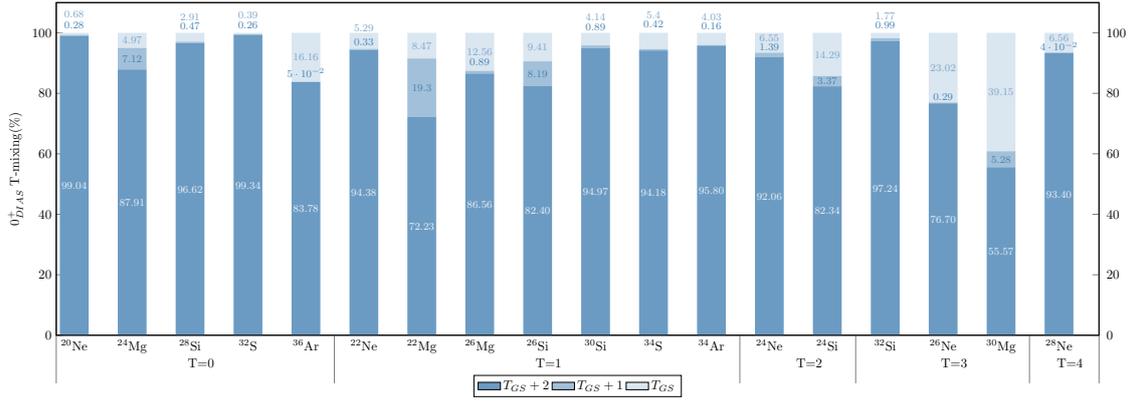

FIGURE 5.16: Isospin mixing in % ($\alpha_i \times 100$) of the $|0^+_{\text{DIAS}}\rangle$ state defined in Eq. (5.14) for all the nuclei studied. The darker color correspond to $\alpha_{T_f+2}$, medium color for $\alpha_{T_f+1}$ and lighter color for $\alpha_{T_f}$.

Overall, there is a tendency to present a larger mixing as the isospin is higher. However, one observes exceptions as for example in $^{28}$Ne. For this nucleus the $0^+$ dimension space is smaller and therefore the position of the $0^+$ in the spectrum ($n_{\text{DIAS}}$) is not so high as in $^{26}$Ne. We have another exception for example in $^{24}$Si and $^{32}$Si, in this case the dimension of the $0^+$ is the same, and $n_{DIAS}$ not too different, but in $^{24}$Si we have more protons than neutrons and therefore it is expected that the Coulomb effect that induces isospin symmetry breaking is higher. Furthermore, it is remarkable that there is preference for the isospin mixing to be higher with the isospin of the ground state than with the isospin of the isobaric $0^+$ state. This can be attributed to a major density of $0^+$ states with isospin $T_{gs}$ close to the DIAS.

In order to gain further insight in the role of the isospin mixing and $\hat{M}^{\gamma\gamma}(M1M1)$ NMEs, we analysed the decomposition of the closure M1M1 NME in terms of its isospin components. Although the $0^+_{\text{GS}}$ has a quite small mixing, $\alpha_{T_{gs}+1,T_{gs}} \sim 0.01$, we can obtain a relatively high $\hat{M}^{\gamma\gamma}(M1M1)$ NME connecting it with the $0^+_{\text{DIAS}}$ components. We find that there are $\hat{M}^{\gamma\gamma}(M1M1)$ NMEs where the dominant contribution comes from $\langle 0^+_{\text{DIAS}}(T = T_{GS} + 2)|M1M1|0^+_{\text{GS}}(T = T_{GS})\rangle$, as in $^{20-24}$Ne, $^{24,26,30}$Mg, $^{24,28,30}$Si, $^{32}$S, while there are other cases where we find either opposite sign contributions whose net effect is to reduce the dominant contribution as in $^{26,28}$Ne, $^{22}$Mg. We can also find the case where both the dominant contribution from the mixing NME between $|0^+_{\text{DIAS}}(T = T_{GS} + 2)\rangle$ and $|0^+_{\text{GS}}(T = T_{GS})\rangle$ adds coherently with the others making an non irrelevant contribution as in $^{32}$Si,

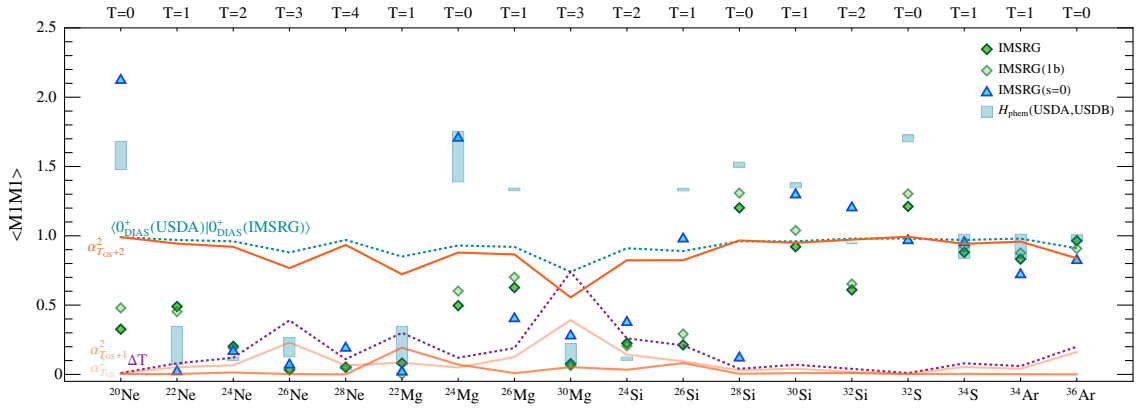

FIGURE 5.17: Closure M1M1 NMEs for the VS-IMSRG and NSM. NSM NMEs have been calculated for the USDA and USDB interactions and are repsented as blue bands. VS-IMSRG(2)($e_{max} = 12$) results are represented in dark green. The two-body contribution that is acquired upon the SRG evolution has been dropped for the light green diamonds to see the effect of such component. The non-evolved NMEs are represented in blue triangles. In blue dashed line is shown the overlap between $0^+_{\text{DIAS}}$ calculated with the EM1.8/2.0 and USDA interactions, while orange lines indicate the mixing coefficients $\alpha_i$ in Eq. 5.14. Finally, the purple dashed line shows the difference between the expected isospin when there is no isospin symmetry breaking and the one obtained in the VS-IMSRG(2)($e_{max} = 12$) calculation.

$^{34}$S, or $^{34}$Ar.

Finally, Figure 5.17 also shows that some $\hat{M}^{\gamma\gamma}(M1M1)$ NMEs are rather small and this can be associated with cancellations between relevant positive and negative contributions to $\hat{M}^{\gamma\gamma}(M1M1)$. This can be seen when we decompose the $\hat{M}^{\gamma\gamma}(M1M1)$ NME as a separate sum of the positive and negative contributions, respectively. This effect is illustrated in Fig. 5.18, where the running sums of the positive ($M1M1(+)$) and negative ($M1M1(-)$) contributions are calculated for $^{20}$Ne and $^{22}$Mg nuclei and represented as a function of the number of intermediate states. Then, a major cancellation is observed in $^{22}$Mg which also has relevant isospin amplitudes $\alpha_{T_f+1}$ and $\alpha_{T_f}$.

The study of $\gamma\gamma$ transitions within the VS-IMSRG method is still an ongoing project and there are several features that still need to be understood. A check that we have not yet done is to analyse the convergence of the NMEs, energies and isospin mixing as a function of the dimension of the truncated space.

## 5.3 Experimental prospects of $\gamma\gamma$(M1M1) decay from DIAS

The connection between $0\nu\beta\beta$ decay and $\gamma\gamma(M1M1)$ transitions from the $0^+(DIAS)$ to the ground state motivates even more the study of this second order electromagnetic process. Isobaric analog states are interesting from the theoretical point of view because they are involved in isospin-breaking phenomena in nuclei. Moreover, the progress in the experimental techniques made it possible to probe our understanding of exotic decays, such as isospin-forbidden multiparticle decays: $\beta$ delayed proton emission, diprotons and $\alpha$ particles [269, 270].

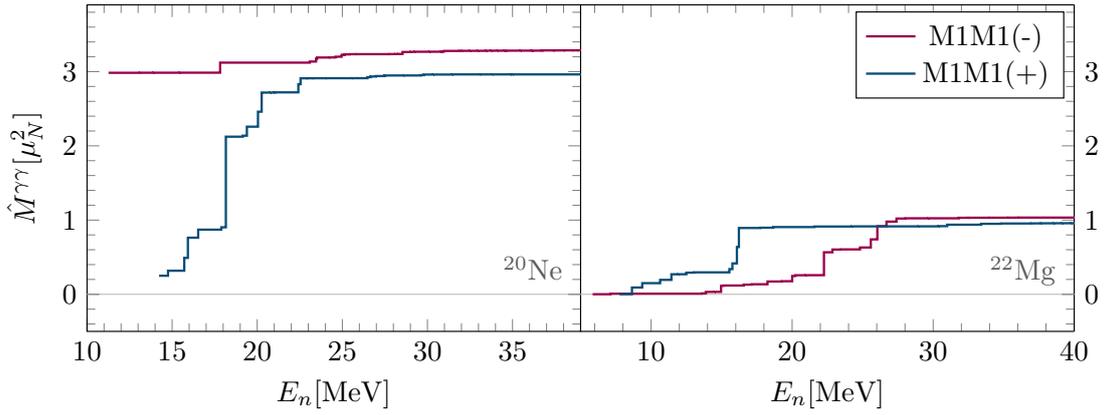

FIGURE 5.18: Positive and negative $\hat{M}^{\gamma\gamma}(M1M1)$ NMEs as a function of the energy of the intermediate state for $^{20}$Ne and $^{22}$Mg nuclei.

On the other hand, $\gamma\gamma$ transitions from DIAS have never been observed. The experimental challenge to populate the initial state together with the small branching ratios measured for competing $\gamma\gamma$ decays [241, 245] encourage a complete characterization of the decay width and branching ratios of the $0^+(DIAS)$ to guide experimentalists in the design and optimization of the detector and data acquisition process.

In this section we will show the main results for the transitions that compete with $\gamma\gamma(M1M1)$ transitions from $0^+(DIAS)$ to the ground state, mainly $\gamma(M1)$ transitions to the isobaric analog states and isospin-forbidden proton decay. The study has been done only for the experimentally relevant $\beta\beta$ final nuclei accessible to the NSM. They represent the first steps in the theoretical characterization of the decay modes of $0^+(DIAS)$, and although its complete description has not been finished yet and has been left for a forthcoming work, they could help in a preliminary analysis of this experimental search.

### 5.3.1  Electromagnetic branching ratios

We will start with the calculation of the transition width for double-magnetic dipole $\gamma\gamma(M1M1)$ transitions from the $0^+_{\text{DIAS}}$ to the ground state. The differential decay width is given by

$$\frac{d\Gamma_{fi}}{d\Omega d\Omega' dE_1} = \frac{E_1 E_2}{(2\pi)^3} \sum_f |\overline{T}_{fi}^{\gamma\gamma}|^2. \tag{5.15}$$

where the transition probability $\overline{T}_{fi}^{\gamma\gamma}$ has been derived in Sec. 4.1 in Eq. (4.85). We will consider the case were the angle between the emitted photons $\theta_{12}$ is measured. Then, after a partial angular integration over $d\Omega'$ and the azimutal angle $d\varphi$, the differential decay width for $\gamma\gamma$-M1M1 $0^+ \to 0^+$ decay is

$$\frac{d\Gamma_{fi}}{dxdE_1} = \frac{2E_1E_2}{3(2\pi)^3} \sum_{K=0,1,2} \hat{K}^2(1+(-1)^K) \begin{pmatrix} 1 & 1 & K \\ -1 & 1 & 0 \end{pmatrix}^2 P_K(x)\mathcal{P}_0^2(M1M1;E_1,E_2), \quad (5.16)$$

where $P_K(x)$ are the Legendre polynomials, $x = \cos\theta_{12}$ and $\mathcal{P}_0(M1M1)$ is the generalized polarizability defined in Eq. (4.2) of Sec. 4.2 for $\gamma\gamma$-M1M1

$$\mathcal{P}_0(M1M1;E_1,E_2) = \frac{2^2\pi\sqrt{3}}{3^2}E_1E_2\left[\alpha_{M1M1}(E_1) + \alpha_{M1M1}(E_2)\right], \quad (5.17)$$

$$\alpha_{M1M1}(E) = \sum_n \frac{(0_f^+||\mathbf{M1}||1_n^+)(1_n^+||\mathbf{M1}||0_i^+)}{E_n - E_i + E}. \quad (5.18)$$

Integrating in $x$ we get

$$\frac{d\Gamma_{fi}}{dE_1} = \frac{2^7\pi E_1^3 E_2^3}{3^5(\hbar c)^6}\left[\alpha_{M1M1}(E_1) + \alpha_{M1M1}(E_2)\right]^2. \quad (5.19)$$

$M^{\gamma\gamma}(M1M1)$ NMEs in Eq. (5.1) were calculated assuming that the two photons are emitted with the same energy $E_1 = E_2 = \frac{1}{2}(E_i - E_f) =: Q_{\gamma\gamma}/2$ and we will define this NME as $\alpha_{M1M1}(Q_{\gamma\gamma}/2) \equiv \bar{\alpha}_{M1M1}$. The approximation $\alpha_{M1M1}(E_1) \approx \bar{\alpha}_{M1M1}$ is good when the energies of the nuclear states involved satisfy $E_n - E_i \gg E_i$. However, we have seen that this condition is not satisfied for the $0^+(\text{DIAS}) \to 0_{\text{gs}}^+$ transition. In order to give a reliable estimation of the $\gamma\gamma$-decay width in the domain where the NMEs have been calculated, we will suppose that we can fix experimentally the energy at which we will detect the photons to $Q_{\gamma\gamma}/2$, and analyse the consequences of this assumption.

We will select the integration interval around $Q_{\gamma\gamma}/2$ to ensure that within that energy range the $\alpha_{M1M1}(E)$ varies smoothly with a maximum variation of 5%. Equation (5.18) can be expanded as

$$\alpha_{M1M1}(E) = \bar{\alpha}_{M1M1}\left[1 + \left(\frac{\Delta\varepsilon}{2\varepsilon_n}\right)^2 + \mathcal{O}\left(\frac{\Delta\varepsilon}{2\varepsilon_n}\right)^4\right], \quad (5.20)$$

where $\Delta\varepsilon = E_1 - E_2$ and $\varepsilon_n = E_n - \frac{1}{2}(E_i + E_f)$. If we truncate the expansion at second order, a variation of $x$% in the NME translates into the upper value $\Delta\varepsilon = 0.2\varepsilon_n^{\min}x^{1/2}$. Values for $\varepsilon_n^{\min}$ are around $\sim 4$ MeV for $^{48}$Ti and $\sim 2$ MeV for $^{76}$Se, $^{82}$Kr, $^{130}$Xe and $^{136}$Ba.

Introducing Eq. (5.20) into the differential decay rate and assuming a smooth variation of 5% in the NME around $Q_{\gamma\gamma}/2$, we can take $\alpha_{M1M1}$ out of the integrand and perform the integration from $E_1 = \frac{1}{2}(Q_{\gamma\gamma} - \Delta\varepsilon)$ to $E_1 = \frac{1}{2}(Q_{\gamma\gamma} + \Delta\varepsilon)$

$$\Gamma_{\gamma\gamma}(M1M1;\varepsilon) = \frac{8\pi}{3^5}\frac{\bar{\alpha}_{M1M1}^2}{(\hbar c)^4}\frac{\Delta\varepsilon}{Q_{\gamma\gamma}}\left[\frac{1}{2}Q_{\gamma\gamma}^7 - \frac{1}{2}\Delta\varepsilon^2 Q_{\gamma\gamma}^5 + \frac{3}{10}\Delta\varepsilon^4 Q_{\gamma\gamma}^3 - \frac{1}{14}\Delta\varepsilon^6 Q_{\gamma\gamma}\right]. \quad (5.21)$$

This expression demands an experimental setup that must be optimized to detect photons with energies around the $1/2E_{\mathrm{DIAS}}$, and therefore a prior measurement of $E_{\mathrm{DIAS}}$ is advisable.

Additionally, as a cross-check we derive the total $\gamma\gamma$-decay width for any multipolarity and compare our results for the branching ratios when data and theoretical matrix elements calculated with other many-body methods are given, as in Ref. [245].

One of the main competing processes that one can think off is the first order single $\gamma$ decay. The *transition probability* per unit time of $\gamma$ decay from an initial nuclear state $i$ to a final nuclear state $f$, $T_{fi}^{(S\lambda\nu)}$, is given by [61]

$$T_{fi}^{(S\lambda\nu)} = \frac{2}{\varepsilon_0 \hbar} \frac{\lambda+1}{\lambda[(2\lambda+1)!!]^2} \left(\frac{E_\gamma}{\hbar c}\right)^{2\lambda+1} |\langle \xi_f J_f m_f | \mathcal{M}_{S\lambda\mu} | \xi_i J_i m_i \rangle|^2, \tag{5.22}$$

where $E_\gamma$ is the transition energy. Since nuclear state projections are not observed separately, an average over the initial substates and a sum over the final substates is assumed, giving a transition probability per unit time

$$T_{fi}^{(S\lambda\nu)} = \frac{2}{\varepsilon_0 \hbar} \frac{\lambda+1}{\lambda[(2\lambda+1)!!]^2} \left(\frac{E_\gamma}{\hbar c}\right)^{2\lambda+1} B(S\lambda; \xi_i J_i \to \xi_f J_f), \tag{5.23}$$

where $\lambda$ is the multipole order, and $S = \mathrm{E,M}$ the multipolar character of the electromagnetic transition. Here, $B(S\lambda; \xi_i J_i \to \xi_f J_f)$ is the *reduced transition probability* defined by

$$B(S\lambda; \xi_i J_i \to \xi_f J_f) \equiv \frac{1}{2J_i+1} |\langle \xi_f J_f |||\xi_i J_i \rangle|^2. \tag{5.24}$$

Furthermore, the single gamma width is

$$\Gamma_\gamma^{(S\lambda)} = \hbar T_{fi}^{(S\lambda)}. \tag{5.25}$$

First we have computed magnetic dipole transitions, from the $0^+(DIAS)$ to lower energy $1_n^+$ states until the gamma branching ratio is of the order of a few per cent. This has been done for nuclei which are of interest for experimental searches of $0\nu\beta\beta$ decay and that can be computed within the NSM. The main results are shown in Table 5.2. There is a clear difference between $^{48}$Ti and the rest of heavier nuclei: while in the first case only one state carries out almost the total width, for heavier systems the contribution to the total width is fragmented in more states, which are different from one interaction to the other.

Due to the hierarchy between electric and magnetic multipoles as discussed in Sec. 4.1.5, it is expected that electric dipole transitions (E1) are important. However, the calculation of electric dipole transitions within the NSM is complicated because since they change parity we need to include at least two major shells in the calculations which increases considerably the basis dimension. This makes the computation of the electric dipole NMEs beyond the goals of this work, but it is left for future research.

The next multipole that would enter into play following the multipolar hierarchy is the electric quadrupole, E2. However, we computed their widths and found that they are two orders of magnitude smaller than in M1 transitions. Therefore we will not study higher multipoles since they have smaller transitions rates and smaller widths, not being a competing transition with magnetic dipole decay.

Taking into account all these elements we have computed the total $\gamma\gamma(M1M1)$ decay width from the DIAS. The results are summarized in Table 5.3 where the $\gamma\gamma/\gamma$ branching ratios for magnetic dipole transitions are also shown.

| Nucleus | $H_{eff}$ | $E_{DIAS}$ (MeV) | $E_{1^+}$ (MeV) | $\Gamma_\gamma[M1]$ ($eV$) | BR |
|---------|-----------|------------------|-----------------|----------------------------|-----|
| $^{48}$Ti | KB3G | 16.09 | 12.2808 | 6.8534 | 92.41 |
| $^{48}$Ti | GXPF1B | 17.33 | 12.6231 | 13.4306 | 96.83 |
| $^{76}$Se | GCN2850 | 19.22 | 12.0189 | 0.7729 | 24.21 |
| | | | 12.2885 | 0.4730 | 14.81 |
| | | | 12.8401 | 0.6310 | 20.04 |
| | | | 13.2032 | 1.1130 | 34.86 |
| $^{76}$Se | JUN45 | 21.23 | 12.9228 | 2.68305 | 68.40 |
| | | | 14.0757 | 0.97741 | 24.92 |
| $^{76}$Se | JJ44B | 23.60 | 13.3455 | 0.5950 | 14.23 |
| | | | 13.5241 | 1.6933 | 40.48 |
| | | | 14.2581 | 0.3442 | 8.23 |
| | | | 15.4180 | 0.9636 | 23.04 |
| $^{82}$Kr | GCN2850 | 21.56 | 13.0702 | 2.2970 | 35.58 |
| | | | 14.0850 | 0.5698 | 8.83 |
| | | | 14.9278 | 2.8805 | 44.62 |
| $^{82}$Kr | JUN45 | 22.90 | 13.6587 | 1.6751 | 33.81 |
| | | | 13.6867 | 0.4181 | 8.44 |
| | | | 14.6360 | 1.3369 | 26.99 |
| | | | 15.6068 | 0.4471 | 9.03 |
| $^{82}$Kr | JJ44B | 26.59 | 19.2733 | 0.6084 | 25.20 |
| | | | 19.6017 | 0.6626 | 27.44 |
| | | | 21.9186 | 0.8187 | 33.91 |
| $^{130}$Xe | GCN5082 | 27.74 | 15.8765 | 0.4150 | 14.60 |
| | | | 17.9250 | 1.0982 | 38.63 |
| | | | 18.3949 | 0.4452 | 15.66 |
| $^{136}$Ba | GCN5082 | 29.14 | 17.5191 | 1.2083 | 9.81 |
| | | | 17.9045 | 2.2233 | 18.05 |
| | | | 18.9608 | 4.4448 | 36.09 |
| | | | 19.2807 | 1.4921 | 12.12 |

TABLE 5.2: Dominant **M1** ($0^+$(DIAS) $\rightarrow 1_n^+$) transitions for final $\beta\beta$ nuclei, calculated with the NSM using different interactions. The only experimental data we have is the energy of the $0^+$(DIAS) for $^{48}$Ti which is $E(0^+$(DIAS)$) = 17.33$MeV, in good agreement with the predicted energy.

| Nucleus | $H_{eff}$ | $\Delta E$ | $\Gamma_\gamma[M1]$(eV) | $\Gamma_{\gamma\gamma}[M1M1]$(eV)($\times 10^{-7}$) | $BR_{\gamma\gamma} \times 10^{-8}$ |
|---------|-----------|------------|-------------------------|------------------------------------------------------|-------------------------------------|
| $^{48}$Ti | KB3G | 3.81 | 6.85 | 2.02 | 2.95 |
| $^{48}$Ti | GXPF1B | 4.71 | 13.43 | 0.55 | 4.09 |
| $^{76}$Se | GCN2850 | 6.01 | 1.11 | 8.70 | 78.18 |
| $^{76}$Se | JUN45 | 8.30 | 2.68 | 15.79 | 58.84 |
| $^{76}$Se | JJ44B | 10.07 | 1.69 | 32.22 | 190.28 |
| $^{82}$Kr | GCN2850 | 6.64 | 2.88 | 11.60 | 40.03 |
| $^{82}$Kr | JUN45 | 9.24 | 1.68 | 12.15 | 72.25 |
| $^{82}$Kr | JJ44B | 4.67 | 0.82 | 27.12 | 33.12 |
| $^{130}$Xe | GCN5082 | 9.81 | 1.10 | 2.23 | 202.57 |
| $^{136}$Ba | GCN5082 | 10.18 | 4.44 | 16.88 | 37.97 |

TABLE 5.3: Predicted values for $\gamma$-M1 and $\gamma\gamma$-M1M1 decay widths calculated with the NSM using different interactions.

### 5.3.2 Isospin-forbidden proton emission

Double isobaric analog states are high energy nuclear states with excitation energy $\sim$ 15MeV that generally lie above single nucleon separation energies, $S_p$ for protons and $S_n$ for neutrons. For those states, particle decay is kinematically allowed opening a particle branch that should be studied in the context of $\gamma\gamma$ decay. Despite $E(DIAS) > S_p, S_n$, so that both proton and neutron emission are open decaying channels, as a first approach to derive the $\gamma\gamma$-particle branching ratio we will focus on proton emission, since due to the Coulomb force this is the more favorable decay chanel.

When a nucleus is unbound with respect to proton emission we could have the decay

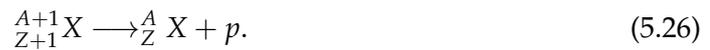

$$^{A+1}_{Z+1}X \longrightarrow ^{A}_{Z}X + p. \tag{5.26}$$

A necessary condition is a positive value for the decay energy $Q_p = B(N, Z+1) + E_i^{exc} - B(N, Z) - E_f^{exc}$, when both initial and final nuclei are in the excited states with energies $E_i^{exc}$ and $E_f^{exc}$, respectively. If the initial nucleus $^{A+1}_{Z+1}X$ is in an excited DIAS with isospin $T_i = T_{gs} + 2$ and the final state is in its ground state (this is the most probable decay) with isospin $T_f = T_{gs} + \frac{1}{2}$, we have an isospin forbidden decay ($\Delta T = \frac{3}{2}$). This kind of transitions proceed via isospin-symmetry breaking, providing a valuable tool to study *isospin mixing* in nuclear states at the same time that the $\gamma$ branch starts to be competitive. An example of this type of transitions are $\beta$-delayed proton emission from isobaric analog states [269, 270].

There are several methods used to compute proton decay widths, which can be divided in three main groups: i) Wentzel-Kramers-Brillouin (WKB) based methods [271, 272], like in the theory of alpha decay proposed independently by Gamow and by Condon and Gurney, ii) wave function matching methods [273, 274], and iii) fission like methods [275]. We will use the semiclassical WKB approximation where the weakly bounded proton can tunnel through the potential barrier of the final nucleus. In this method, the decay width of the proton-emitting state is given by [276]

$$\Gamma_p = SF \frac{\mathcal{N}}{4\mu} \exp\left[-2\int_{r_1}^{r_2}|k(r)|dr\right], \tag{5.27}$$

where $\mathcal{N}$ is a normalization factor, $r_i$ with $r_0 < r_1 < r_2$ are the classical turning points defined by $V(r_i) = Q_p$, and $V(r)$ is the radial part of the proton-daughter nucleus interaction potential. Here, $SF$ is the spectroscopic factor, which represents the proton preformation amplitude. The proton emission process can be described by a wave function that is a superposition of different channel wave functions

$$\Phi_P(\boldsymbol{r}, \boldsymbol{s}, \xi_P) = \sum_c A^{(c)}(r)\Phi_D^{(c)}\Phi_p^{(c)}, \tag{5.28}$$

where $A^{(c)} = \langle\Phi_P|\Phi_D^{(c)}\Phi_p^{(c)}\rangle$ is the preformation amplitude. Furthermore, the function $k(r)$ is the wave number defined by

$$\hbar k(r) = \sqrt{\frac{2\mu}{\hbar^2}|E - V(r)|}, \tag{5.29}$$

where $E$ is the energy of the emitted proton $Q_p$, and $\mu$ represents the reduced mass, $\mu = M(A)/A$. The potential barrier is the sum of the nuclear ($V_N$), Coulomb ($V_C$) and centrifugal terms

$$V(r) = V_N(r) + V_C(r) + \frac{\hbar^2}{2\mu r^2}l(l+1). \tag{5.30}$$

The nuclear potential is approximated by the sum of the Woods-Saxon potential [61]

$$V_{WS} = \frac{-V_0}{1 + e^{\frac{(r-R)}{a}}}, \tag{5.31}$$

with parameters $a = r_0 A^{1/3} = 1.27A^{1/3}$fm, $a = 0.67$fm and $V_0 = \left(51 + 33\frac{N-Z}{A}\right)$ MeV, and the spin-orbit potential

$$V_{LS}(r) = V_{LS}^{(0)}\left(\frac{r_0}{\hbar}\right)^2 \frac{1}{r}\frac{d}{dr}\left[\frac{1}{1 + e^{\frac{(r-R)}{a}}}\right]. \tag{5.32}$$

On the other hand, the Coulomb potential is approximated by the potential generated by a uniformly charged sphere of radius $R$

$$V_C(r) = \frac{Ze^2}{4\pi\varepsilon_0}\begin{cases} \frac{3-(r/R)^2}{2R}, & r \leq R, \\ \frac{1}{r}, & r > R. \end{cases} \tag{5.33}$$

Finally, the normalization constant $\mathcal{N}$ which is related with the assaulting frequencies, is given by

$$\mathcal{N}^{-1} = \int_{r_0}^{r_1} \frac{dr}{k(r)} \cos^2 \left[ \int_{r_0}^{r} k(r')dr' - \frac{\pi}{4} \right] \simeq \frac{1}{2} \int_{r_0}^{r_1} \frac{dr}{k(r)}, \tag{5.34}$$

where the cosine is well approximated by its average value.

All the terms excluding the spectroscopic factor determine what is known as the *single-particle* width, thus the total width is defined as

$$\Gamma_p = \Gamma_{sp} SF. \tag{5.35}$$

Due to its simplicity this method has been used for spherical proton emitters, and in its domain of applicability it gives predictions in good agreement with other methods such as the distorted-wave Born approximation and the two-potential approach [276].

We have implemented the WKB approximation to derive the proton single-particle width and checked it with literature values [270]. The authors in Ref. [270] calculated $\Gamma_{sp}$ from the scattering cross section with a Woods-Saxon potential with a potential depth that could be modified to reproduce known proton energies. This is for example implemented in `wspot` [277], an online code that calculates the widths of unbound states resonances whose results are in good agreement with the WKB approach when the latter can be applied.

Finally, the spectroscopic factor has been obtained experimentally for $^{32}$S where there is a non negligible $\gamma$-branch [278] of the lowest isobaric analog state $J^{\pi}, T = 0^{+}, 2$ at 12.047 MeV excitation energy. The total width of this state is $\Gamma = 40(15)$eV [279, 280], while the $\gamma$ width $\Gamma_{\gamma} \simeq 2$eV. Likewise, this state has an isospin-nonconserving proton decay width of $\Gamma_p = 38$eV. The theoretical prediction for the single-particle width for a proton emission to the ground state of $^{31}$P with energy $Q_p = 3.29$ MeV is $\Gamma_{sp} = 1.05$MeV, giving an empirical spectroscopic factor of $SF_{\text{exp}} = 3.7 \times 10^{-5}$. This spectroscopic factor will be used to predict the proton width in $^{48}$Ti.

Table 5.4 shows the results for the calculated proton single-particle width of DIAS, for the proton emission in $^{32}$S to the ground state $J^{\pi} = \frac{1}{2}^{+}$ of $^{31}$P, and the widths for the proton emission in $^{48}$Ti to the ground state $J^{\pi} = \frac{7}{2}^{-}$ and first excited states with $J^{\pi} = \frac{3}{2}^{-}$ and $J^{\pi} = \frac{5}{2}^{-}$ in $^{47}$Sc. The orbit quantum numbers $nl_j$ characterise the single-particle orbital that the valence proton occupy in the initial nucleus. In this table are shown the dominant $\gamma$-M1, and $\gamma\gamma$-M1M1 widths used to obtain the branching ratios, both $BR_{\gamma} = \Gamma_{\gamma\gamma}/\Gamma_{\gamma}$ and $BR_p = \Gamma_{\gamma\gamma}/\Gamma_p$. Although particle emission has a higher transition probability, the small spectroscopic factor $SF \sim 10^{-5}$ prevents that the particle branch will be much higher than the single gamma one, being the branching ratio only one order of magnitude smaller. This prevents the particle emission dominance in the total width of the DIAS.

| Nucleus | $H_{eff}$ | $\Delta T$ | $Q_p$ | Orbit | $\Gamma_{sp}$(MeV) | $\Gamma_\gamma$[M1](eV) | $\Gamma_{\gamma\gamma}$[M1M1](eV) | $BR_\gamma \times 10^{-8}$ | $BR_p \times 10^{-9}$ |
|---|---|---|---|---|---|---|---|---|---|
| $^{32}$S | USDA | $\frac{3}{2}$ | 3.29 | $1s_{1/2}$ | 1.02 | 1.27 | 2.24 | 17.6 | 6.01 |
| $^{32}$S | USDB | $\frac{3}{2}$ | 3.29 | $1s_{1/2}$ | 1.02 | 1.31 | 2.24 | 18.6 | 6.50 |
| $^{48}$Ti | KB3G | $\frac{3}{2}$ | 5.93 | $0f_{7/2}$ | 0.52 | 6.85 | 2.02 | 2.95 | 10.3 |
| $^{48}$Ti | GXPF1B | $\frac{3}{2}$ | 5.93 | $0f_{7/2}$ | 0.52 | 13.43 | 0.55 | 4.09 | 0.28 |
| $^{48}$Ti | KB3G | $\frac{3}{2}$ | 5.13 | $1p_{3/2}$ | 1.60 | 6.85 | 2.02 | 2.95 | 3.41 |
| $^{48}$Ti | GXPF1B | $\frac{3}{2}$ | 5.13 | $1p_{3/2}$ | 1.60 | 13.43 | 0.55 | 4.09 | 0.93 |
| $^{48}$Ti | KB3G | $\frac{3}{2}$ | 4.64 | $0f_{5/2}$ | 6.12 | 6.85 | 2.02 | 2.95 | 8.92 |
| $^{48}$Ti | GXPF1B | $\frac{3}{2}$ | 4.64 | $0f_{5/2}$ | 6.12 | 13.43 | 0.55 | 4.09 | 2.43 |

TABLE 5.4: Predicted values for proton, $\gamma(M1)$ and $\gamma\gamma(M1M1)$ decay widths for different proton orbits an interactions from the $0^+(DIAS)$. For the particle branching ratio $BR_p$ the spectroscopic factor $SF = 3.7 \times 10^{-5}$ has been used.

Apart from single $\gamma$ decay and proton emission, there are other mechanisms through which $0^+$ states could decay into other $0^+$ state such as internal conversion and internal pair creation. Although their contribution could be relevant, it is not dominant and can be separated experimentally. Therefore, we have left the study of these processes together with the electric dipole transitions for future work.

## 5.4　$0\nu\beta\beta$ decay from $2\nu\beta\beta$ decay data

In this section we will discuss the main results and conclusions of the study of the correlation between $0\nu\beta\beta$ decay and $2\nu\beta\beta$ decay published in Ref. [13]. There are several reasons to be interested in this analysis, but we will highlight two of them. First, since $2\nu\beta\beta$ decay half-lives have been measured, a correlation between $0\nu\beta\beta$-decay $2\nu\beta\beta$-decay NMEs can lead to a prediction of $0\nu\beta\beta$-decay NMEs based on $2\nu\beta\beta$ data. Second, the correlation can be used to give a theoretical uncertainty on $0\nu\beta\beta$-decay NME, that is based on $2\nu\beta\beta$ decay data and on systematic calculations that follow the same correlation.

In this paper we studied the relation between the NMEs of the two decaying modes of $\beta\beta$ decay for several dozens of nuclei across a wide range of nuclear masses, from calcium to xenon. In order to have further insight the analysis has been done using both the NSM and QRPA, where different effective interactions and values of the isoscalar particle-particle pairing $g_{pp}^{T=0}$ were used.

Figure 5.19 from Ref. [13] shows the results of the calculated $0\nu\beta\beta$-decay $2\nu\beta\beta$-decay NMEs, for the standard component of the light neutrino exchange mechanism ($M_L^{0\nu}$) as in our study of the $\gamma\gamma$-$0\nu\beta\beta$ correlation. This figure shows a very good correlation for both QRPA and the NSM, though in NSM the correlation depends on the nuclear mass. As in $M^{\gamma\gamma}(M1M1)$ the different slopes are due to the typical energies of the intermediate states which most contribute to the $2\nu\beta\beta$-decay NME. In Fig. 5.19 a quenching factor $q$ has been introduced, since the calculated 2νββ-decay NMEs overestimate the experimental results [91]. Studies based on $\beta$- and $\beta\beta$-decays give the shell-model quenching ranges [13]: $q = 0.65 - 0.77$ for $46 \leq A \leq 60$, $q = 0.55 - 0.64$ for $72 \leq A \leq 84$, and $q = 0.42 - 0.72$ for $124 \leq A \leq 136$. While in the pnQRPA method, the typical value $q = 0.79$ ($g_A^{eff} = 1.0$) is assumed. Furthermore, the empirical 2νββ-decay NMEs from Ref. [49] can be combined with the 68% CL prediction bands of the correlation shown in Fig. 5.19 to obtain 0νββ-decay NMEs with uncertainties. Figure 5.20 from Ref. [13] shows the comparison between the NSM(red) and pnQRPA(blue) 0νββ-decay NMEs, including previous theoretical predictions. The error bars marked by horizontal lines, show the 0νββ-decay NMEs calculated from the 68% CL prediction bands from the best fit to the correlation combined with the much smaller uncertainties in the empirical NMEs. Additionally, the error bars marked with triangles show the combination of the uncertainties from the 68% CL prediction bands and the uncertainty in the quenching $q$.

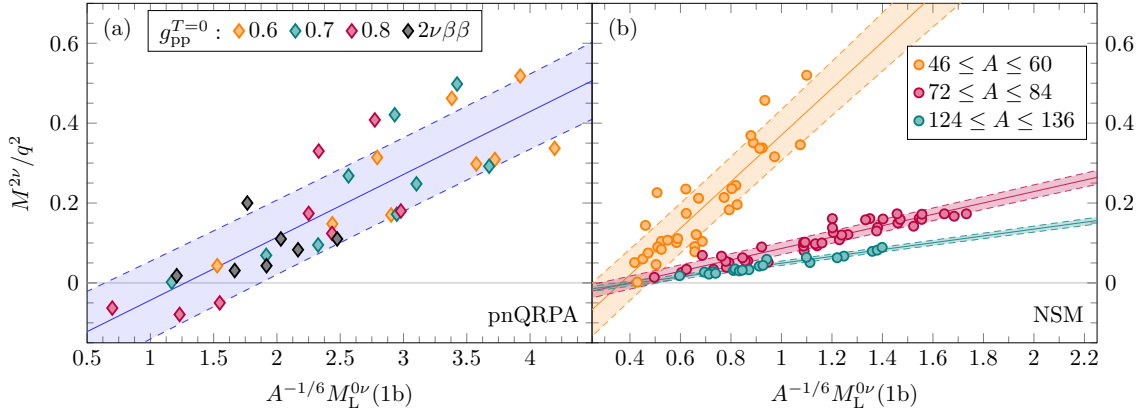

FIGURE 5.19: Figure taken from Ref. [13]. 2νββ- ($M^{2\nu}$) vs standard 0νββ-decay ($M_L^{0\nu}$) NMEs obtained with (a) QRPA with different isoscalar pairing $g_{pp}^{T=0}$ values, and (b) nuclear shell model (NSM) with different interactions for three regions of the nucleon-number $A$. $M_L^{0\nu}$ results are multiplied by $A^{-1/6}$, and the denominator $q^2$ notes the need to quench $M^{2\nu}$ values. Shell-model quenching ranges are: $q = 0.65 - 0.77$ for $46 \leq A \leq 60$, $q = 0.55 - 0.64$ for $72 \leq A \leq 84$, and $q = 0.42 - 0.72$ for $124 \leq A \leq 136$. In the pnQRPA method, the typical value $q = 0.79$ ($g_A^{eff} = 1.0$) is assumed. Solid and dashed lines correspond to linear fits and their 68% CL prediction band, respectively.

These results obtained with the 0νββ-2νββ correlation show that although QRPA NMEs are larger than the NSM ones, they are consistent when considering error bars. Moreover, they are in good agreement with previous calculations in both models. It is important to note that although our predicted uncertainty is comparable with the band obtained for a handful of literature values, the uncertainty derived from the correlation includes information from systematic calculations in tens of nuclei using different interactions.

Finally, in Ref. [13] we also studied the effect of adding two-body currents and the short-range operator into 0νββ-decay NMEs. We also found a good correlation in this case, although the uncertainties in the NMEs increase due to the uncertainty that is present in the calculation of the additional contributions.

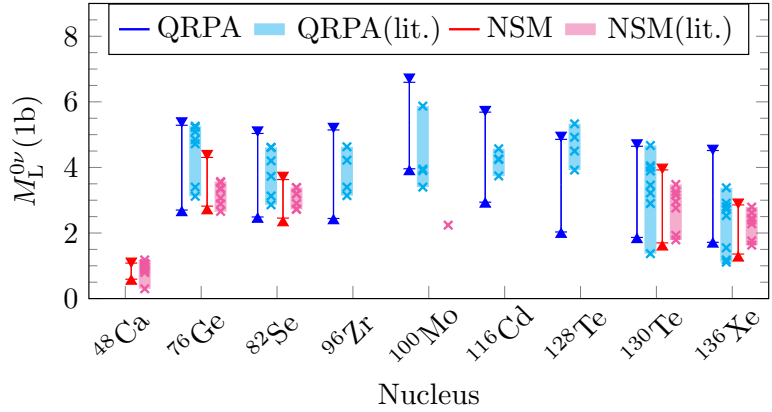

FIGURE 5.20: Standard $0\nu\beta\beta$-decay NMEs obtained from the correlations in Fig. 5.19. The narrow error bars come from the 68% CL bands of the linear fits, while the wide ones also contain uncertainties in the NME calculations. Bands (crosses) show the literature NME ranges (individual values), shell model (NSM) in red [111, 114, 262, 281–283], QRPA in blue [84, 114, 284–287].

## 5.5   $2\nu\beta\beta$ decay of $^{136}$Xe to the first excited $0^+$ state in $^{136}$Ba

In this section we will summarize the main results and conclusions of the study of the $2\nu\beta\beta$ decay of $^{136}$Xe to the first excited state in $^{136}$Ba published in Ref. [14].

The $2\nu\beta\beta$ decay half-lives depends quadratically on the NME, $M^{2\nu}$, which in turn encodes the information on the nuclear structure of the nuclear states involved in the transition. Therefore, one motivation is to benchmark theoretical predictions derived from different nuclear many-body calculations against $2\nu\beta\beta$ decay measurement, and test in this way nuclear structure methods. This provides valuable insight since with the same many-body methods we compute the $0\nu\beta\beta$-decay NMEs, and both transitions share the initial and final nuclear states, so a proper description of their nuclear structure is needed. Additionally, both transition operators are sensitive to spin and isospin operators, and could have correlated NMEs as discussed previously in Sec. 5.4. Therefore, tests of $M^{2\nu}$ could help in reduce the uncertainty on $0\nu\beta\beta$-decay NMEs.

The $2\nu\beta\beta$ decay to the excited states have been measured without previous theoretical prediction in $^{100}$Mo [49, 288] and $^{150}$Nd [49, 289]. They are being explored in other nuclei as in $^{76}$Ge [290, 291], $^{82}$Se [292], $^{130}$Te [293], and $^{136}$Xe [55, 294, 295]. In the work of Ref. [14] we predicted the half-life of the $2\nu\beta\beta$ decay of $^{136}$Xe to the first $0_1^+$ excited state in $^{136}$Ba with different many-body methods, namely the QRPA framework, the NSM, the IBM, and an effective field theory (EFT$_\beta$) for $\beta$ and $\beta\beta$ decays. In all of these calculations similar strategies as in the computation of $2\nu\beta\beta$ decay to the ground state have been followed.

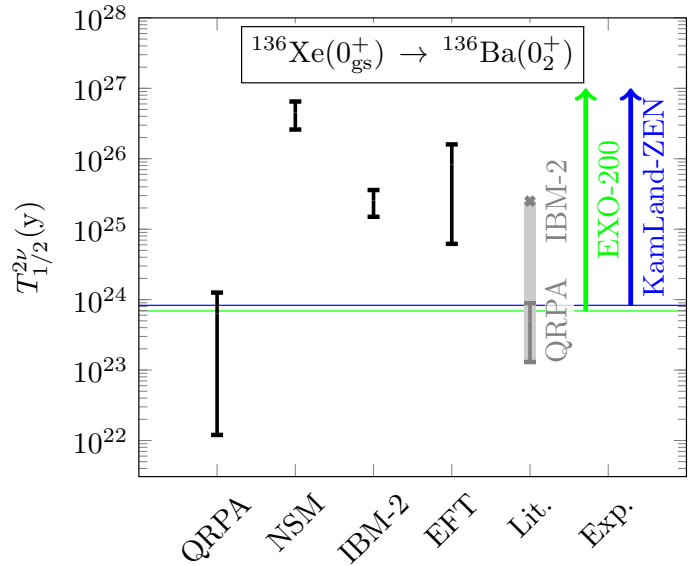

FIGURE 5.21: Figure taken from Ref. [14].$^{136}$Xe $2\nu\beta\beta$ decay half-life to the $^{136}$Ba $0_2^+$ state obtained (black bars) with the QRPA, nuclear shell model (NSM), IBM-2, and EFT. The results are compared with literature values (Lit.) [298, 299] (in gray) and experimental limits [294, 295] (horizontal lines with arrows).

The NSM results in Ref. [14] represent the first predictions for this decay using the same Hamiltonians as in previous $^{136}$Xe studies [91, 93, 114, 296, 297]. In the other methods the calculated NMEs improve previous predictions.

Figure 5.21 from Ref. [14] shows our predictions together with the ones of previous studies [298, 299]. The experimental limits in Ref. [294, 295], which were the data available previous to the publication of Ref. [14], are also shown for comparison. Although all the results are consistent with the experimental data available, a large uncertainty is observed from nuclear theory. In Fig. 5.21 we see that only a small part of the predicted QRPA value is inside the allowed region. On the other hand, the EFT$_\beta$ and IBM-2 results are compatible and predict a half-live an order of magnitude higher than current experimental lower limits. Meanwhile the NSM predicts the highest half-lives among all the calculated, indicating that the observation of this decay may demand improvements in the current experimental sensitivities by two orders of magnitude. We note that, from the latest experimental lower limits to the half-live obtained at 90% CL reported in Ref. [55], there are some tensions with the predicted values in QRPA.

Finally, this sparse theoretical predictions of the half-live point out the importance of the experimental measurement of $^{136}$Xe $2\nu\beta\beta$ decay to the $0_2^+$ state in $^{136}$Ba, since they will provide useful information to test the many-body methods used to calculate $0\nu\beta\beta$-decay NMEs.

# Chapter 6

# Conclusions and Outlook

The main goal of this thesis has been the study of a suitable second-order electromagnetic transition with the simultaneous emission of two-photons, $\gamma\gamma$ decay, which captures the main physics features that are present in the $0\nu\beta\beta$ decay. As quantum processes that occur within the atomic nucleus, the physics is encoded in the matrix elements. First, these matrix elements depend on the structure of the nuclear wave functions describing both the initial and final states. Therefore, the closer the initial and final states, the more related we expect to be the nuclear matrix elements. This is why we study the $\gamma\gamma$ decay in the $\beta\beta$ final nucleus, thus the final state of both transitions is the same. Moreover, isospin symmetry assures a good correspondence between $\gamma\gamma$-$\beta\beta$ initial states if we take as initial $\gamma\gamma$ state the double isobaric analog state of the $\beta\beta$ initial nucleus. Furthermore, the spin-tensor structure of the transition operator also plays an important role. Thus, the closest electromagnetic operator to the spin-isospin operator is the magnetic dipole operator. On the other hand, the reason of choosing an electromagnetic transition is because its measurement would allow a better determination of the $\gamma\gamma(M1M1)$ NME, which could be more accessible than the $0\nu\beta\beta$-decay NME.

As part of this thesis, I derived the transition amplitude and the differential decay width for a general two photon decay reproducing expressions in the literature [10]. Additionally, I checked the expressions of the total $\gamma\gamma$-decay width when applied to other multipolarity two-photon decay where data is available [245], and reproduced the experimental branching ratio from the reported best fit values of the mixed polarizabilities.

We found a good correlation between $\gamma\gamma(M1M1)$ and $0\nu\beta\beta$-decay NMEs obtained with the nuclear shell model using at least two different nuclear interactions [12]. The calculations include nineteen nuclei comprising titanium, chromium and iron isotopes with mass number ranging $46 \leq A \leq 60$ and twenty five heavier nuclei including zinc, germanium, selenium, krypton, tellurium, xenon and barium isotopes with mass number varying $72 \leq A \leq 136$. The correlation depends on the mass number, being slightly higher for nuclei in the $pf$ shell than for heavier nuclei in $pfg$ and $sdgh$ configuration spaces. However, this distinct behaviour is due to the energy denominator that defines the $\gamma\gamma(M1M1)$ operator.

This connection between $\gamma\gamma(M1M1)$ and $0\nu\beta\beta$-decay NMEs suggests a new avenue to predict $0\nu\beta\beta$-decay NMEs from a measurement of $\gamma\gamma(M1M1)$ decay from $0^+(DIAS)$, and also gives an uncertainty which is based on many sytematic calculations. Therefore, the correlation between $0\nu\beta\beta$ and $\gamma\gamma$ decay not only constitutes an alternative tool to improve our understanding of $0\nu\beta\beta$ decay guided by an experimental input, but it also motivates the experimental measurement of $\gamma\gamma(M1M1)$ decay from $0^+(DIAS)$ which constitutes a fascinating nuclear structure measurement.

In connection with the previous study, we also analysed the correlation between $0\nu\beta\beta$- and $2\nu\beta\beta$-decay NMEs, with the later being an observable for which there is data available. The results of this work have been published in Ref. [13]. Analogously to the $\gamma\gamma$ decay we calculated with the nuclear shell model both NMEs for the same nuclei and using the same effective interactions. We observed a very good linear correlation between the standard component of the $0\nu\beta\beta$-decay and $2\nu\beta\beta$-decay NMEs. Here, the data available from $2\nu\beta\beta$ decay half-life measurements is used together with the best linear fit and 68%CL prediction bands to give a value of the $0\nu\beta\beta$-decay NME with an uncertainty based on experimental data and the systematics of many NMEs. Two main conclusions are derived from this study. One is that the predicted NMEs by the two nuclear many-body methods (nuclear shell model and QRPA) are consistent when considering uncertainties. The other is that although the uncertainty in the predicted $0\nu\beta\beta$-decay is comparable with the band spanned by a handful of literature values, the uncertainty derived from this correlation includes in a more systematic way many NME calculations which incorporate also the effect of using different interactions.

Further, we give the first prediction of the half-life of the $2\nu\beta\beta$ decay of $^{136}$Xe to the first excited $0^+_2$ state of $^{136}$Ba calculated with the nuclear shell model in Ref. [14] before its experimental measurement. Additionally, the half-live has been predicted by other many-body methods, so that future data could test those methods and give valuable information on the nuclear structure of the nuclear states involved and therefore be helpful for the calculation of $0\nu\beta\beta$-decay NMEs.

Finally, we have studied the theoretical analysis of the main transitions from $0^+(DIAS)$ that can compete with $\gamma\gamma(M1M1)$ transitions. Since the $0^+(DIAS)$ are high energy nuclear states that lie above the nucleon separation energies, proton emission is an open decaying channel. However, it is an isospin forbidden decay, allowing the $\gamma$-decay branch to be competitive. We have computed the branching ratios of $\gamma\gamma(M1M1)$ with $\gamma(M1)$ transitions in all relevant isotopes for $\beta\beta$ decay searches and found that $\Gamma_{\gamma\gamma}/\Gamma_\gamma \simeq 10^{-6} - 10^{-8}$. In the case of proton emission we only reported the branching ratios $\Gamma_{\gamma\gamma}/\Gamma_p \simeq 10^{-8} - 10^{-9}$ for $^{48}$Ti, for which a phenomenological spectroscopic factor needed to calculate $\Gamma_p$ has been obtained. A fully theoretical analysis of the processes that can compete with $\gamma\gamma$-M1M1, which includes the calculation of the electric dipole transitions and internal pair conversion, is left for a future project in collaboration with the experimental group at the Laboratori Nazionaali di Legnaro [153].

In the same way that we found a correlation between $0\nu\beta\beta$-decay and $2\nu\beta\beta$-decay NMEs, there is a correlation between $2\nu\beta\beta$-decay and $\gamma\gamma$-M1M1 decay NMEs which could be used to obtain $\gamma\gamma$-M1M1 NMEs and transition probabilities from $2\nu\beta\beta$ decay data. This would give valuable information since it provides $\gamma\gamma(M1M1)$-decay NMEs with an uncertainty that comes from the correlation, and also it provides a consistency check of our individual calculations with a given effective interaction.

Although we have started with the analysis of $\gamma\gamma$ decay within the VS-IMSRG, there is still work to do to understand the interplay between isospin mixing and some convergence problems we have encountered in the $\gamma\gamma$-M1M1 operator. Another interesting fact that remains to be understood is why for some nuclei there are higher cancellations between positive and negative contributions to the magnetic dipole operator. Once we have understood these questions then we will go through the study of $\gamma\gamma(M1M1)$ and $0\nu\beta\beta$-decay NMEs and their possible correlation within the VS-IMSRG method.

Finally, a future project could include the evaluation of the two-nucleon current contributions to the double magnetic dipole and $0\nu\beta\beta$ decays. Another possible avenue would be trying to understand the origin of the different correlations obtained for $\gamma\gamma$-$\beta\beta$ NMEs obtained with the NSM and QRPA many-body methods.

# Appendix A

# Two neutrino double beta decay formulae

## A.1   Fermion wave functions

The emitted electrons in $2\nu\beta\beta$ decay are not free fermions but feel the Coulomb field of the final nucleus, therefore the scattering wave is not a plane wave but it can be expressed as a superposition of Coulomb-distorted spherical waves [300]

$$\psi_s(p_e, \mathbf{x}) = \sum_{\kappa, m, \mu} (4\pi) i^{l_\kappa} (l_\kappa \, m \, \tfrac{1}{2} \, s | j_\kappa \, \mu) Y^*_{l_\kappa m}(\hat{\mathbf{p}}) e^{-i\Delta_\kappa} \begin{pmatrix} G_\kappa(r) \chi^\mu_\kappa(\hat{\mathbf{r}}) \\ i F_\kappa(r) \chi^\mu_{-\kappa}(\hat{\mathbf{r}}) \end{pmatrix}, \qquad (A.1)$$

where $(l_\kappa \, m \, \tfrac{1}{2} \, s | j_\kappa \, \mu)$ is the Clebsch-Gordan coefficient, $Y^*_{l_\kappa m}(\hat{\mathbf{p}})$ the spherical harmonic for the unit vector $\hat{\mathbf{p}} = \mathbf{p}/|\mathbf{p}|$, and $\Delta_\kappa$ is a phase introduced to satisfy boundary conditions at $r \to \infty (V_C \to 0)$. The functions $G_\kappa$ and $F_\kappa$ depending on the quantum number $\kappa = (l-j)(2j+1)$ are the solutions of coupled first order differential equations derived from relativistic Dirac equation in presence of the external electrostatic potential.

There are several approaches to obtain radial wave functions $G_\kappa(r)$ and $F_\kappa(r)$ as described in [73]. They differ from the electrostatic potential chosen to solve the relativistic Dirac equation, that is either consider a uniform charge distribution, a pointlike distribution or a uniform charge distribution together with the screening effect of atomic electrons, which is important for heavier nuclei.

The anti-neutrino wave function is

$$v_s(p) e^{-i\mathbf{p}\cdot\mathbf{r}} = \sum_{\kappa, m, \mu} \frac{4\pi}{\sqrt{2}} i^{-l_\kappa} (l_\kappa \, m \, \tfrac{1}{2} \, s | j_\kappa \, \mu) Y^*_{l_\kappa m}(\hat{\mathbf{p}}) \begin{pmatrix} -i f_\kappa(r) \chi^\mu_{-\kappa}(\hat{\mathbf{r}}) \\ g_\kappa(r) \chi^\mu_{-\kappa}(\hat{\mathbf{r}}) \end{pmatrix}, \qquad (A.2)$$

with $g_\kappa = j_{l_\kappa}(pr)$ and $f_\kappa = \mathrm{sgn}(\kappa)j_{l_{-\kappa}}(pr)$. The leading order contribution comes from considering that the four leptons are emitted in $s_{1/2}$-wave

$$\psi_s(p_e) = \begin{pmatrix} g_{-1}(E,R)\chi_s \\ f_{+1}(E,R)(\boldsymbol{\sigma}\cdot\hat{\boldsymbol{p}})\chi_s \end{pmatrix}, \quad \psi_s(p_{\bar{\nu}}) = \sqrt{\frac{E_{\bar{\nu}}+m_\nu}{2E_{\bar{\nu}}}} \begin{pmatrix} \frac{(\boldsymbol{\sigma}\cdot\hat{\boldsymbol{p}})}{E_{\bar{\nu}}+m_\nu}\chi_s \\ \chi_s \end{pmatrix}, \qquad (A.3)$$

where the functions $g_{-1}(E)$ and $f_{+1}(E)$ depend on the electron energy and are obtained by evaluating the radial wave function at the nucleus radius $R = 1.2 A^{1/3}$. In this approximation the total angular momentum of the four lepton can be 0, 1 or 2, therefore $0^+ \to 0^+$ and $0^+ \to 2^+$ $2\nu\beta\beta$ decays can occur.

## A.2   Neutrino phase space integrals

The differential decay rate for $2\nu\beta\beta$ $0^+ \to 0^+$ decay can be calculated using the second order Fermi's Golden Rule

$$d\Gamma^{2\nu} = 2\pi \sum_{spin} |\mathcal{R}^{2\nu}|^2 \delta(E_{e_1} + E_{e_2} + E_{\bar{\nu}_1} + E_{\bar{\nu}_2} + E_f - E_i) d\Omega_{e_1} d\Omega_{e_2} d\Omega_{\nu_1} d\Omega_{\nu_2}, \qquad (A.4)$$

with $d\Omega_i = d^3\boldsymbol{p}_i/(2\pi)^3$. After integrating over the phase-space of the outgoing neutrinos, the differential decay rate as a function of the electrons energies ($0 \le E_{e_i} \le Q + m_e$) and the angle $\theta$ between the electrons momenta [70]

$$\frac{d\Gamma^{2\nu}}{dE_{e_1}dE_{e_2}d\cos\theta} = c_{2\nu}\left(A^{2\nu}(E_{e_1},E_{e_2}) + B^{2\nu}(E_{e_1},E_{e_2})\cos\theta\right), \qquad (A.5)$$

with

$$c_{2\nu} = \frac{G_F^4 \cos^2\theta_C m_e^9}{8\pi^7}. \qquad (A.6)$$

The functions $A^{2\nu}$ and $B^{2\nu}$ are phase-space integrals depending on electron energies obtained by integrating over the neutrino phase space

$$A^{2\nu}(E_{e_1},E_{e_2}) = \frac{1}{m_e^{11}} \int_{m_\nu}^{E_i - E_f - E_{e1} - E_{e1}} \mathcal{A}^{2\nu} p_{\bar{\nu}_1} E_{\bar{\nu}_1} p_{\bar{\nu}_2} E_{\bar{\nu}_2} dE_{\bar{\nu}_1} dE_{\bar{\nu}_2}, \qquad (A.7)$$

$$B^{2\nu}(E_{e_1},E_{e_2}) = \frac{1}{m_e^{11}} \int_{m_\nu}^{E_i - E_f - E_{e1} - E_{e1}} \mathcal{B}^{2\nu} p_{\bar{\nu}_1} E_{\bar{\nu}_1} p_{\bar{\nu}_2} E_{\bar{\nu}_2} dE_{\bar{\nu}_1} dE_{\bar{\nu}_2}. \qquad (A.8)$$

They encode the nuclear and leptonic matrix elements as

$$
\begin{aligned}
\mathcal{A}^{2\nu} = \frac{1}{4} \Bigg\{ & \left| g_V^2 (M_F^K + M_F^L) - g_A^2 (M_{GT}^K + M_{GT}^L) \right|^2 \\
& + 3 \left| g_V^2 (M_F^K - M_F^L) + \frac{1}{3} g_A^2 (M_{GT}^K - M_{GT}^L) \right|^2 \Bigg\} \\
& \times [g_{-1}^2(E_{e_1}, R) + f_1^2(E_{e_1}, R)][g_{-1}^2(E_{e_2}, R) + f_1^2(E_{e_2}, R)],
\end{aligned}
\tag{A.9}
$$

$$
\begin{aligned}
\mathcal{B}^{2\nu} = \Bigg\{ & \left| g_V^2 (M_F^K + M_F^L) - g_A^2 (M_{GT}^K + M_{GT}^L) \right|^2 \\
& - \left| g_V^2 (M_F^K - M_F^L) + \frac{1}{3} g_A^2 (M_{GT}^K - M_{GT}^L) \right|^2 \Bigg\} \\
& \times f_1^2(E_{e_1}, R) f_1^2(E_{e_2}, R) g_{-1}(E_{e_1}, R) g_{-1}(E_{e_2}, R).
\end{aligned}
\tag{A.10}
$$

The matrix elements $M_\alpha^{K,L}$, $\alpha = $ F, GT, are defined as

$$
M_\alpha^{K,L} = m_e \sum_{n,\eta} M_\alpha(n) \frac{\epsilon_{\alpha,n}}{\epsilon_{\alpha,n}^2 + (-1)^\eta \epsilon_{K,L}^2},
\tag{A.11}
$$

with $\eta = 1,2$ and $J_n^\pi = 0_n^+ (1_n^+)$ for Fermi and Gamow-Teller operators respectively, and the energies $\epsilon_{\alpha,n} = E_n(J_n^\pi) - (E_i + E_f)/2$, $\epsilon_{K(L)} = (E_{e2(e1)} + E_{\nu2} - E_{e1(e2)} - E_{\nu1})/2$, $\epsilon_{K(L)} \in (-Q/2, Q/2)$.

In order to simplify the differential $2\nu\beta\beta$ decay rate, the energies $\epsilon_{K,L}$ are usually neglected. This allows one to completely decouple the nuclear matrix elements (NMEs) from the leptonic phase-space integral. Due to the isospin symmetry, the Fermi transition is highly suppressed and only the Gamow-Teller NME is consider

$$
M^{2\nu} = M_{GT} = m_e \sum_n \frac{\langle 0_f^+ || \sum_i \tau_i^- \boldsymbol{\sigma}_i || 1_n^+ \rangle \cdot \langle 1_n^+ || \sum_j \tau_j^- \boldsymbol{\sigma}_j || 0_f^+ \rangle}{E_n(1_n^+) - \frac{1}{2}(E_i + E_f)}.
\tag{A.12}
$$

On the other hand, the phase-space integrands are now

$$
\mathcal{A}^{2\nu} = \frac{1}{4} |M^{2\nu}|^2 [g_{-1}^2(E_{e_1}, R) + f_1^2(E_{e_1}, R)][g_{-1}^2(E_{e_2}, R) + f_1^2(E_{e_2}, R)],
\tag{A.13}
$$

$$
\mathcal{B}^{2\nu} = |M^{2\nu}|^2 f_1^2(E_{e_1}, R) f_1^2(E_{e_2}, R) g_{-1}(E_{e_1}, R) g_{-1}(E_{e_2}, R).
\tag{A.14}
$$

Since $M^{2\nu}$ does not depend on the lepton energies, it can be factorized out of the phase-space integral and the decay half-life is

$$
\left[ T_{1/2}^{2\nu} \right]^{-1} = G_{2\nu} g_A^4 |M^{2\nu}|^2,
\tag{A.15}
$$

where $G_{2\nu}$ is the phase-space factor [73, 74].

# Appendix B

# Isospin symmetry in Nuclei

The isospin symmetry is the symmetry between neutrons and protons (or between quarks up and down) with respect the strong interaction. After the discovery of the neutron by Chadwick, Heisenberg introduced the isospin degree of freedom similar to the spin, as a way to mathematically describe neutrons and protons as different components of the same object, the nucleon. He was guided by the similar proton and neutron masses ($m_p c^2 = 938.27203(8)$MeV and $m_n c^2 = 939.56536(8)$MeV), and both have the same spin of $1/2$. Moreover, experiments show that if one neglects the electromagnetic interaction, the nuclear interaction between protons is close to the interaction between neutrons.

Nucleons are then characterized by an isospin quantum number $t = 1/2$, similar to the spin, and are represented by two-component spinors spanning and abstract vector space where the isospin operator $\boldsymbol{t}$ acts. The neutron and the proton are the two eigenstates of the third component of the isospin operator, $t_3$:

$$\psi_n(\boldsymbol{r}) = \psi(\boldsymbol{r}) \begin{pmatrix} 1 \\ 0 \end{pmatrix}, \quad \psi_p(\boldsymbol{r}) = \psi(\boldsymbol{r}) \begin{pmatrix} 0 \\ 1 \end{pmatrix}, \tag{B.1}$$

with eigenvalues $m_t = \pm 1/2$. The three components of the isospin operator $\boldsymbol{t}$, $t_i$, are the generators of the $SU(2)$ algebra

$$[t_i, t_j] = i\epsilon_{ijk} t_k, \tag{B.2}$$

and the square of the isospin operator, $\boldsymbol{t}^2$, commutes with all generators $[\boldsymbol{t}^2, t_i] = 0$. The ladder operators, $t_\pm$

$$t_\pm = t_1 \pm i t_2, \tag{B.3}$$

transform a proton into a neutron and vice versa. From charge conservation and the approximated charge-independence of the strong interaction, Wigner [301] introduced the total isospin for an $A$ system of nucleons

$$\boldsymbol{T} = \sum_{k=1}^{A} \boldsymbol{t}(k), \tag{B.4}$$

and analogously for its components

$$T_\pm = \sum_{k=1}^{A} t_\pm(k), \qquad T_3 = \sum_{k=1}^{A} t_3(k), \tag{B.5}$$

with eigenvalues $T(T+1)$ and $M_T = 1/2(N-Z)$ of the operators $\mathbf{T}^2$ and $T_3$, respectively. A charge independent Hamiltonian would satisfy

$$[H_{\mathrm{nucl}}, \mathbf{T}] = 0, \tag{B.6}$$

or

$$[H_{\mathrm{nucl}}, T_3] = [H_{\mathrm{nucl}}, T_\pm] = 0. \tag{B.7}$$

As a consequence, the $A$ nucleon eigenstates of $H_{\mathrm{nucl}}$ have good isospin. Therefore $T$ represents an additional quantum number to label an $A$ nucleon state apart from the total angular momentum and parity $J^\pi$. This way, we have degenerate multiples $(J^\pi, T)$ for nuclei with the same $A$ and $M_T = -T, -T+1, \ldots, T$, called the *isobaric analog states* (IAS).

However, the isospin symmetry is approximate and a dynamical breaking of the isospin symmetry can explain the splitting of the isobaric multiplets [302], known as Coulomb displacement energies. The main source of isospin symmetry breaking in the nucleus comes from the Coulomb interaction. This can be seen if the Coulomb interaction between two nucleons, $V_C$, is written in terms of isospin operators $\mathbf{t}$. After doing it, $V_C$ is expanded by a sum of an isoscalar ($V_0^{(0)}$), an isovector ($V_0^{(1)}$) and an isotensor ($V_0^{(2)}$) term [302]

$$V_c = V_0^{(0)} + V_0^{(1)} + V_0^{(2)}. \tag{B.8}$$

All terms are the 0-component of rank-q ($q = 0, 1, 2$) since the electric charge is conserved. Therefore, the effect of the Coulomb interaction on a given state $|\xi T M_T\rangle$, with $\xi$ denoting the additional quantum numbers, can be estimated using perturbation theory at first-order, with the energy shift of this state given by the diagonal matrix element $\langle \xi T M_T | V_C | \xi T M_T \rangle$

$$E_C(\xi, T, M_T) = E^{(0)}(\xi, T) + E^{(1)}(\xi, T) M_T + E^{(2)}(\xi, T)(3M_T^2 - T(T+1)), \tag{B.9}$$

where $E^{(q)}(\xi, T)$ are functions of $T$ and the reduced matrix elements of $V_0^{(q)}$ in the isospin space. Thus, in lowest-order of perturbation theory, the diagonal matrix elements of the Coulomb interaction are given by a quadratic form in $M_T$ by Eq. (B.9). However, the expansion in $M_T$ in Eq. (B.9) is an approximation of the true Coulomb interaction, since it represents the diagonal part where the isovector and isotensor parts have been neglected. In this case the Coulomb interaction is assumed not to mix states $|\xi T M_T\rangle$ with different values of $T$ ($T = M_T, M_T + 1, \ldots$).

To find the total energy of a specific nuclear state, we need to include the nuclear interaction, and if it is at most of two-body character, the same arguments here apply. Then, when such dependence is rewritten for nuclear masses, we find the isobaric multiplet mass equation (IMME) [303]

$$M(\xi, T, M_T) = a(\xi, T) + b(\xi, T)M_T + c(\xi, T)M_T^2, \tag{B.10}$$

where $M$ represents the atomic mass excess (mass excess is the difference between the experimental nuclear mass and $A$ atomic mass units expressed in energy units). Then, the excitation spectra of different nuclei belonging to the same isopin multiplet, that is the same $T$ but different $M_T$ (the isobaric analog states), are identical but the corresponding states do not have the same binding energy.

Finally, note that every symmetry brings with it selection rules for transitions operators. Transition operators can be expressed as tensor operators in the isospin space. In the case of electromagnic transitions, if they are given by one-body operators as assumed in Sec. 4.1.2, for electromagnetic operators of multipolarity $L$ one has the contribution of an isoscalar and an isovector term $O_{ML} = O_{ML}^{(0)} + O_{ML}^{(1)}$. Therefore, transitions with $\Delta T = 1$ are only driven by the isovector component $O_{ML}^{(1)}$, while for transitions with $\Delta T = 0$ both terms contribute, except transitions between two $T = 0$ estates where only the isovector component contributes.

# Appendix C

# Second Quantization

The name of "second quantization" is just to refer to an alternative formulation of the usual coordinate space quantum mechanics, which in turn is very useful for dealing with the many-body problem. A more explicit way to refer to this formulation is the *occupation number representation*.

## C.1 Occupation number representation for fermions

The wave function of a system of $N$ fermions is completely antisymmetric under the exchange of any two particles ($i \leftrightarrow j$)

$$\Psi(\boldsymbol{r}_1, \boldsymbol{r}_2, \ldots, \boldsymbol{r}_i, \ldots, \boldsymbol{r}_j, \ldots, \boldsymbol{r}_N) = -\Psi(\boldsymbol{r}_1, \boldsymbol{r}_2, \ldots, \boldsymbol{r}_j, \ldots, \boldsymbol{r}_i, \ldots, \boldsymbol{r}_N). \tag{C.1}$$

If the Hamiltonian of the total system is separable, that is there are no interaction terms that depend on more than one particle coordinate, then the total many-body wave function can be written in terms of the single-particle states $\psi_\alpha$

$$\Psi(\boldsymbol{r}_1, \ldots, \boldsymbol{r}_N) = \frac{1}{\sqrt{N!}} \sum_P \text{sign}(P) P\{\psi_{\alpha_1}(\boldsymbol{r}_i) \ldots \psi_{\alpha_A}(\boldsymbol{r}_N)\}, \tag{C.2}$$

which is just an $N$-particle Slater determinant. Here the sum runs over all $N!$ permutations of the $N$ indices, and $\text{sign}(P)$ is $-1$ for an odd and $+1$ for an even permutation. However, since the information of which particle occupies which single particle state is meaningless from the point of view of quantum mechanics, the relevant information is encoded in how many particles populate the state $\psi_\alpha$, called the *occupation number $n_\alpha$*. The many-particle state is

$$|\Psi\rangle = |n_1, n_2, \ldots, n_N\rangle. \tag{C.3}$$

The values of $n_\alpha$ are 0 or 1 if the single particle state $\psi_\alpha$ is empty or occupied. The wave function $\psi_\alpha(\boldsymbol{x})$ is the coordinate representation of the state vector $|\alpha\rangle$, $\langle \boldsymbol{x}|\alpha\rangle$.

The Hilbert space of these abstract vectors characterized by varying particle numbers is called the *Fock space*, that is the infinite direct sum of the antisymmetrized tensor product of single-particle Hilbert spaces

$$\mathcal{F}(\mathcal{H}) = \bigoplus_{n=0}^{\infty} AS(\mathcal{H}^{\otimes n}). \tag{C.4}$$

In this space are defined operators that change particle number, such as the creation and annihilation operators $c_\alpha^\dagger$ and $c_\alpha$, respectively. The operator $c_\alpha^\dagger$ creates a fermion in the state $|\alpha\rangle$, and $c_\alpha$ destroys a fermion in the state $|\alpha\rangle$, therefore they satisfy the anticommutation relations

$$\{c_\alpha, c_\beta^\dagger\} = \delta_{\alpha\beta}, \quad \{c_\alpha, c_\beta\} = 0, \quad \{c_\alpha^\dagger, c_\beta^\dagger\} = 0. \tag{C.5}$$

Their action on the Fock state $|n_1, n_2, \ldots\rangle$ is defined as

$$c_\alpha^\dagger |n_1, \ldots, n_\alpha, \ldots\rangle = \begin{cases} \zeta_\alpha |n_1, \ldots, n_\alpha + 1, \ldots\rangle, & \text{if } n_\alpha = 0, \\ 0, & \text{if } n_\alpha = 1, \end{cases} \tag{C.6}$$

$$c_\alpha |n_1, \ldots, n_\alpha, \ldots\rangle = \begin{cases} \zeta_\alpha |n_1, \ldots, n_\alpha - 1, \ldots\rangle, & \text{if } n_\alpha = 1, \\ 0, & \text{if } n_\alpha = 0 \end{cases}, \tag{C.7}$$

with the phase factor $\zeta_\alpha = \pm 1$. The particle vacuum is defined as

$$|0\rangle \equiv |0, 0, \ldots\rangle, \tag{C.8}$$

and $c_\alpha |0\rangle = 0$ for all $\alpha$. The $N$-particle state is given by

$$|n_1, n_2, \ldots, n_N\rangle = c_{\alpha_1}^\dagger c_{\alpha_2}^\dagger \ldots c_{\alpha_N}^\dagger |0\rangle = \Pi_\alpha (c_\alpha^\dagger)^{n_\alpha} |0\rangle. \tag{C.9}$$

## C.2   Matrix Elements

In order to make many-body calculations in an efficient way one works in the occupation number representation. Since we use the harmonic oscillator basis as our basis of single particle orbitals the creation and annihiliation operators of a nucleon in an orbit $a$, $c_a$ and $c_a^\dagger$, satisfy Eq. (C.5). The label $a$ is a shorthand notation for specifying the harmonic oscillator orbit and is defined as $a = (n_a, l_a, j_a, m_a, t_{z,a})$, where $n_a, l_a, j_a, m_a$, and $t_{z,a}$ are the principal quantum number, orbital angular momentum, total angular momentum, $z$-component of $j_a$ and $z$-component of the isospin which distinguishes between neutrons and protons, respectively. In this basis the proton and neutron states have $t_z = -1/2(1/2)$, respectively.

In terms of the creation and annihilation operators, a general $n$-body operator $O^{(n)}$ is defined by

$$O^{(n)} = \frac{1}{(n!)^2} \sum_{a_1' \ldots a_n'} \sum_{a_1 \ldots a_n} O_{a_1' \ldots a_n' a_1 \ldots a_n} c_{a_1'}^\dagger \ldots c_{a_n'}^\dagger c_{a_1} \ldots c_{a_n}, \tag{C.10}$$

with $O_{a_1' \ldots a_n' a_1 \ldots a_n} = \langle a_1' \ldots a_n' | \hat{O}^{(n)} | a_1 \ldots a_n \rangle$.

### C.2.1 One-body operators

In the occupation number representation a one-body operator can be expressed as

$$T = \sum_{a,b} t_{ab} c_a^\dagger c_b, \tag{C.11}$$

with $t_{ab} \equiv \langle a | T | b \rangle$.

In the same way, a spherical tensor of rank $L$, $\boldsymbol{T}_{LM}$, is given by

$$\boldsymbol{T}_{LM} = \sum_{a,b} \langle a | \boldsymbol{T}_{LM} | b \rangle c_a^\dagger c_b. \tag{C.12}$$

Using the Wigner-Eckart Theorem, $\boldsymbol{T}_{LM}$ can be rewritten as

$$\boldsymbol{T}_{LM} = \hat{L}^{-1} \sum_{a,b} \langle j_a || \boldsymbol{T}_{LM} || j_b \rangle [c_a^\dagger \tilde{c}_b]_{LM}, \tag{C.13}$$

where $\tilde{c}_a = (-1)^{j_a + m_a} c_{-a}$, $c_{-a} = c_{j_a, -m_a}$, and

$$[c_a^\dagger \tilde{c}_b]_{LM} = \sum_{m_a, m_b} (j_a m_a j_b m_b | LM) c_a^\dagger \tilde{c}_b, \tag{C.14}$$

and the matrix element $\langle j_a || \boldsymbol{T}_{LM} || j_b \rangle$ is called the *reduced one-body* or *single-particle matrix element* [61].

Finally, we can obtain the matrix element between general initial and final states $|\xi_i J_i M_i\rangle$ and $|\xi_f J_f M_f\rangle$, respectively. Here $\xi_{i,f}$ label additional quantum numbers needed to fully characterize the quantum system. Then, we have

$$\langle \xi_f J_f || \boldsymbol{T}_{LM} || \xi_i J_i \rangle = \hat{L}^{-1} \sum_{a,b} \langle j_a || \boldsymbol{T}_{LM} || j_b \rangle \langle \xi_f J_f || [c_a^\dagger \tilde{c}_b]_L || \xi_i J_i \rangle. \tag{C.15}$$

The reduced nuclear matrix elements in Eq. (C.15) is called the *transition amplitude* or *decaying amplitude*, when it refers to a decay process. The reduced nuclear matrix element $\langle \xi_f J_f || [c_a^\dagger \tilde{c}_b]_L || \xi_i J_i \rangle$ is called the *reduced one-body transition density* (OBTD). While the reduced one-body nuclear matrix element are calculated analytically, the reduced one-body transition densities encode the many-body properties of the initial and final nuclear states, being objects calculated numerically within a many-body method. They only know from the transition operator its rank and one-body character.

# Resum

En aquest treball, un procés de física conegut des de fa molt temps, la desintegració doble-gamma ($\gamma\gamma$), s'ha revisat des d'una nova perspectiva: proporcionant informació valuosa en els elements de matriu nuclear (NMEs) de la desintegració doble beta sense neutrins ($0\nu\beta\beta$). La desintegració $0\nu\beta\beta$, encara no descoberta, té un potencial físic únic ja que és sensible a la manera de descriure els neutrins i, de manera més fonamental, a les simetries de la natura. Preguntes bàsiques que continuen sent grans desafiaments oberts per a la física. Però com a grans reptes, continuen empenyent els límits del nostre coneixement tant a nivell teòric com experimental.

El descobriment de les oscil·lacions de sabor dels neutrins va demostrar que els neutrins són partícules massives, exigint una extensió del Model Estàndard de la física de partícules. Aquests experiments són sensibles a les diferències de quadrat de massa dels neutrins i els angles de barreja, amb dades actuals que recolzen l'existència d'almenys dos neutrins actius amb masses no nul·les. No obstant això, no proporcionen informació sobre l'escala de massa absoluta. D'altra banda, els efectes de la matèria durant la propagació dels neutrins han mesurat un dels dos signes per a les dues diferències de massa al quadrat. Així, hi ha dues possibilitats d'ordenació de masses, ja siga que els neutrins estiguen en la jerarquia normal de masses (NH) o en la jerarquia invertida de masses (IH), com es mostra a la Fig. R1.

L'origen de la massa dels neutrins i l'escala absoluta estan intrínsecament relacionats amb la naturalesa dels neutrins, ja siguen fermions de Dirac o de Majorana, i amb la violació de les simetries globals exactes del Model Estàndard. La petitesa de les masses dels neutrins en comparació amb altres fermions apunta cap a un mecanisme de generació de masses diferent al de la resta del contingut de partícules del Model Estàndard. Un argument comúment utilitzat en contra que els neutrins tinguin només un terme de massa de Dirac és que requereix la introducció d'un acoblament de Yukawa extremadament petit de l'ordre de $\mathcal{O}(10^{-14})$, que és aproximadament de 10 ordres de magnitud més petit que els acoblaments de Yukawa d'altres fermions. No obstant això, hi ha un mecanisme general per explicar la petitesa de la massa dels neutrins més enllà del Model Estàndard, que es basa en l'observació que els neutrins es poden descriure com un espinor de Majorana de dos components. En concret, un fermió representat per un espinor de Majorana és el

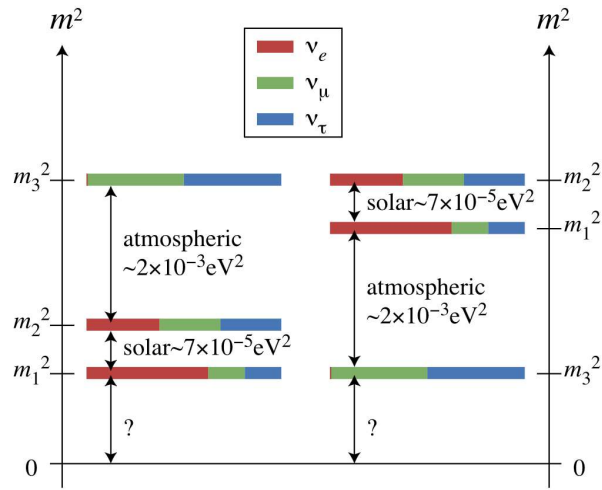

<small>Fɪɢ. R1: Dues configuracions d'ordenacions de quadrats de massa dels neutrins. A l'esquerra: jerarquia normal (NH). A la dreta: jerarquia invertida (IH). La probabilitat que un estat d'autoestat de massa de neutrí particular $\nu_i$ amb massa $m_i$ continga un estat de base de massa de lepton carregat particular $\nu_l$ ($l = e, \mu, \tau$) es representa amb colors. Figura extreta de Ref. [304].</small>

seu propi estat de conjugació de càrrega, el que permet escriure un terme de massa diferent construït només a partir de camps de neutrins levogir conegut com *terme de massa de Majorana*. Una possible explicació de la petitesa de les masses dels neutrins que apareixen en aquest terme de massa prové del marc de la teoria efectiva de camps i de la inclusió de l'operador de dimensió-5 d'ordre més baix construït amb el contingut de camps del Model Estàndard, conegut com l'operador de Weinberg. Així, la petitesa de les masses dels neutrins apunta cap a l'existència d'una nova escala de la física fonamental.

Un dels processos més prometedors per a esclarir la naturalesa dels neutrins és la desintegració doble beta sense neutrins ($0\nu\beta\beta$), una transició feble de segon ordre on dos neutrons dins d'un nucli atòmic es transformen en dos protons amb l'emissió de dos electrons i sense antineutrins, violant la conservació del nombre leptònic. Per tant, una observació d'aquesta desintegració mostraria que el neutrí té un component de Majorana, i per tant, que el neutrí és la seua pròpia antipartícula. Així doncs, el descobriment de la desintegració $0\nu\beta\beta$ podria sondejar models que busquen explicar l'asimetria matèria-antimatèria en l'univers, i proporcionar informació valuosa sobre la massa del neutrí.

Aquestes implicacions deixen clar el potencial físic de la desintegració $0\nu\beta\beta$ com a descobriment innovador, el qual ha motivat una àvida recerca experimental. Tot i que encara no s'ha observat la desintegració $0\nu\beta\beta$, amb el límit inferior actual de la semivida experimental d'aproximadament $10^{26}$yr, hi ha un programa ambiciós que té com a objectiu augmentar el límit fins a $10^{28}$yr gràcies al desenvolupament de nova tecnologia de detectors, la selecció cuidadosa de materials radiopurs i augmentant la massa de l'isòtop de desintegració $\beta\beta$ fins a l'escala de tones i més enllà. En el mecanisme més simple, la desintegració $0\nu\beta\beta$ és causada per l'intercanvi de neutrins de Majorana lleugers. En aquest cas, la semivida depèn de forma quadràtica del que es coneix com la *massa efectiva de neutrins $m_{\beta\beta}$*, un paràmetre que combina les masses dels neutrins $m_i$, els elements de la matriu de mescla de Pontecorvo-Maki-Nakawa-Sato (PMNS) ($U_{ei}$) i dos fases de Majorana. Depenent de l'ordenació de les masses dels neutrins, hi ha regions permeses distintes per a $m_{\beta\beta}$, que els experiments poden sondejar si la desintegració $0\nu\beta\beta$ està causada per l'intercanvi de neutrins lleugers.

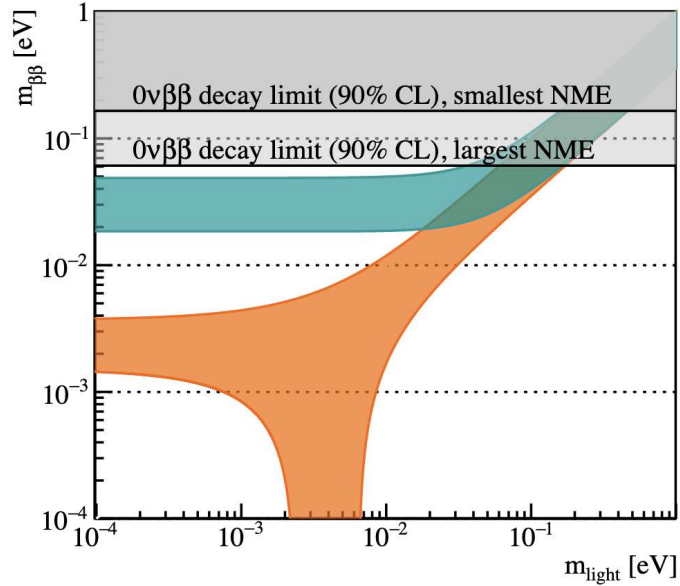

FIG. R2: Espai de paràmetres màximament permès per a $m_{\beta\beta}$ com a funció de $m_{\text{light}}$. Les àrees ocre i blava ombrejades mostren l'espai de paràmetres permès per a NH i IH, respectivament. Mentre que les àrees grises mostren l'espai de paràmetres exclòs pels experiments de desintegració $0\nu\beta\beta$. Figura presa de Ref. [62]

Els futurs detectors a escala de tones tenen com a objectiu cobrir tota la regió de masses IH, $m_{\beta\beta} > (18.4 \pm 1.3)$meV. Per tant, un descobriment serà possible si l'espectre de masses està invertit o si $m_{\text{lightest}} > 50$meV, independentment de l'ordre. En cas d'un senyal positiu a escala de tones, les tècniques desenvolupades en aquests experiments o en els seus programes de recerca i desenvolupament (R&D) paral·lels serien de vital importància per a mesures de precisió amb l'objectiu d'establir el mecanisme de desintegració.

Al mateix temps que una àvida cerca experimental de la desintegració $0\nu\beta\beta$ està en curs, en el costat teòric tant la comunitat de física de partícules com la teoria nuclear enfronten un programa desafiador ja que una interpretació adequada dels resultats experimentals requereix prediccions teòriques fiables. L'expressió general per a la semivida de la desintegració $0\nu\beta\beta$ és:

$$\left[T_{1/2}^{0\nu}\right]^{-1} = \sum_i G_i g_i^4 |\mathcal{M}_i|^2 f_i(\Lambda) + \text{interference terms}, \tag{R1}$$

que es factoritza com a producte de quatre factors: **i)** el factor d'espai de fase $G_i$, que té en compte la cinemàtica de la desintegració i depèn principalment de $Q_{\beta\beta}$ i del nombre atòmic del nucli inicial; **ii)** $g_i$ és l'element de matriu hadrònic que descriu el coupling al nucleó; **iii)** l'element de matriu nuclear $\mathcal{M}$ que codifica l'estructura nuclear dels estats nuclears inicials i finals així com l'operador de transició escrit en termes dels graus de llibertat dels nucleons i, finalment, **iv)** $f_i$ és el paràmetre sense dimensions de nova física distintiu de cada mecanisme de violació del nombre leptònic (LNV). D'una banda, la comunitat de física de partícules està centrada en la reducció dels processos de LNV responsables de la desintegració. D'aquesta manera, es proposen recerques d'altres sondes com ara experiments de neutrins d'energia baixa, col·lisionadors d'alta energia, observacions astrofísiques i cosmològiques per complementar els resultats dels experiments de $0\nu\beta\beta$. Mentrestant, la comunitat de teoria nuclear es centra en la part hadrònica i de molts cossos que entra en l'element de matriu nuclear. Aquestes dues peces estan lligades però les seues millores principals es poden treballar per separat.

L'assumpció estàndard és que la font dominant de LNV a baixes energies és la massa de neutrins de Majorana ($-\overline{\Psi}M_\nu\Psi^C/2$), on el spinor conjugat de càrrega $\Psi^C = C\overline{\Psi}^T$, i el vector $\Psi$ inclou en la seua versió minimal només els neutrins del Model Estàndard $\Psi^T = (\nu_{L,e}, \nu_{L,\mu}, \nu_{L,\tau})$. Llavors, la semivida de desintegració $0\nu\beta\beta$ en aquest cas és

$$\left[T_{1/2}^{0\nu}\right]^{-1} = G_{01}g_A^2 |\mathcal{M}_{\text{ligth}}^{0\nu}|^2 \frac{m_{\beta\beta}^2}{m_e^2}, \tag{R2}$$

on $m_{\beta\beta}$ és el *paràmetre de massa efectiva* esmentat anteriorment i es defineix com

$$m_{\beta\beta} = \left| \sum_{i=1}^{3} |U_{ei}|^2 e^{i\varphi_i} m_i \right|. \tag{R3}$$

És la *massa Majorana efectiva* de $\nu_e$, que és l'element $ee$ de la matriu de masses $M_\nu = U\text{diag}(m_1, m_2, m_3)U^T$. Ací, $U$ és la matriu de barreja de neutrins de Pontecorvo-Maki-Nakawaga-Sakata (PMNS) i $\varphi = \{\varphi_1, \varphi_2, 1\}$ són coneguts com a fases de Majorana. A més, $\mathcal{M}_{\text{light}}^{0\nu}$ és l'EMN de l'intercanvi de neutrins lleugers i $G_{01}$ és el factor d'espai de fase. És clar a partir de l'Eq. (R2) que si la comunitat de física experimental poguera mesurar la desintegració doble beta sense neutrins junt amb les millores en la mesura de la matriu de masses PMNS, i si la comunitat de teoria nuclear pot determinar tant el factor d'espai de fase com l'EMN, llavors es podria determinar la massa absoluta del neutrí. No obstant això, tant la part experimental com la teoria enfronten grans desafiaments com discutirem detalladament en aquesta tesi.

Encara que $f_i$ i $M^{0\nu\beta\beta}$ són dos ingredients fonamentals, sense observació de $0\nu\beta\beta$ i mentre les prediccions teòriques del paràmetre de nova física siguin desconegudes, els EMN són una entrada que s'ha de calcular dins d'un model de estructura nuclear. De

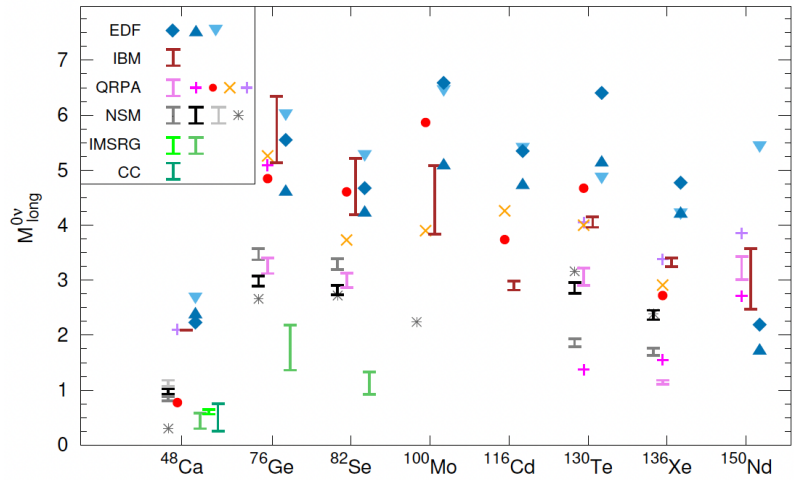

F<small>IG</small>. R3: Elements de matriu nuclear de $M^{0\nu}$ per al component estàndard de l'intercanvi de neutrins lleugers de diferents mètodes de molts cossos. Figura presa de Ref. [62].

fet, un dels problemes més urgents amb els quals s'ha enfrontat la comunitat de teoria nuclear en l'última dècada és la discrepància d'un factor d'aproximadament 3-5 en les prediccions d'EMN de diferents mètodes de molts cossos i per diferents grups, com es mostra a la Fig. R3. Per tant, tant una observació de $0\nu\beta\beta$ com un límit exigiran un càlcul precís de l'EMN amb una dependència mínima del model i una incertesa quantificada per extreure una informació quantitativa de les masses dels neutrins o dels paràmetres de LNV d'altres models més enllà del Model Estàndard.

Les comunitats de física nuclear i de partícules han fet grans progressos durant els últims anys. Aquestes millores s'han centrat en el desenvolupament del marc de teoria de camps efectius (EFT) que descriu aquesta desintegració des de l'escala d'alta energia en què emergeix la nova física fins a l'escala d'energia baixa on té lloc la desintegració nuclear $0\nu\beta\beta$. En paral·lel, els primers càlculs dels EMN de la desintegració $0\nu\beta\beta$ més fiables des de mètodes *ab initio* de molts cossos han esdevingut recentment factibles. El marc EFT també proporciona una manera de millorar sistemàticament els hamiltonians d'interacció arrelats en la teoria fonamental de la cromodinàmica quàntica i, quan es combinen amb un mètode *ab initio* de molts cossos, permeten quantificar la incertesa dels EMN de la desintegració $0\nu\beta\beta$, un aspecte teòric clau.

Actualment, la quantificació de la incertesa dels EMN de la desintegració $0\nu\beta\beta$ és un camp en creixement, on diversos enfocaments es troben sota recerca activa. L'objectiu principal d'aquesta tesi ha estat investigar un d'aquests enfocaments, que és el càlcul d'observables nuclears relacionades amb la desintegració $0\nu\beta\beta$ com una manera d'ajudar a determinar i reduir les incerteses teòriques en els EMN de la desintegració $0\nu\beta\beta$. Hi ha dues observables encara no mesurades que mostren una bona correlació amb els EMN de la desintegració $0\nu\beta\beta$: les transicions electromagnètiques doble Gamow-Teller i doble dipol magnètic, $\gamma\gamma$(M1M1). Aquest últim ha estat l'objecte principal d'estudi en aquesta tesi. Hem proposat que la mesura del decaïment doble dipol magnètic $\gamma\gamma$ des de l'estat analògic isobàric doble del $0_{gs}^+$ del nucli inicial de la $\beta\beta$ fins al $0_{gs}^+$ del nucli final de la desintegració $\beta\beta$ podria establir el valor i reduir la incertesa en els EMN de la desintegració $0\nu\beta\beta$, ja que els EMN dels dos processos estan molt ben correlacionats.

De la mateixa manera, també hem explorat la validesa d'aquest enfocament per predir i quantificar les incerteses dels EMN en el cas on hi ha dades disponibles per a l'observable nuclear relacionada amb la desintegració $0\nu\beta\beta$, és a dir, per a la desintegració permesa pel Model Estàndard $2\nu\beta\beta$. La relació trobada entre els EMN de les desintegracions $2\nu\beta\beta$ i $0\nu\beta\beta$ tant en NSM com en QRPA va permetre l'estimació dels EMN de la desintegració $0\nu\beta\beta$ a partir dels valors mesurats de la semivida de $2\nu\beta\beta$. Dins d'aquest enfocament, els EMN de $0\nu\beta\beta$ poden ser predits utilitzant tant les dades com els càlculs sistemàtics a través de desenes de nuclis utilitzant diferents interaccions nuclears. D'aquesta manera, també permet proporcionar les seues incerteses teòriques basades en els errors sistemàtics que són capturats per tots els EMN que segueixen la mateixa correlació.

L'interès d'estudiar les transicions electromagnètiques en els nuclis és que proporcionen una eina potent per provar les propietats de l'estructura nuclear, així com per donar informació útil sobre la dinàmica dels graus de llibertat rellevants que entren en joc a una energia determinada. A més, la naturalesa electromagnètica de la interacció permet una interpretació clara de les dades experimentals.

La raó de triar un observable electromagnètic com a possible observable relacionat amb $0\nu\beta\beta$ és perquè la seua mesura permet una determinació clara dels EMN una vegada es dissenya una configuració experimental adequada. Un punt de partida raonable seria triar un operador amb una dependència de l'espín similar a la contribució dominant als EMN de la desintegració $0\nu\beta\beta$, que és l'operador de Gir Gamow-Teller d'espín-isospin ($\boldsymbol{\sigma} \cdot \boldsymbol{\tau}$). Per tant, l'operador de dipol magnètic $\boldsymbol{M1}$, definit com a

$$\boldsymbol{M1} = \mu_N \sqrt{\frac{3}{4\pi}} \sum_{i=1}^{A} (g_i^l \mathbf{l}_i + g_i^s \mathbf{s}_i), \tag{R4}$$

sembla el millor candidat. A més, ja que la dependència de l'estructura nuclear dels EMN ve dels matrius de densitat de transició, es pot esperar una estructura d'espín-isospin més propera de les dues densitats de transició per a una correspondència millorada entre els dos EMN. Per aconseguir aquest objectiu, la simetria d'isospin garanteix una gran similitud entre els decaïments $\gamma\gamma$ i $\beta\beta$ si triem el decaïment de l'estat analògic isobàric doble (DIAS) de l'estat inicial de la $\beta\beta$ com es representa a la Figura R4. Llavors, els decaïments $\gamma\gamma$ i $\beta\beta$ comprenen els nuclis

$$\gamma\gamma: \qquad {}_{Z}^{A}Y_N^* \to {}_{Z}^{A}Y_N + 2\gamma \tag{R5}$$

$$0\nu\beta\beta: \qquad {}_{Z-2}^{A}X_{N+2} \to {}_{Z}^{A}Y_N + 2e^- \tag{R6}$$

amb $N, Z$ el nombre de neutrons i protons. $^A_Z Y^*_N$ representa el nucli final $\beta\beta$ excitat en el DIAS. Aquest és un estat amb isospín $T = T_z + 2$, amb $T_z$ la tercera component de l'isospin del nucli final $\beta\beta$, $T_z = (N - Z)/2$. Per resumir, el decaïment $\gamma\gamma$ involucra

$$|0^+_i\rangle_{\gamma\gamma} \equiv |0^+_i\rangle_{\beta\beta}(\text{DIAS}) = \frac{T^- T^-}{K^{1/2}} |0^+_i\rangle_{\beta\beta}, \tag{R7}$$

$$|0^+_f\rangle_{\gamma\gamma} \equiv |0^+_f\rangle_{\beta\beta}, \tag{R8}$$

amb $K$ una constant de normalització i $T^- = \sum_i^A t_i^-$ l'operador de disminució de l'isospin nuclear, que només canvia $T_z$. A més, per connectar $|0^+_i\rangle_{\gamma\gamma}$ i $|0^+_f\rangle_{\gamma\gamma}$ aquests estats intermedis han de ser els estats analògics isobàrics del nucli intermig amb isospín $T = T_z + 1$, com es mostra a la Fig. R4, que assumeix cap barreja d'isospín.

Encara que l'EMN del dipol elèctric puga ser dominant, les distribucions angulars de les transicions $\gamma\gamma$(E1E1), $\gamma\gamma$(M1M1) i de la seua interferència són diferents i, en principi, podrien ser separades experimentalment. D'altra banda, l'operador M1 connecta estats amb la mateixa paritat i un canvi en el moment angular d'una unitat, de manera que el conjunt complet d'estats intermedis són tots els estats amb spin-paritat $J^\pi = 1^+$. A més, per connectar $|0^+_i\rangle_{\gamma\gamma}$ i $|0^+_f\rangle_{\gamma\gamma}$ aquests estats intermedis han de ser els estats analògics isobàrics del nucli intermig amb isospín $T = T_z + 1$, com es mostra a la Fig. R4, que assumeix cap barreja d'isospín. Finalment, l'amplitud de transició per a la desintegració

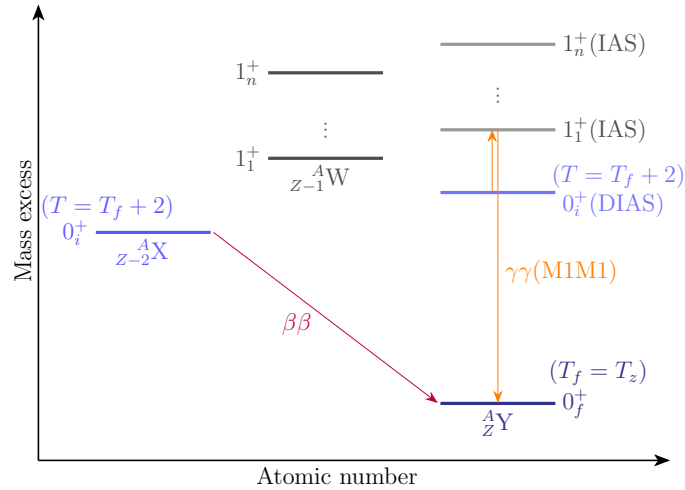

FIG. R4: Els estats nuclears involucrats en les desintegracions $\gamma\gamma$(M1M1) i beta. En blau clar es mostren l'estat beta inicial, $0^+_i$, i el seu estat doble anàleg isobàric $0^+_i$(DIAS), l'estat inicial $\gamma\gamma$. En blau fosc, l'estat final comú beta i $\gamma\gamma$ $0^+_f$. En gris es mostren tots els estats isobàrics analògs $1^+_n$ intermedis, que connecten $0^+_i$(DIAS) i $0^+_f$ a través del component isovector de l'operador dipolar magnètic $\mathbf{M1}$.

$\gamma\gamma$(M1M1) és proporcional a l'EMN $M^{\gamma\gamma}$(M1M1) definit per

$$M^{\gamma\gamma}(M1M1) = \sum_n \frac{\langle 0^+_f ||\mathbf{M1}|| 1^+_n\rangle \langle 1^+_n ||\mathbf{M1}|| 0^+_i\rangle}{E_n - (E_i + E_f)/2}, \tag{R9}$$

on $E_{i,f,n}$ són les energies dels estats nuclears inicial, final i intermedi. Per a fer els EMN de $\gamma\gamma$ independents de les energies dels fotons, restringirem el cas on els dos fotons compareteixen igualment l'energia disponible en la transició, $k_0 = k_0' = Q_{\gamma\gamma}/2$. Com que la probabilitat de transició és simètrica sota $k_0 \leftrightarrow k_0'$, és raonable començar analitzant aquesta situació, que a més és la més probable. Afortunadament, aquesta condició es pot realitzar experimentalment.

Els EMN de $\gamma\gamma$(M1M1) i de desintegració $0\nu\beta\beta$ s'han calculat dins del model de capa nuclear utilitzant els codis del model de capa ANTOINE [200, 202] i NATHAN [202]. Els resultats es mostren a la Figura R5, i cobreixen una gran quantitat de nuclis amb un nombre de massa comprès entre $46 \leq A \leq 136$. Cobreixen tres espais de configuració diferents generats pels següents orbitals de partícules individuals d'oscil·lador harmònic, tant per a protons com per a neutrons, i amb interaccions isospín-simètriques $40f_{7/2}$, $1p_{3/2}$, $0f_{5/2}$ i $1p_{1/2}$ (espai de configuració $pf$): $^{46-58}$Ti, $^{50-58}$Cr i $^{54-60}$Fe amb les interaccions efectives KB3G [169] i GXPF1B [170].

- $40f_{7/2}$, $1p_{3/2}$, $0f_{5/2}$ i $1p_{1/2}$ (espai de configuració $pf$): $^{46-58}$Ti, $^{50-58}$Cr i $^{54-60}$Fe amb les interaccions efectives KB3G [169] i GXPF1B [170].

- $1p_{3/2}$, $0f_{5/2}$, $1p_{1/2}$ i $0g_{9/2}$ ($pfg$ espai de configuració): $^{72-76}$Zn $^{74-80}$Ge, $^{76-82}$Se, $^{82,84}$Kr amb les GCN2850 [171], JUN45 [172] i JJ4BB [173] interaccions.

- $1d_{5/2}$, $0g_{7/2}$, $2s_{1/2}$, $1d_{3/2}$ i $0h_{11/2}$ ($sdgh$ espai de configuració): $^{124-132}$Te, $^{130-134}$Xe i $^{134,136}$Ba amb les GCN5082 [171] and QX [174] interaccions.

La Figura R5 il·lustra la relació entre els NMEs de desintegració $0\nu\beta\beta$ estàndard i els NMEs de dipol magnètic doble $M^{\gamma\gamma}$(M1M1). Mostra una correlació lineal molt bona tant per als nuclis de la capa $pf$, que inclou disset nuclis que contenen isòtops de titani, crom i ferro. A més, la Figura R5 presenta una correlació molt bona tant per als nuclis de les capes $pfg$ com de $sdgh$. Les dades inclouen NMEs per a vint-i-cinc nuclis, incloent isòtops de zinc, germani, seleni, criptó, tel·luri, xenó i bari. Aquests NMEs $M^{\gamma\gamma}$(M1M1) s'han calculat utilitzant els $g$-factors orbitals de gir de partícula lliures, però observem gairebé la mateixa correlació quan fem servir els factors $g$ efectius que estan en un acord lleugerament millor amb els moments i transicions dipolars magnètics experimentals, prenent $g^{s,\text{eff}}i = 0.9g_i^s$, $g^{l,\text{eff}}p = g_p^l + 0.1$, $g_n^{l,\text{eff}} = g_n^l - 0.1$ a la capa $pf$ [261], i $g_i^{s,\text{eff}} = 0.7g_i^s$ per als nuclis $pfg$ [172]. Un tret important és que la correlació és independent de la interacció efectiva utilitzada. La Figura R5 mostra que la correlació lineal és diferent per als nuclis més lleugers que per als més pesats, i això és degut al denominador energètic en $M^{\gamma\gamma}$(M1M1).

La relació entre la desintegració $0\nu\beta\beta$ i les transicions $\gamma\gamma$ M1M1 des de l'estat DIAS fins a l'estat fonamental permet quantificar la incertesa als NMEs de desintegració $0\nu\beta\beta$ a partir d'una mesura hipotètica d'aquesta transició $\gamma\gamma$. Aquesta incertesa i valor mitjà dels

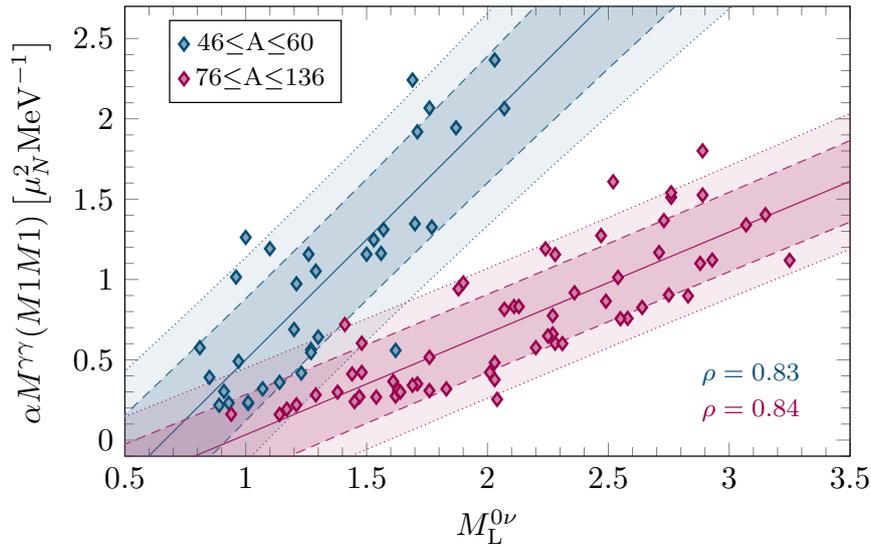

FIG. R5: NMEs estàndard de $0\nu\beta\beta$ ($M_L^{0\nu}$), de la Ref. [147] vs NMEs de doble dipol magnètic ($M^{\gamma\gamma}(M1M1)$) obtinguts amb diferents interaccions per a cada espai de configuració. El factor $\alpha$ prové de comparar els elements de matriu reduïts en l'espai de l'isospí. Les línies sòlides corresponen a la millor ajustada lineal, mentre que les línies puntejades i puntejades representen les bandes de predicció al 68(90)% de nivell de confiança, respectivament.

NMEs de desintegració $0\nu\beta\beta$ predits a partir del millor ajust lineal a $M^{0\nu} = a + bM^{\gamma\gamma}$ es basen en càlculs sistemàtics de diverses interaccions efectives en un ampli rang de la taula nuclear. Aquest tipus d'estudis constitueixen una eina alternativa per millorar la nostra comprensió dels NMEs de desintegració $0\nu\beta\beta$ guiat pels condicionants experimentals. La Figura R6 compara la dispersió actual en els NMEs de la desintegració estàndard $0\nu\beta\beta$ [62] representada per bandes ombrejades, amb els seus valors i incerteses derivats de la correlació i la banda de predicció al 90% de CL de la Figura R5, representats per punts amb barres d'error. A més, també mostrem a la Figura R6 els resultats NSM de la literatura com a creus. Hi ha una bona concordança entre els valors predits a partir de la correlació i els NMEs d'altres mètodes. Els NMEs obtinguts de la correlació NSM són compatibles amb els valors NSM anteriors reportats a la Ref. [62].

Per tant, basant-nos en la correlació obtinguda a partir de desenes de nuclis i diferents interaccions, i assumint un valor hipotètic per al $M^{\gamma\gamma}$(M1M1) experimental amb un error, podríem estimar $M_L^{0\nu}$ amb una incertesa que millora les restriccions respecte a la dispersió en els resultats combinats de molts cossos. Més important encara, això representa una incertesa NSM obtinguda a partir de càlculs sistemàtics per a desenes de nuclis utilitzant diverses interaccions, que és més fiable que una col·lecció de resultats, com s'estimava anteriorment en la literatura.

En relació amb l'estudi anterior, també hem analitzat la correlació entre els NMEs de desintegració $0\nu\beta\beta$ i $2\nu\beta\beta$, sent aquest últim un observable per al qual hi ha dades disponibles. Els resultats d'aquest treball s'han publicat a Ref. [13]. De manera anàloga a la desintegració $\gamma\gamma$, vam calcular amb el model nuclear de capa els dos NMEs per als mateixos nuclis i utilitzant les mateixes interaccions efectives. Vam observar una molt bona correlació lineal entre la component estàndard dels NMEs de desintegració $0\nu\beta\beta$ i $2\nu\beta\beta$. Aquí, les dades disponibles de les mesures de la semivida de la desintegració $2\nu\beta\beta$ s'utilitzen juntament amb el millor ajust lineal i les bandes de predicció al 68% CL per donar un valor dels NMEs de desintegració $0\nu\beta\beta$ amb una incertesa basada en dades experimentals i la sistemàtica de molts NMEs. Es deriven dues conclusions principals d'aquest estudi. Una és que els NMEs predits pels dos mètodes de molts cossos nuclears (model de capa nuclear i QRPA) són consistents quan es consideren les incerteses. L'altra és que tot i que la incertesa en la desintegració $0\nu\beta\beta$ predita és comparable amb la banda compresa per uns quants valors de la literatura, la incertesa derivada d'aquesta correlació inclou de manera més sistemàtica molts càlculs de NME que incorporen també l'efecte d'utilitzar diferents interaccions.

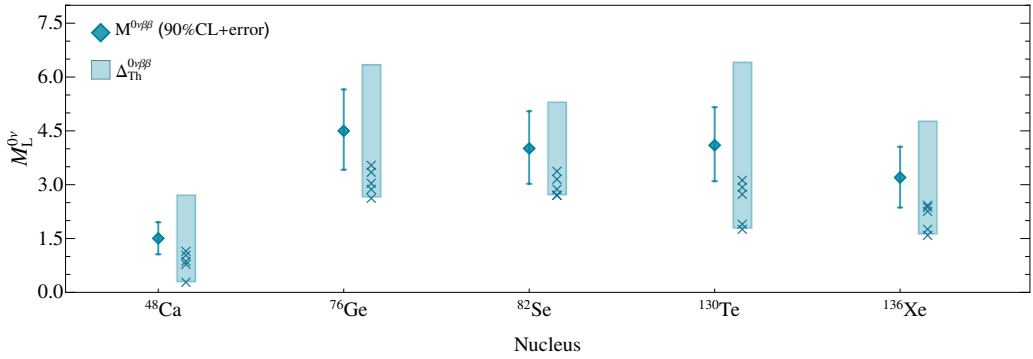

Fig. R6: Els punts de dades amb barres d'error representen els NME de la desintegració beta estàndard ($M^{0\nu}L$) obtinguts de la correlació a la Figura R5 i les bandes de prediccions al 90% de CL assumint un error en la mesura experimental de $\pm15\%$, així com la seva dispersió actual ($\Delta^{0\nu\beta\beta}$Th) reportada a la Ref. [62]. Les creus representen $M^{0\nu}_L$ de NSM, també extretes de la Ref. [62].

A més, la desintegració $2\nu\beta\beta$ de $^{136}$Xe a l'estat excitat primer de $^{136}$Ba ha estat calculada abans de la seva mesura experimental. A la vista del pla experimental per mesurar la semivida d'aquesta desintegració, i a causa de la informació de l'estructura nuclear codificada en aquest observable, el càlcul de molts cossos utilitzat per diferents models nuclears podria ser provat, particularment el mètode de molts cossos utilitzat en aquesta tesi. A més, la semivida ha estat predita per altres mètodes de molts cossos, de manera que les futures dades podrien provar aquests mètodes i proporcionar informació valuosa sobre l'estructura nuclear dels estats nuclears involucrats i, per tant, ser útils per al càlcul dels NMEs de desintegració $0\nu\beta\beta$.

Un objectiu addicional ha estat iniciar la caracterització teòrica dels primers passos cap a la mesura de la proposta desintegració $\gamma\gamma$ doble dipol magnètic des de l'estat doble analògic isobàric. En particular, hem estudiat els principals modes de descomposició que poden competir amb aquest procés: desintegració gamma simple i emissió de protons, i hem calculat les corresponents relacions de ramificació.

Finalment, hem estudiat l'anàlisi teòrica de les principals transicions des de $0^+(DIAS)$ que poden competir amb les transicions $\gamma\gamma$(M1M1). Com que els $0^+(DIAS)$ són estats nuclears d'alta energia que es troben per sobre de les energies de separació dels nucleons, l'emissió de protons és un canal de descomposició obert. No obstant això, és una desintegració prohibida en isospín, el que permet que la branca de decaïment $\gamma$ siga competitiu. Hem calculat les relacions de ramificació de $\gamma\gamma$(M1M1) amb les transicions $\gamma$(M1) en tots els isòtops rellevants per a les cerques de desintegració $\beta\beta$ i hem trobat que $\Gamma_{\gamma\gamma}/\Gamma_\gamma \simeq 10^{-6} - 10^{-8}$. En el cas de l'emissió de protons, només hem informat les relacions de ramificació $\Gamma_{\gamma\gamma}/\Gamma_p \simeq 10^{-8} - 10^{-9}$ per a $^{48}$Ti, per al qual s'ha obtingut un factor espectroscòpic fenomenològic necessari per calcular $\Gamma_p$. Una anàlisi totalment teòric dels processos que poden competir amb $\gamma\gamma$(M1M1), que inclou el càlcul de les transicions de dipol elèctric i la conversió interna de parelles, es deixa per a un futur projecte en col·laboració amb el grup experimental del Laboratori Nazionaali di Legnaro [153].

De la mateixa manera que vam trobar una correlació entre els NMEs de desintegració $0\nu\beta\beta$ i $2\nu\beta\beta$, hi ha una correlació entre els NMEs de desintegració $2\nu\beta\beta$ i $\gamma\gamma$-M1M1 que podria utilitzar-se per obtenir els NMEs i les probabilitats de transició de $\gamma\gamma$(M1M1) a partir de les dades de desintegració $2\nu\beta\beta$. Això proporcionaria informació valuosa ja que proporciona NMEs de desintegració $\gamma\gamma$(M1M1) amb una incertesa que prové de la correlació, i també proporciona una verificació de la consistència de les nostres càlculs individuals amb una interacció efectiva donada.

La bona correlació trobada utilitzant el NSM a Ref. [12] per a la transició $0^+_{DIAS} \rightarrow 0^+_{gs}$, va encoratjar a iniciar l'estudi de la desintegració $\gamma\gamma$ dins del mètode *ab initio* no pertorbatiu conegut com a grup de renormalització de similitud en medi de l'espai valència (VS-IMSRG). A més, els autors de Ref. [150] també exploren aquesta correlació dins del mètode QRPA. Tot i que hem començat amb l'anàlisi de la desintegració $\gamma\gamma$ dins del VS-IMSRG, encara hi ha treball a fer per entendre la interacció entre la barreja d'isospín i alguns problemes de convergència que hem trobat en l'operador $\gamma\gamma$(M1M1). Un altre fet interessant que resta per entendre és per què per a alguns nuclis hi ha una cancel·lació més alta entre les contribucions positives i negatives a l'operador de dipol magnètic. Una vegada entenguem aquestes qüestions, passarem a l'estudi dels NMEs de desintegració $\gamma\gamma$(M1M1) i $0\nu\beta\beta$, i la seua possible correlació dins del mètode VS-IMSRG.

# List of Publications

During this thesis the following articles were published related to the content of the research:

The following paper is based on a project done during 2022 MITP Talent Summer School on "Effective field theories in light nuclei: from structure to reactions":

The following papers based on results presented in this thesis are in preparation: